\documentclass[11pt,a4paper]{article}
\pdfoutput=1

\usepackage[]{Packages}
\RequirePackage{placeins}
\RequirePackage{ragged2e}

\usepackage[mathscr]{euscript}
\allowdisplaybreaks[1]

\newcommand{\OfficialTitle}{Aspects of hot Galilean field theory}

\preprint{YITP-SB-14-27}

\title{\vspace{2cm}
  {\huge   \textbf{\OfficialTitle}}
}

\author{
  \begin{minipage}{.8\linewidth}
    \vspace{1cm}
    \begin{center}
      {\small \textbf{Kristan Jensen}}
    \end{center}
    \vspace{1cm}
    \begin{minipage}{\linewidth}
      {\itshape \footnotesize \begin{center}
       C.N. Yang Institute for Theoretical Physics \\  \vspace{.3cm}
        SUNY Stony Brook, Stony Brook, NY 11794-3840 \end{center}
      }
    \end{minipage}
    \vspace{2cm}
  \end{minipage}
}

\date{\today}

\begin{document}

\setstretch{1.1}

\numberwithin{equation}{section}

\begin{titlepage}

  \maketitle

  \thispagestyle{empty}


  \abstract{\RaggedLeft We reconsider general aspects of Galilean-invariant thermal field theory. Using the proposal of our companion paper, we recast non-relativistic hydrodynamics in a manifestly covariant way and couple it to a background spacetime. We examine the concomitant consequences for the thermal partition functions of Galilean theories on a time-independent, but weakly curved background. We work out both the hydrodynamics and partition functions in detail for the example of parity-violating normal fluids in two dimensions to first order in the gradient expansion, finding results that differ from those previously reported in the literature. As for relativistic field theories, the equality-type constraints imposed by the existence of an entropy current appear to be in one-to-one correspondence with those arising from the existence of a hydrostatic partition function. Along the way, we obtain a number of useful results about non-relativistic hydrodynamics, including a manifestly boost-invariant presentation thereof, simplified Ward identities, the systematics of redefinitions of the fluid variables, and the positivity of entropy production.}

\end{titlepage}

\clearpage

\numberwithin{equation}{section}
\newcommand{\beq}{\begin{equation}}
\newcommand{\eeq}{\end{equation}}

\newcommand{\nG}{\Gamma}
\newcommand{\II}{\mathrm{I}\hspace{-0.8pt}\mathrm{I}}
\newcommand{\TM}{\mathcal{TM}}
\newcommand{\tG}{\tilde{\Gamma}}
\newcommand{\oG}{\mathring{\Gamma}}
\newcommand{\oR}{\mathring{R}}
\newcommand{\oD}{\mathring{D}}
\newcommand{\oT}{\mathring{T}}
\newcommand{\M}{\mathcal{M}}
\newcommand{\N}{\mathcal{N}}
\newcommand{\og}{\mathring{g}}
\renewcommand{\v}{V}

\def\KJ#1{\textbf{[KJ:#1]}}

\tableofcontents

\section{Introduction and summary} 

Quantum field theory at nonzero temperature is a rich subject. Enough cannot be said about its practical importance in most arenas of experimental physics, but in this work we will be concerned with its theoretical status. The basic observable in thermal field theory, the thermal partition function, encodes a host of local and non-local information about the field theory. Its variations give thermal correlators and therefore characterize transport. The insertions of non-local operators like Polyakov loops can serve as order parameters for phase transitions in theories with no continuous symmetries, like the deconfinement transition in pure Yang-Mills theory. More generally still, the partition function is sensitive to the topology of the underlying spacetime on which the hot theory is placed.\footnote{For those enamored of supersymmetric partition functions on compact Euclidean manifolds, it is worth noting that many of these manifolds, like $\mathbb{S}^3$ or $\mathbb{S}^1\times \mathbb{S}^3$, are contractible circle bundles. The SUSY partition function on such a space can be viewed as a thermal partition function, where the Hamiltonian appearing in the Boltzmann weight generates translations along the circle.}

There is a universal sector of couplings on which the thermal partition function can depend. Namely, we can put the theory on a fixed, non-trivial spacetime as encoded by some geometry, and if the theory has any global symmetry currents, then we can couple these to background gauge fields. By geometry, we mean an ordinary metric for relativistic field theories, and someone else which we describe below for non-relativistic systems. Not only are these couplings universal, the conjugate stress tensor and symmetry currents are vital for the study of low-energy physics at $T>0$. In a typical hot many-body system, the low-energy physics is dominated by the dynamics of conserved quantities, whose existence necessitate collective gapless excitations known as the hydrodynamic modes. Indeed, fluid mechanics may be regarded as an (unconventional) effective theory of the hydrodynamic modes in terms of the one-point functions of the stress tensor and symmetry currents. Moreover, the correlations of the stress tensor and currents encode thermal transport, magnetic response, \&c.

It is then natural to study the dependence of the thermal partition function on the spacetime geometry and any background gauge fields. This dependence is constrained by symmetries. Up to anomalies, the partition function does not depend on the choice of coordinates nor on the choice of gauge. This independence leads to Ward identities for the stress tensor and currents, as well as constraints on the functional form of the partition function. Over the past decade, largely inspired by results obtained for theories with a holographic dual, these symmetries have been used to great effect in relativistic field theory at $T>0$ in two complementary ways.

\begin{enumerate}
\item \emph{Hydrodynamics.} Hydrodynamics is the near-universal effective description of thermal field theory, describing the evolution of the collective modes corresponding to the relaxation of conserved quantities. As of the time at which this article was written, relativistic hydrodynamics is an effective theory not quite, but approaching the same footing as Wilsonian effective field theory. In it, the one-point functions of the stress tensor and symmetry currents are expressed in a gradient expansion of the classical hydrodynamic variables and spacetime background~\cite{Bhattacharyya:2008jc,Baier:2007ix}. The Ward identities provide equations of motion which determine the hydrodynamic variables (and so the one-point functions) as functionals of the spacetime background. The constitutive relations are constrained by gauge-invariance, Lorentz invariance, and a local version of the second Law, which amounts to demanding that entropy always increases locally for physical fluid flows. This entropy condition imposes non-trivial equality-type and inequality-type relations on the resulting transport. An example of an equality-type relation is the Einstein relation between electromagnetic and thermal conductivities, whilst an example of an inequality-type relation is the requirement that viscosity is non-negative. At the time of its inception, the entropy condition was physically motivated (see e.g. Landau's discussion~\cite{LL6}), but now is mostly understood from field theory considerations (see especially the next item).

There has been even more success, to which we can hardly do justice. Instead we summarize a few highlights. By paying careful attention to the entropy argument, Son and Surowka~\cite{Son:2009tf} have shown that quantum anomalies modify hydrodynamics at first order in the gradient expansion. The authors of~\cite{Bhattacharya:2011tra} worked out the systematics of hydrodynamic field redefinitions as well as the full theory of first-order dissipative superfluid hydrodynamics. By turning on a slowly varying background metric and/or gauge fields and solving the fluid equations on that background, one accesses the full suite of retarded hydrodynamic correlation functions at the level of linear~\cite{Baier:2007ix} and non-linear~\cite{Moore:2010bu} response.

\item \emph{The hydrostatic partition function.} The thermal partition function in a time-independent, but slowly varying background dramatically simplifies compared to the full partition function of a garden-variety interacting field theory. At $T>0$, most field theories have finite correlation length. In that case the hydrostatic partition function can be expressed locally in a gradient expansion of the background fields~\cite{Jensen:2012jh,Banerjee:2012iz}. This is a more general (and perhaps more precise) version of Luttinger's argument~\cite{Luttinger:1964zz} relating local temperature to an external gravitational potential. The end result is that one can parameterize the thermal partition function in the hydrostatic limit to any finite order in gradients, independent of the details of the microscopic field theory. Matching the response to hydrodynamics, one finds that the two can be matched only if the equality-type conditions are satisfied. This puts the local second Law (and so the interrelations it imposes on transport) on a fairly solid footing, insofar as most of its predictions can instead be derived from the hydrostatic partition function. \end{enumerate}

By way of comparison with its relativistic cousin, non-relativistic fluid mechanics is poorly understood. By non-relativistic fluid mechanics, we mean the textbook hydrodynamics of a Galilean-invariant system as presented in, say, Landau and Lifshitz~\cite{LL6}. At present, it is not known how to systematically enforce the Galilean boost symmetry beyond first order in gradients (or even to enforce the boost symmetry at first order in gradients for a system with broken parity). As far as correlation functions go, the state of the art is the canonical technique of Kadanoff and Martin~\cite{KM}, which only accesses a subset of hydrodynamic correlators. That subset does not include the correlators that characterize Hall transport. It is not understood how to couple the fluid mechanics to a background spacetime. Even the fact that physics ought to be invariant under field redefinitions of the hydrodynamic variables (that is, the act of transforming from one ``hydrodynamic frame'' to another) is rather murky in the textbook treatment. Many of these problems are tied to the fact that the Galilean boost symmetry is not manifest.

We endeavor to modernize non-relativstic fluid mechanics and thereby remedy all of the deficiencies just mentioned.

The crux of our construction is that we do the most obvious thing imaginable: we manifest all of the symmetries of the problem. In our earlier companion paper~\cite{Jensen:2014aia}, largely inspired by Son's work~\cite{Son:2005rv,Son:2008ye,Son:2013rqa,Geracie:2014nka}, we deduced the local symmetries of a Galilean field theory by putting such a theory on a curved background.\footnote{Of course, Son and collaborators were not the only authors who considered this problem. See especially the study of~\cite{Christensen:2013rfa,Christensen:2013lma} which obtained many relevant results for non-relativistic QFT and NC geometry by way of holography. See~\cite{Duval:1983pb,Duval:1984cj,Andreev:2013qsa,Banerjee:2014nja,Brauner:2014jaa} for other work in purely field theory terms. See also~\cite{Gromov:2014vla,Bradlyn:2014wla} which studied the local symmetries of non-relativistic, non-Galilean theories.} These symmetries include a local version of the Galilean boost invariance. While we describe those results in some detail in the next Section, we present the highlights here so that we can summarize what we find for non-relativistic fluid mechanics and thermal partition functions. 

\subsection{Summary of results for fluid mechanics}

Rather than coupling to a metric as one does for a relativistic field theory, a $d$-dimensional Galilean field theory couples to a version of  ``Newton-Cartan'' (NC) geometry. The background fields therein are $(n_{\mu},h_{\mu\nu},A_{\mu})$, where $A_{\mu}$ is a $U(1)$ gauge field which couples to the particle number current, $n_{\mu}$ effectively defines a local time direction, and $h_{\mu\nu}$ is a rank$-(d-1)$ spatial metric. We demand that
\beq
\gamma_{\mu\nu} = n_{\mu}n_{\nu}+h_{\mu\nu}\,.
\eeq
is positive-definite. One can thereby obtain the contravariant data $(v^{\mu},h^{\mu\nu})$ satisfying
\beq
n\cdot v = 1\,, \qquad h^{\mu\nu}n_{\nu}=0\,, \qquad h_{\mu\nu}v^{\nu} = 0\,, \qquad h_{\mu\rho}h^{\nu\rho}=P^{\nu}_{\mu}=\delta_{\mu}^{\nu} - v^{\nu}n_{\mu}\,.
\eeq
Because there is no underlying metric, one has to carefully distinguish between lower-index (covariant) and upper-index (contravariant) tensors. Throughout this work, we will sometimes raise indices, always doing so with the ``spatial co-metric'' $h^{\mu\nu}$.

In constructing a Galilean theory, one obviously demands invariance under $U(1)$ gauge transformations and changes of coordinates. In addition, we impose a local boost symmetry, known as invariance under ``Milne boosts'' in a subset of the NC literature~\cite{Duval:1983pb}. Under it, $h_{\mu\nu}$ and $A_{\mu}$ shift as
\beq
h_{\mu\nu} \to h_{\mu\nu} - (n_{\mu}\psi_{\nu} + n_{\nu}\psi_{\mu}) + n_{\mu}n_{\nu}\psi^2\,, \qquad A_{\mu} \to A_{\mu} + \psi_{\mu} - \frac{1}{2}n_{\mu}\psi^2\,,
\eeq
where $\psi_{\mu}$ is an arbitrary one-form satisfying $v^{\mu}\psi_{\mu}=0$. Here $\psi^2 = \psi_{\mu}\psi^{\mu}=\psi_{\mu}\psi_{\nu}h^{\mu\nu}$. As far as upper index data goes, $v^{\mu}$ shifts as 
\beq
v^{\mu} \to v^{\mu} + \psi^{\mu}\,, 
\eeq
and $h^{\mu\nu}$ is invariant. One can readily check that the simplest Galilean field theory,
\beq
S = \int d^dx \sqrt{\gamma} \left\{ \frac{i v^{\mu}}{2}\left( \Psi^{\dagger}D_{\mu}\Psi - (D_{\mu}\Psi^{\dagger})\Psi\right) - \frac{h^{\mu\nu}}{2m}D_{\mu}\Psi^{\dagger}D_{\nu}\Psi + V(\Psi^{\dagger}\Psi)\right\}\,,
\eeq
with $D_{\mu}\Psi = (\partial_{\mu} - i m A_{\mu})\Psi$, is invariant under all of these symmetries.

It is easy to derive Ward identities from these symmetries~\cite{Jensen:2014aia,Geracie:2014nka}. One first defines the various symmetry currents through variation of the partition function with respect to the various background fields, e.g. the number current $J^{\mu}$ is conjugate to $A_{\mu}$. In our conventions, there is also an energy current $\mathcal{E}^{\mu}$, a spatial momentum current $\mathcal{P}_{\mu}$, and a spatial stress tensor $T_{\mu\nu}$. The latter are spatial insofar as $\mathcal{P}_{\mu}v^{\mu}=0$ and $T_{\mu\nu}v^{\nu} = 0$. The $U(1)$ and coordinate reparameterization invariance lead to conservation equations for $J^{\mu}$, the energy current, and $T_{\mu\nu}$, while the Milne Ward identity is
\beq
\mathcal{P}_{\mu}=h_{\mu\nu}J^{\nu}\,.
\eeq
This establishes the folklore theorem that momentum equals particle number current.

Now for hydrodynamics. Revisiting thermal field theory in Subsection~\ref{S:euclidean}, we show that the flat-space thermal equilibria of a Galilean theory are characterized by a temperature $T$, a chemical potential $\mu$ for particle number, and a boost-invariant fluid velocity $u^{\mu}$ satisfying $u^{\mu}n_{\mu}=1$. In flat space with $n = dx^0$, the velocity is just $u^{\mu}\partial_{\mu} = \partial_0 + u^i \partial_i$ where $u^i$ is the usual spatial fluid velocity one finds in Landau~\cite{LL6}. In fluid mechanics, one promotes these variables $(T,\mu,u^{\mu})$ to classical fields known as the fluid variables. We can also put the theory on a slowly varying spacetime background. The one-point functions $(J^{\mu},\mathcal{E}_{\mu},\mathcal{P}_{\mu},T_{\mu\nu})$ are then given in terms of $(T,\mu,u^{\mu})$, the spacetime background $(n_{\mu},h_{\mu\nu},A_{\mu})$, and derivatives of both. These expressions are known as constitutive relations. In supplying the constitutive relations, one works in a gradient expansion, and the term $n^{th}$ order hydrodynamics refers to the case where the constitutive relations have been specified to $\mathcal{O}(\partial^n)$. The fluid variables are then eliminated by demanding that $J^{\mu}, \mathcal{E}^{\mu}$, \&c satisfy the Ward identities, which are in one-to-one correspondence with the $(T,\mu,u^{\mu})$. In this way, the Ward identities give equations of motion for the fluid variables.

There are two important steps that remain. The first is to specify the constitutive relations in such a way that they are $U(1)$-invariant and transform as they ought under boosts and changes of coordinates. The second is to impose a local version of the second Law for fluid flows that solve the Ward identities. The trick in both cases is to manifest the Milne symmetry.

While $(\mathcal{E}^{\mu},\mathcal{P}_{\mu},T_{\mu\nu})$ all transform under the boosts, we find invariant combinations in hydrodynamics. First, we can form a boost-invariant spacetime stress tensor $\mathcal{T}^{\mu\nu}$
\beq
\mathcal{T}^{\mu\nu} = T^{\mu\nu} + v^{\mu}\mathcal{P}^{\nu} + v^{\nu}\mathcal{P}^{\mu} + v^{\mu}v^{\nu}n_{\rho}J^{\rho}\,,
\eeq
where indices are raised with $h^{\mu\nu}$. $\mathcal{T}^{\mu\nu}$ contains all of the components of $(J^{\mu},T_{\mu\nu})$. However, just using the currents and the NC data, there is no way to obtain a boost and $U(1)$-invariant version of the energy current. In fluid mechanics, we have more options. Using that $u^{\mu}$ is boost-invariant, we show that the energy current can be expressed as
\beq
\mathcal{E}^{\mu} = \tilde{\mathcal{E}}^{\mu} + \left( u_{\nu} - \frac{1}{2}n_{\nu}u^2\right)\mathcal{T}^{\mu\nu}\,,
\eeq
where $u_{\mu}=h_{\mu\nu}u^{\nu}$, $u^2 = u_{\mu}u^{\mu}$, and $\tilde{\mathcal{E}}^{\mu}$ is boost-invariant. The Ward identities can be expressed in a completely boost-invariant way in terms of $\mathcal{T}^{\mu\nu}$, $\tilde{\mathcal{E}}^{\mu}$, and an appropriate, boost-invariant definition of a covariant derivative $\tilde{D}_{\mu}$. The connection $\tilde{\Gamma}^{\mu}{}_{\nu\rho}$ is given in~\eqref{E:milneD}, and the Ward identities in~\eqref{E:ward2}

One can then specify constitutive relations in a way that automatically implements the boost symmetry by expressing $\mathcal{T}^{\mu\nu}$ and $\tilde{\mathcal{E}}^{\mu}$ in a basis of boost and $U(1)$ invariant tensors built from the fluid variables and spacetime background. For example, ideal non-relativistic hydrodynamics, that is fluid mechanics to $\mathcal(\partial^0)$, is just
\beq
\label{E:summaryIdealHydro}
\mathcal{T}^{\mu\nu} = P h^{\mu\nu} + \rho u^{\mu}u^{\nu} + \mathcal{O}(\partial)\,, \qquad \tilde{\mathcal{E}}^{\mu} = \varepsilon u^{\mu} + \mathcal{O}(\partial)\,,
\eeq
where $P$ is the pressure, $\rho$ the number density, and $\varepsilon$ the energy density satisfying
\beq
dP = s dT + \rho d\mu\,, \qquad \varepsilon = - P + T s + \mu \rho\,.
\eeq
The physical energy current $\mathcal{E}^{\mu}$ constructed from this data is
\beq
\mathcal{E}^{\mu} = \left( \varepsilon + \frac{1}{2}\rho u^2\right) u^{\mu} + P\, P_{\nu}^{\mu} u^{\nu} + \mathcal{O}(\partial)\,.
\eeq
In flat space with $n = dx^0$, $h_{\mu\nu}dx^{\mu}dx^{\nu} = \delta_{ij}dx^idx^j$ and $u^{\mu}\partial_{\mu}=\partial_0 + u^i\partial_i$, this energy current is the usual one found in e.g. Chapter 1 of Landau~\cite{LL6},
\beq
\mathcal{E}^0 = \varepsilon + \frac{1}{2}\rho u^2 + \mathcal{O}(\partial)\,, \qquad \mathcal{E}^i = \left( \varepsilon + P + \frac{1}{2}\rho u^2\right)u^i + \mathcal{O}(\partial)\,.
\eeq
One can also deduce the number current and spatial stress tensor from $\mathcal{T}^{\mu\nu}$, and in flat space they coincide with the textbook expressions. That is,~\eqref{E:summaryIdealHydro} repackages standard ideal fluid mechanics in a way that manifests all of the symmetries.

With this machinery in hand, we discuss various systematic aspects of fluid mechanics in Section~\ref{S:NRhydro}, including the local second Law and the role of field redefinitions of the $(T,\mu,u^{\mu})$. We go on to construct first-order normal fluid mechanics for parity-preserving theories in Subsection~\ref{S:1stOrder}. Our result ends up just being a covariant version of the textbook presentation~\cite{LL6}. The one-derivative transport is all dissipative and includes a bulk viscosity $\zeta$, a shear viscosity $\eta$, and a thermal conductivity $\kappa$.

Life gets more interesting when we move on to the fluid mechanics of parity-violating systems in two spatial dimensions in Section~\ref{S:2dFluids}. (Some time ago, we and collaborators studied the corresponding problem for relativistic fluids in~\cite{Jensen:2011xb}. Our results here are eerily similar upon translation.) We have in mind fluids of chiral molecules, anyonic fluids, \&c, in which parity is broken even in the absence of a magnetic field. We may as well stress here that our analysis only holds for systems subjected to a $\mathcal{O}(\partial)$ magnetic field, and so does not apply to quantum Hall states. We solve the positivity of entropy condition to deduce the constraints on the constitutive relations. We then find that there are several non-dissipative transport coefficients which are allowed by the local second Law. 

For a particular definition of the fluid variables known as the Eckhart frame, the constitutive relations are
\begin{subequations}
\label{E:summary1stOrder}
\begin{align}
\begin{split}
\tilde{\mathcal{E}}^{\mu} & = \varepsilon u^{\mu} + \eta^{\mu}\,,
\\
\mathcal{T}^{\mu\nu}& = \rho u^{\mu}u^{\nu} + \mathcal{P} h^{\mu\nu} + \tau^{\mu\nu}\,,
\end{split}
\end{align}
where $\eta^{\mu}$ and $\tau^{\mu\nu}$ are transverse to $n_{\mu}$ and $\tau^{\mu\nu}$ is traceless. Then we find
\begin{align}
\begin{split}
\mathcal{P} &= P - \zeta \vartheta + \tilde{\chi}_B \mathcal{B} + \tilde{\chi}_n \mathcal{B}^n + \mathcal{O}(\partial^2)\,,
\\
\eta^{\mu} & = - \kappa  U^{\mu} - \tilde{\kappa} \varepsilon^{\mu\nu\rho}n_{\nu}U_{\rho} - \frac{\varepsilon + P}{\rho}\left( \tilde{\chi}_E \varepsilon^{\mu\nu\rho}n_{\nu}E_{\rho} + \tilde{\chi}_T \varepsilon^{\mu\nu\rho}n_{\nu}\partial_{\rho}T\right)+ \mathcal{O}(\partial^2) \,,
\\
\tau^{\mu\nu}& = - \eta \sigma^{\mu\nu} - \tilde{\eta} \tilde{\sigma}^{\mu\nu} + \mathcal{O}(\partial^2)\,,
\end{split}
\end{align}
\end{subequations}
and we have to unpack the notation. We use $u^{\mu}$ to define a boost-invariant version of $A_{\mu}$,
\beq
\tilde{A}_{\mu} = A_{\mu} + u_{\mu} - \frac{1}{2}n_{\mu} u^2\,,
\eeq
whose field strength is $\tilde{F}_{\mu\nu}$, along with a boost-invariant derivative $\tilde{D}_{\mu}$ through a boost-invariant connection $\tilde{\Gamma}^{\mu}{}_{\nu\rho}$ in~\eqref{E:milneD}. Then the various tensors in~\eqref{E:summary1stOrder} are
\begin{subequations}
\begin{align}
E_{\mu} &= \tilde{F}_{\mu\nu}u^{\nu}\,, & \mathcal{B} &= \frac{1}{2}\varepsilon^{\mu\nu\rho}n_{\mu}\tilde{F}_{\nu\rho}\,,
\\
E^n_{\mu} & = \left( \partial_{\mu}n_{\nu}-\partial_{\nu}n_{\mu}\right)u^{\nu}\,, & \mathcal{B}^n & = \varepsilon^{\mu\nu\rho}n_{\mu}\partial_{\nu}n_{\rho}\,,
\\
\vartheta &= \tilde{D}_{\mu}u^{\mu}\,,& \sigma^{\mu\nu} &= \frac{1}{2}\left( \tilde{D}^{\mu}u^{\nu} + \tilde{D}^{\nu}u^{\mu} - \frac{2}{d-1}h^{\mu\nu}\vartheta\right)\,,
\\
U_{\mu}& = \left( \partial_{\mu}+E^n_{\mu} \right) T\,, &\tilde{\sigma}^{\mu\nu} &= \frac{1}{2}\left( \varepsilon^{\mu\rho\sigma}n_{\rho} \sigma_{\sigma}^{\nu} + \varepsilon^{\nu\rho\sigma}n_{\rho}\sigma_{\sigma}^{\mu}\right)\,.
\end{align}
\end{subequations}
Here $\varepsilon^{\mu\nu\rho}= \frac{\epsilon^{\mu\nu\rho}}{\sqrt{\gamma}}$ with $\epsilon^{\mu\nu\rho}$ the epsilon symbol, the shear tensor $\sigma^{\mu\nu}$ has the property that $\sigma^{\mu\nu}n_{\nu}=0$, and in the last line the index of $\sigma^{\mu\nu}$ is lowered with $h_{\mu\nu}$.

The local second Law enforces
\beq
\zeta \geq 0\,, \qquad \kappa\geq 0\,, \qquad \eta \geq 0\,.
\eeq
The dissipationless transport coefficients in~\eqref{E:summary1stOrder} are the Hall viscosity $\tilde{\eta}$, anomalous thermal Hall conductivity $\tilde{\kappa}$, a magnetic susceptibility $\mathcal{M}$ and an ``energy magnetic'' susceptibility $\mathcal{M}_n$, all of which are unconstrained by our analysis
\beq
\tilde{\eta} \in \mathbb{R}\,, \qquad \tilde{\kappa}\in \mathbb{R}\,, \qquad \mathcal{M}\in \mathbb{R}, \qquad \mathcal{M}_n \in \mathbb{R}\,.
\eeq
The remaining response coefficients ($\tilde{\chi}_B, \tilde{\chi}_n, \tilde{\chi}_E,\tilde{\chi}_T)$ are determined by $\mathcal{M}$ and $\mathcal{M}_n$ as
\begin{align}
\begin{split}
\label{E:tildeChis}
\tilde{\chi}_B & = \left( \frac{\partial P}{\partial\varepsilon}\right)_{\rho}\left( T \frac{\partial\mathcal{M}}{\partial T}+ \mu \frac{\partial \mathcal{M}}{\partial \mu}-\mathcal{M}\right) + \left( \frac{\partial P}{\partial\rho}\right)_{\varepsilon} \frac{\partial\mathcal{M}}{\partial\mu}\,,
\\
\tilde{\chi}_n & =\left(  \frac{\partial P}{\partial\varepsilon}\right)_{\rho}\left( T \frac{\partial\mathcal{M}_n}{\partial T}+ \mu \frac{\partial\mathcal{M}_n}{\partial\mu}-2\mathcal{M}_n\right) + \left( \frac{\partial P }{\partial\rho} \right)_{\varepsilon}\left( \frac{\partial\mathcal{M}_n}{\partial\mu}+ \mathcal{M}\right)\,,
\\
\tilde{\chi}_E & = - \left\{ \frac{\partial\mathcal{M}}{\partial\mu} + R \left( \frac{\partial \mathcal{M}_n}{\partial\mu}+ \mathcal{M}\right)\right\}\,,
\\
T \tilde{\chi}_T & = - \left\{ T \frac{\partial\mathcal{M}}{\partial T}+ \mu\frac{\partial\mathcal{M}}{\partial\mu}-\mathcal{M} + R \left( T\frac{\partial\mathcal{M}_n}{\partial T} + \mu \frac{\partial\mathcal{M}_n}{\partial\mu}-2\mathcal{M}_n\right)\right\}\,,
\end{split}
\end{align}
where $R = \frac{\rho}{\varepsilon + P}$, and derivatives with respect to $T$ and $\mu$ are taken at fixed $\mu$ or $T$. Were it not for the entropy condition, the $\tilde{\chi}$'s would be unconstrained, however the entropy current imposes the equality-type relations~\eqref{E:tildeChis} among them. $\mathcal{M}$ and $\mathcal{M}_n$ can be understood as the response of the partition function to a magnetic field $\mathcal{B}$ or ``energy magnetic field'' $\mathcal{B}^n$ in the source-free thermal state. In this sense, $\mathcal{M}$ and $\mathcal{M}_n$ are the magnetization and ``energy magnetization'' of the state in zero magnetic field.

The Hall viscosity $\tilde{\eta}$ and anomalous thermal Hall conductivity $\tilde{\kappa}$ characterize dissipationless, out-of-equilbrium transport. However, the parameters $\mathcal{M}$ and $\mathcal{M}_n$ are hydrostatic data. The corresponding response coefficients $(\tilde{\chi}_B,\tilde{\chi}_n,\tilde{\chi}_E,\tilde{\chi}_T)$ contribute to zero-frequency, low-momentum correlation functions. While we have the tools available, we postpone the computation of Kubo formulae and hydrodynamic correlators for future work.

There is a great deal of physics to unpack in the first-order transport~\eqref{E:summary1stOrder}. In the interest of brevity, we limit ourselves to the following observation. In gapped relativistic field theory at $T=0$, the partition function can be expressed locally in a gradient expansion of the background fields. One can have Chern-Simons terms, and the corresponding Chern-Simons levels are physical modulo an integer, as the addition of local counterterms can only modify the level by an integer. Moreover, the level does not depend analytically on coupling constants that do not close the gap. An electromagnetic Chern-Simons term then leads to a Hall conductivity and zero-field magnetization which are equal and moreover do not depend smoothly on coupling constants. However, at $T>0$ both of these properties simply disappear. This is already visible in hydrodynamics~\cite{Jensen:2011xb} -- the Hall conductivity and magnetization are independent transport coefficients which can vary with $T$ and $\mu$ -- but one can also understand it from effective field theory. At $T>0$ there is no longer a gap owing to the hydrodynamic modes, and so the partition function can no longer be written in a gradient expansion of the background. The local object is the Schwinger-Keldysh effective action for the hydrodynamic modes. So much for the Hall conductivity. In a screened phase, the magnetization can be computed from the hydrostatic partition function, which can be written locally on the two-dimensional spatial slice. So the magnetization is no longer related to a Chern-Simons level, and it typically varies smoothly with coupling constants. For example, the magnetization of the theory of a massive Dirac fermion goes like $\tanh(m/T)$.

A similar set of statements seems to hold in non-relativistic field theory. In gapped Galilean theories at $T=0$, there are also Chern-Simons terms in the partition function~\cite{Wen:1992ej} which govern the electromagnetic Hall conductivity and the Hall viscosity~\cite{WZtoHall}. As in the relativistic case, the levels are physical modulo an integer and do not depend smoothly on coupling constants. However, in hydrodynamics we see that see that the Hall response can smoothly depend on coupling constants and the state. We can understand this from thermal field theory in the same way as above.

There has been some previous work studying parity-violating non-relativistic fluid mechanics in two spatial dimensions. The Hall viscosity was identified long ago~\cite{Avron:1995fg,1997physics..12050A} in the context of quantum Hall states. More recently there have been three papers which have studied this problem more systematically~\cite{Kaminski:2013gca,Banerjee:2014mka,Geracie:2014zha}. We postpone a detailed comparison until the Appendix, wherein we find disagreement with the results of Kaminski and Moroz~\cite{Kaminski:2013gca} and trivial agreement with Banerjee, et al~\cite{Banerjee:2014mka} when the only global symmetry is particle number. Geracie and Son~\cite{Geracie:2014zha} have recently studied the magnetohydrodynamics of the lowest Landau level, which is not analytically related to our work for the usual reason that the $B\to 0$ and low energy limits do not commute.

\subsection{Summary of results for thermal partition functions}

Above, we referred to the results obtained in~\cite{Jensen:2012jh,Banerjee:2012iz} for the hydrostatic partition function of relativistic field theory. By ``hydrostatic,'' we mean the thermal partition function of a theory in a time-independent background which varies over long wavelengths. In Section~\ref{S:thermal} we perform the corresponding analysis for Galilean field theories.

\begin{figure}[t]
\begin{center}
\includegraphics[width=3in]{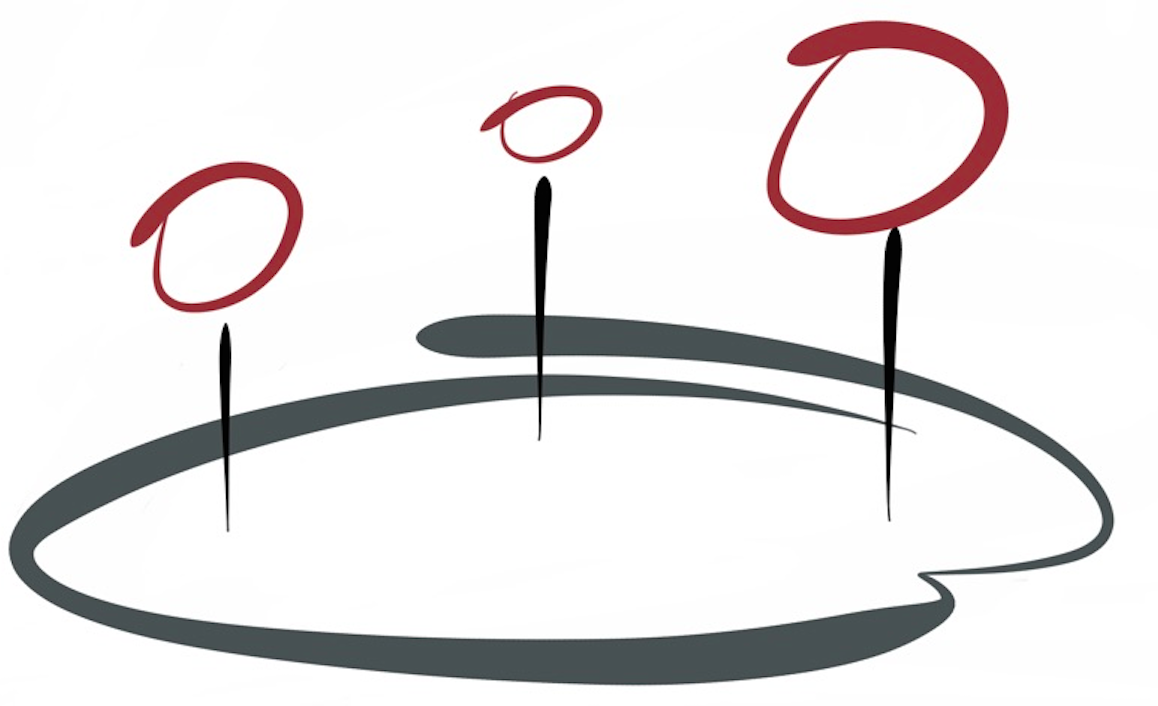}
\caption{\label{F:fiber} A depiction of the Euclidean manifolds from which we construct thermal field theory. There is a thermal circle at each point on the spatial slice, whose size and orientation vary smoothly as one moves around in space. The spacetime is invariant under uniform translation along the circle. The total structure is that of a (thermal) circle bundle.}
\end{center}
\end{figure}

We set the stage in Section~\ref{S:euclidean}, where we review how the usual sum over states in the thermal ensemble is related to Euclidean field theory in flat space. The idea is simple: given a real-time, time-independent background, one can form a thermal partition function using the operator $\mathcal{H}_{\tau}$ that generates time translations in that background. For theories with a functional integral description, this partition function can be mapped to a functional integral over an analytic continuation of the original spacetime. In coordinates where time is $x^0$, one just analytically continues it as $x^0 = - i \tau_E$ and then compactifies $\tau_E \sim \tau_E + \beta$. These ``Euclidean'' spacetimes have the topology of a fiber bundle, wherein the thermal circle may be fibered non-trivially over a spatial base. See Fig.~\ref{F:fiber}.

We mostly work with a covariant version of this construction, like that in~\cite{Jensen:2012jh}, but we also show how the NC background and symmetries work in an explicitly time-independent gauge as in~\cite{Banerjee:2012iz}. Each approach has its relative advantages and disadvantages, but both are equivalent when it comes to describing the hydrostatic partition function. In each case, one can form a local temperature $T$, local chemical potential $\mu$, and local fluid velocity $u^{\mu}$ from the spacetime: $T$ is the inverse integral of $n_{\mu}$ around the thermal circle, $\mu/T$ is the logarithm of the boost-invariant holonomy of $A_{\mu}$ around the circle, and $u^{\mu}$ is a normalized, analytically continued tangent vector to the circle.

Most field theories at $T>0$ are screened, meaning that the correlation length is finite. This in turn implies that the thermal partition function, as a functional of time-independent couplings, can be expressed locally in a gradient expansion. It can be written either in an explicitly time-independent gauge on the spatial slice, or covariantly on the Euclidean spacetime provided one writes terms using the background fields as well as $(T,\mu,u^{\mu})$. Using this result, we go on in Subsections~\ref{S:Z0derivative} and~\ref{S:Z1derivative} to parameterize the most general possible zero and one-derivative terms that may appear in the hydrostatic partition function, as well as the corresponding stress tensor and energy current. We term this stress tensor and energy current the hydrostatic response -- it is dissipationless response which is encoded in Euclidean field theory. Anyway, the zero-derivative part just recovers ideal non-relativistic hydrodynamics in curved spacetime~\eqref{E:summaryIdealHydro}. The one-derivative part vanishes in a parity-preserving theory, but can be non-vanishing when parity is violated. This recovers the usual statement that there is no parity-preserving dissipationless response in first-order hydrodynamics.

In two spatial dimensions, there are two functions of state which characterize the one-derivative term in the partition function. By computing the stress tensor and energy current and matching to hydrodynamics, we find that these functions are exactly $\mathcal{M}$ and $\mathcal{M}_n$ that appeared in our summary~\eqref{E:summary1stOrder} and~\eqref{E:tildeChis} of hydrodynamics in this setting. The Hall viscosity $\tilde{\eta}$ and anomalous thermal Hall conductivity $\tilde{\kappa}$ characterize out-of-equilibrium dissipationless transport, and therefore are unconstrained by hydrostatics.

Recall that the $\tilde{\chi}$'s would be free functions of state were it not for the entropy condition, which determines them in terms of $\mathcal{M}$ and $\mathcal{M}_n$ via~\eqref{E:tildeChis}. Combining this with the previous paragraph, hydrodynamics matches the partition function only if one demands the equality conditions mandated by the local second Law. We conjecture that this is true in general. Just as in the relativistic setting (see especially~\cite{Bhattacharyya:2013lha}), this provides strong evidence that the local second Law is bona fide rather than ad hoc. 

In three spatial dimensions, we find that there are three terms which can appear at first order in gradients. Each of these is a Chern-Simons on the spatial slice, so the couplings are pure numbers rather than functions of state. These Chern-Simons coefficients lead to one-derivative hydrostatic response which is exactly fixed by the Chern-Simons couplings and thermodynamics. Presumably this implies that, rather like relativistic hydrodynamics with anomalies~\cite{Son:2009tf}, non-relativistic hydrodynamics may be modified at first order in gradients by transport which is determined by pure numbers.

In Subsection~\ref{S:superfluid}, we describe how all of this technology can be used to describe the zero-energy, low-momentum effective action in a superfluid phase. This is the hydrostatic effective action for the Goldstone mode of the superfluid. We classify the effective action at zeroth order in gradients, from which we obtain a covariant version of ideal superfluid hydrodynamics. Finally, in Subsection~\ref{S:spin} we sketch out how the introduction of a magnetic moment, which modifies the action of the Milne boosts, can be incorporated in the hydrostatic partition function.

\subsection{The plan}

The outline for the rest of this article is as follows. We summarize prerequisite material in Section~\ref{S:primer}, including the salient results of our previous work in Subsections~\ref{S:NCgeometry} and~\ref{S:symmetries}. In Subsection~\ref{S:milneD}, we develop the technology to express derivatives, one-point functions, and Ward identities in a manifestly boost-invariant way.

Section~\ref{S:Tneq0} lays the foundation for the rest of the paper. We revisit the Euclidean thermal field theory of Galilean systems in Subsection~\ref{S:euclidean}. We go on to set up the constitutive relations in a gradient expansion and classify boost-invariant tensors with one derivative.

The heart of our analysis is found in Section~\ref{S:NRhydro}, where we establish the hydrodynamics of normal fluids at zeroth and first order in the gradient expansion. Along the way we obtain a number of useful results, including a discussion of hydrodynamic field redefinitions, the form of the canonical entropy current, and a modern presentation of the local second Law.

As a non-trivial application of these tools, in Section~\ref{S:2dFluids} we classify the hydrodynamics of parity-violating normal fluids at first order in gradients. The parity-violating response is dissipationless, and to constrain it we employ a version of the local second Law known as the adiabaticity condition (which in the relativistic case first appeared in~\cite{Loganayagam:2011mu}). We thereby classify the parity-violating transport we already summarized in~\eqref{E:summary1stOrder}.

Section~\ref{S:thermal} is a paper within the paper. We systematically study the thermal partition function in the hydrostatic limit to first order in gradients, and compute the corresponding hydrostatic response. We also show that this technology characterizes hydrostatics in a superfluid phase. We conclude with some future directions in Section~\ref{S:discuss} and compare with previous work in the Appendix.

Without any further ado, let us put our hands to the plow and not turn back.

\section{A primer to Galilean-invariant field theory in curved spacetime} 
\label{S:primer}

We begin by recapitulating the proposal made in~\cite{Jensen:2014aia} by which Galilean theories ought to be coupled to $\M$. Newton-Cartan geometry is an integral ingredient of that proposal, and since this geometry is rather esoteric to most theorists, we also review it here. Refs.~\cite{Kunzle1972,Duval:1984cj,2009JPhA...42T5206D} were rather useful when learning Newton-Cartan geometry and we refer them to the interested reader. 

To our knowledge, the first place where the torsional Newton-Cartan geometry was properly discussed was in the holographic analysis of~\cite{Christensen:2013rfa}. The results for the defining data, connection, \&c obtained there agree with those below up to translation. However, the Milne boosts are absent from their discussion.

\subsection{Newton-Cartan geometry and Milne boosts} 
\label{S:NCgeometry}

There are various notions of Newton-Cartan geometry. Here we present the one relevant for the proposal of~\cite{Jensen:2014aia}. Given a smooth orientable $d$-dimensional manifold $\M$, we define a Newton-Cartan geometry on it by equipping $\M$ with various tensor fields. There are various equivalent ways of parameterizing the tensor data. One is to equip $\M$ with $(n_{\mu},h_{\mu\nu},A_{\mu})$, where $A_{\mu}$ is a $U(1)$ connection. Here $n_{\mu}$ is a nowhere-vanishing one-form and $h_{\mu\nu}$ is a symmetric rank-$(d-1)$ tensor. $n_{\mu}$ and $h_{\mu\nu}$ are almost, but not quite arbitrary. We demand that $h_{\mu\nu}$ is positive semi-definite and that the tensor
\beq
\gamma_{\mu\nu} \equiv n_{\mu}n_{\nu}+h_{\mu\nu}\,,
\eeq
is rank$-d$ and so invertible. Equivalently, we demand that $\gamma_{\mu\nu}$ is positive-definite and so equips $\mathcal{M}$ with a Riemannian metric.\footnote{This metric is, sadly, rather useless in most of Newton-Cartan geometry, as it varies under the boost symmetry we discuss below. Its determinant, however, is boost-invariant.} Denoting its inverse as $\gamma^{\mu\nu}$, we define the contravariant tensors $v^{\mu}$ and $h^{\mu\nu}$ via 
\beq
v^{\mu} =\gamma^{\mu\nu}n_{\nu}\,, \qquad h^{\mu\nu} = \gamma^{\mu\nu}-v^{\mu}v^{\nu}\,
\eeq
which then satisfy (on account of $\gamma^{\mu\nu}n_{\mu}n_{\nu}=1$)
\beq
\label{E:NCdefn}
n_{\mu}v^{\mu} = 1\,, \qquad h_{\mu\nu}v^{\nu} = 0\,, \qquad h^{\mu\nu}n_{\nu} = 0\,, \qquad h^{\mu\rho}h_{\nu\rho} = P_{\nu}^{\mu} = \delta^{\mu}_{\nu} - v^{\mu}n_{\nu}\,.
\eeq
The one-form $n_{\mu}$ effectively defines a local time direction, and $h_{\mu\nu}$ may be regarded as a metric on spatial slices. Alternatively, one can treat $(v^{\mu},h^{\mu\nu},A_{\mu})$ as the defining data of Newton-Cartan geometry, form the invertible $\gamma^{\mu\nu}=v^{\mu}v^{\nu}+h^{\mu\nu}$, and then reconstruct $n_{\mu}$ and $h_{\mu\nu}$.\footnote{We thank Andreas Karch for discussions on this point.}

One immediate benefit of the Newton-Cartan structure is that it provides a local expression for the volume form on $\M$, by regarding $\gamma_{\mu\nu}$ as a Riemannian metric. Defining 
\begin{equation*}
\gamma = \text{det}(\gamma_{\mu\nu})\,,
\end{equation*}
the volume form on $\M$ is
\beq
\text{vol}(\M) = \frac{1}{d!}\varepsilon_{\mu_1\hdots \mu_d}dx^{\mu_1}\wedge \hdots \wedge dx^{ \mu_d}\,, \qquad \varepsilon_{\mu_1\hdots \mu_d} = \sqrt{\gamma} \epsilon_{\mu_1\hdots\mu_d}\,,
\eeq
with $\epsilon_{\mu_1\hdots \mu_d}$ the epsilon symbol satisfying $\epsilon_{01\hdots d-1} = +1$. More concisely, the covariant measure appearing in integrals is $d^dx \sqrt{\gamma}$. 

One can uniquely define a covariant derivative from this data, in analogy with the uniqueness of the Levi-Civita connection in Riemannian geometry. Our convention is that the covariant derivative acts on e.g. a $(1,1)$ tensor $\mathfrak{T}^{\mu}{}_{\nu}$ as
\beq
D_{\mu} \mathfrak{T}^{\nu}{}_{\rho} = \partial_{\mu} \mathfrak{T}^{\nu}{}_{\rho} + \Gamma^{\nu}{}_{\sigma\mu} \mathfrak{T}^{\sigma}{}_{\rho} - \Gamma^{\sigma}{}_{\rho\mu} \mathfrak{T}^{\nu}{}_{\sigma}\,.
\eeq
To fix the derivative, one first demands the covariant constancy of $(n_{\mu},h^{\nu\rho})$,
\beq
\label{E:nhConstant}
D_{\mu}n_{\nu} = 0\,, \qquad D_{\mu}h^{\nu\rho} = 0\,,
\eeq
along with that the geodesic acceleration and curl of $v^{\mu}$ are determined via
\beq
\label{E:Dv}
v^{\nu}D_{\nu} v^{\mu} = - F^{\mu}{}_{\nu}v^{\nu}\,, \qquad D^{\mu}v^{\nu} - D^{\nu}v^{\mu} = F^{\mu\nu}\,,
\eeq
where $F_{\mu\nu}$ is the field strength of $A_{\mu}$ and indices are raised with $h^{\mu\nu}$. Finally, one demands that the torsion of $\Gamma$ is completely temporal, i.e.
\beq
T^{\mu}{}_{\nu\rho} \equiv \Gamma^{\mu}{}_{\nu\rho} - \Gamma^{\mu}{}_{\rho\nu}\,, \qquad h_{\mu\sigma} T^{\sigma}{}_{\nu\rho} = 0\,.
\eeq
The unique connection satisfying these criteria is
\begin{align}
\begin{split}
\label{E:gamma}
\Gamma^{\mu}{}_{\nu\rho} & = v^{\mu}\partial_{\rho}n_{\nu} + \frac{1}{2}h^{\mu\sigma} \left( \partial_{\nu}h_{\rho\sigma} + \partial_{\rho}h_{\nu\sigma} - \partial_{\sigma} h_{\nu\rho}\right) + h^{\mu\sigma} n_{(\nu}F_{\rho)\sigma}\,,
\\
T^{\mu}{}_{\nu\rho} & = v^{\mu}F^n_{\rho\nu}\,,
\end{split}
\end{align}
where (square) round brackets indicate (anti-)symmetrization
\beq
A^{(\mu\nu)} = \frac{1}{2}\left( A^{\mu\nu} + A^{\nu\mu}\right)\,, \qquad A^{[\mu\nu]} = \frac{1}{2}\left( A^{\mu\nu}-A^{\nu\mu}\right)\,,
\eeq
and we have defined
\beq
F^n_{\mu\nu} = \partial_{\mu}n_{\nu} - \partial_{\nu}n_{\mu}\,.
\eeq
See~\cite{Jensen:2014aia} for details (the case with $dn = 0$ was worked out in~\cite{Kunzle1972}).

Using this connection, we can decompose the derivative of $v^{\mu}$ into scalars, vectors, and tensors of the residual rotational symmetry which fixes $v^{\mu}$ as
\beq
\label{E:Dv}
D_{\mu}v^{\nu} = - n_{\mu}E_v^{\nu} +\frac{1}{2}(B_v)_{\mu}{}^{\nu} +h_{\mu\rho} \sigma_v^{\nu\rho}+\frac{P_{\mu}^{\nu}}{d-1}\vartheta_v\,,
\eeq
where
\begin{align*}
(E_v)_{\mu}&=F_{\mu\nu}v^{\nu}\,, & (B_v)_{\mu\nu} &= P_{\mu}^{\rho}P_{\nu}^{\sigma}F_{\rho\sigma}\,,
\\
\vartheta_v& = D_{\mu}v^{\mu}\,, & \sigma_v^{\mu\nu} &= \frac{1}{2}\left( D^{\mu}v^{\nu}+D^{\nu}v^{\mu}-\frac{2}{d-1}h^{\mu\nu}\vartheta_v\right)\,,
\end{align*}
and indices are raised with $h^{\mu\nu}$. Both $E_v$ and $B_v$ are transverse to $v^{\mu}$, $\sigma_v$ is transverse to $n_{\mu}$, and $\sigma_v$ is traceless. Moreover,
\begin{equation*}
\sigma^{\mu\nu}_v = \frac{1}{2}h^{\mu\rho}h^{\nu\sigma}\pounds_v h_{\rho\sigma} - \frac{h^{\mu\nu}}{d-1}\vartheta_v\,,
\end{equation*}
where $\pounds_v$ is the Lie derivative along $v^{\mu}$. Using the definition of $\vartheta_v$ and $\sigma_v^{\mu\nu}$, this last equation is equivalent to
\begin{equation*}
2D^{(\mu}v^{\nu)} = h^{\mu\rho}h^{\nu\sigma}\pounds_v h_{\rho\sigma}\,.
\end{equation*}
This decomposition will be useful later.

The other crucial ingredient in~\cite{Jensen:2014aia} is a shift symmetry, which is known in the Newton-Cartan literature as invariance under ``Milne boosts''~\cite{Duval:1983pb,Duval:1984cj}. Under the boost, the velocity $v^{\mu}$ is shifted by a transverse vector in such a way as to preserve the defining relation $n_{\mu}(v')^{\mu} = 1$. Accordingly, $h_{\mu\nu}$ is also shifted to preserve e.g. $(h')_{\mu\nu}(v')^{\nu}=0$. The most general such shift may be parameterized by a transverse one-form $\psi_{\mu}$ via
\beq
\label{E:hvMilne}
(v')^{\mu} = v^{\mu} + h^{\mu\nu}\psi_{\nu}\,, \qquad (h')_{\mu\nu} = h_{\mu\nu} - \left( n_{\mu} P_{\nu}^{\rho} + n_{\nu}P_{\mu}^{\rho}\right)\psi_{\rho} + n_{\mu}n_{\nu} h^{\rho\sigma} \psi_{\rho}\psi_{\sigma}\,.
\eeq
The gauge field $A_{\mu}$ also shifts under Milne boosts. One way of deducing its variation is to consider the simplest Galilean-invariant theory, that of a free field $\Psi$ carrying charge $m$ under particle number with action
\beq
\label{E:Sfree}
S_{free} = \int d^dx \left\{ \frac{i}{2}\Psi^{\dagger}\overleftrightarrow{D}_0 \Psi - \frac{\delta^{ij}}{2m}D_i\Psi^{\dagger}D_j \Psi\right\}\,,
\eeq
with $\Psi^{\dagger}\overleftrightarrow{D}_{\mu}\Psi \equiv \Psi^{\dagger}D_{\mu}\Psi - (D_{\mu}\Psi^{\dagger})\Psi$ and $D_{\mu}\Psi = \partial_{\mu}\Psi - i m A_{\mu} \Psi$. There is an obvious way of putting this theory on $\M$, namely to take the action to be
\beq
\label{E:Scov}
S_{cov} = \int d^dx \sqrt{\gamma}\left\{ \frac{iv^{\mu}}{2}\Psi^{\dagger}\overleftrightarrow{D}_{\mu}\Psi - \frac{h^{\mu\nu}}{2m}D_{\mu}\Psi^{\dagger}D_{\nu}\Psi\right\}\,.
\eeq
This action is manifestly invariant under coordinate reparameterization and $U(1)$ gauge transformations. What about Milne boosts? It is easy to show that the measure $\sqrt{\gamma}$ is Milne-invariant. It then remains to fix the variation of $A_{\mu}$ under Milne boosts by demanding that $S_{free}$ is invariant under the shifts of $(v^{\mu},h_{\mu\nu})$ in~\eqref{E:hvMilne}, which gives\footnote{This shift symmetry is present even in the flat-space theory. In that case it acts as \begin{align*}
\partial_0 &\to \partial_0 +\psi^i\partial_i\,,
\\
A_0 & \to A_0 - \frac{1}{2}\psi^i\psi_i\,,
\\
A_i &\to A_i + \psi_i\,,
\end{align*}
for $\psi_i$ a spatial covector with an arbitrary dependence on space and time.}
\beq
(A')_{\mu} = A_{\mu} + P_{\mu}^{\nu}\psi_{\nu} - \frac{1}{2}n_{\mu} h^{\nu\rho}\psi_{\nu}\psi_{\rho}\,.
\eeq
The connection $\Gamma$ defined in~\eqref{E:gamma} is not invariant under the boost. It varies as
\beq
\label{E:deltaGammaMilne}
\Delta_{\psi}\Gamma^{\mu}{}_{\nu\rho} = h^{\mu\sigma} \left\{ \left( \partial_{[\rho}n_{\nu]}P_{\sigma}^{\alpha} + \partial_{[\sigma}n_{\nu]}P_{\rho}^{\alpha} + \partial_{[\sigma}n_{\rho]}P_{\nu}^{\alpha}\right)\psi_{\alpha} + \frac{\psi^2}{2} \left( n_{\nu} \partial_{[\rho} n_{\sigma]} + n_{\rho} \partial_{[\nu}n_{\sigma]}\right)\right\}\,,
\eeq
where $\Delta_{\psi}$ denotes the additive variation under a Milne boost. Unfortunately, there is no way to make the connection $\Gamma$ invariant under both $U(1)$ gauge transformations and Milne boosts just using the Newton-Cartan data.

\subsection{Symmetries and Ward identities}
\label{S:symmetries}

We are now in a position to state the proposal of~\cite{Jensen:2014aia}. In putting a Galilean-invariant theory on $\M$, one should couple it to the Newton-Cartan data $(n_{\mu},h_{\mu\nu},A_{\mu})$ in such a way that it is invariant under reparameterizations of coordinates, $U(1)$ gauge transformations, and the Milne boosts. The Milne boost invariance effectively imposes a local version of the Galilean boost invariance. 

To see this, we first define the various symmetry currents via variation of the generating functional $W$ of correlation functions. In what follows, our discussion closely parallels the derivation of Ward identities in~\cite{Geracie:2014nka} and our previous work. 

To make contact with the literature, it is convenient to regard $W$ as a functional of an overcomplete set of background fields, namely $(n_{\mu},h^{\mu\nu},v^{\mu},A_{\mu})$ (recall that $n_{\mu}$ is algebraically determined by $v^{\mu}$ and $h^{\mu\nu}$). Then $W$ is a functional $W[n_{\mu},h^{\mu\nu},v^{\mu},A_{\mu}]$. The particle number current $J^{\mu}$, momentum density $\mathcal{P}_{\mu}$, energy current $\mathcal{E}^{\mu}$, and spatial stress tensor $T_{\mu\nu}$ are given via~\cite{Geracie:2014nka,Jensen:2014aia}
\beq
\label{E:defineCurrents}
\delta W = \int d^dx \sqrt{\gamma} \left\{ \delta A_{\mu} J^{\mu} - \delta n_{\mu} \mathcal{E}^{\mu} - \delta \bar{v}^{\mu} \mathcal{P}_{\mu} - \frac{1}{2}\delta \bar{h}^{\mu\nu}T_{\mu\nu}\right\}\,.
\eeq
Here we have implicitly taken the variations of $n_{\mu}$ to be arbitrary, so that some of the variations of $(v^{\mu},h^{\mu\nu})$ are fixed as
\begin{align}
\begin{split}
\label{E:deltavh}
\delta v^{\mu} & = - v^{\nu}\delta n_{\nu} v^{\mu} + P_{\nu}^{\mu} \delta \bar{v}^{\nu}\,,
\\
\delta h^{\mu\nu}& = - \left( v^{\mu}h^{\nu\rho} + v^{\nu}h^{\mu\rho}\right)\delta n_{\rho} + P^{\mu}_{\rho}P^{\nu}_{\sigma} \delta \bar{h}^{\rho\sigma}\,,
\end{split}
\end{align}
with $\delta \bar{v}^{\mu}$ and $\delta \bar{h}^{\mu\nu}$ unconstrained. Note that the momentum and spatial stress tensor defined this way are completely transverse to $v^{\mu}$. The invariance of $W$ under the symmetries of the problem leads to various Ward identities for the currents. In particular, invariance under Milne boosts gives the folklore equality of momentum and particle number currents
\beq
\label{E:milneWard}
\mathcal{P}_{\mu} = h_{\mu\nu}J^{\nu}\,.
\eeq

Using this relation, the Ward identities corresponding to invariance under $U(1)$ gauge transformations and coordinate reparameterizations can be efficiently written in terms of a spacetime stress tensor~\cite{Jensen:2014aia}
\beq
\label{E:calT}
\mathcal{T}^{\mu\nu} \equiv T^{\mu\nu} + \mathcal{P}^{\mu}v^{\nu} + \mathcal{P}^{\nu}v^{\mu} + v^{\mu}v^{\nu}n_{\rho}J^{\rho}\,,
\eeq
where indices are raised with $h^{\mu\nu}$. In terms of $\mathcal{T}^{\mu\nu}$ and the covariant derivative defined from~\eqref{E:gamma}, the remaining Ward identities become~\cite{Jensen:2014aia}
\begin{align}
\begin{split}
\label{E:ward}
\left( D_{\nu}-\mathcal{G}_{\nu}\right)\mathcal{E}^{\mu} & = \mathcal{G}_{\mu} \mathcal{E}^{\mu} -h_{\rho(\mu}D_{\nu)}v^{\rho}\mathcal{T}^{\mu\nu} \,,
\\
\left( D_{\nu}-\mathcal{G}_{\nu}\right) \mathcal{T}^{\mu\nu} & = - (F^n)^{\mu}{}_{\nu}\mathcal{E}^{\nu}\,,
\end{split}
\end{align}
where
\beq
\mathcal{G}_{\mu} \equiv T^{\nu}{}_{\mu\nu} = - F^n_{\mu\nu}v^{\nu}\,.
\eeq
The covariant divergence appearing here is somewhat strange, but at least for vectors it is just the ordinary notion of a divergence with a volume element $\sqrt{\gamma}$,
\beq
\label{E:dontWorryAboutDivergence}
\left( D_{\mu}-\mathcal{G}_{\mu}\right)\mathfrak{v}^{\mu} = \frac{1}{\sqrt{\gamma}}\partial_{\mu}\left( \sqrt{\gamma}\mathfrak{v}^{\mu}\right)\,,
\eeq
where $\mathfrak{v}^{\mu}$ is any vector field.

Note that the longitudinal component of the stress Ward identity~\eqref{E:ward} is the conservation of particle number,
\beq
n_{\mu} \left( D_{\nu} -\mathcal{G}_{\nu}\right) \mathcal{T}^{\mu\nu} = \left( D_{\mu} - \mathcal{G}_{\mu}\right)J^{\mu} = 0\,.
\eeq
Here we have used~\eqref{E:milneWard}, the constancy of $n_{\mu}$, and $n_{\mu}(F^n)^{\mu}{}_{\nu} = 0$.

The spacetime stress tensor $\mathcal{T}^{\mu\nu}$ not only simplifies the Ward identities, but it has the property of being invariant under Milne boosts~\cite{Jensen:2014aia}. The Milne variation of the energy current can be deduced from this, the variation of the connection under Milne boosts~\eqref{E:deltaGammaMilne}, and the Ward identities~\eqref{E:ward}. The variation is~\cite{Jensen:2014aia}
\beq
\label{E:deltaEMilne}
(\mathcal{E}')^{\mu} = \mathcal{E}^{\mu} - \left( P_{\nu}^{\rho}\psi_{\rho} - \frac{1}{2}n_{\nu}\psi^2\right)\mathcal{T}^{\mu\nu}\,.
\eeq
This will be especially useful when we turn our attention to hydrodynamics.

\subsection{A Milne covariant derivative, and a boost-invariant energy current} 
\label{S:milneD}

At this point it is rather difficult to form tensorial invariants under all of the symmetries of the problem. Both the connections $A_{\mu}$ and $\Gamma^{\mu}{}_{\nu\rho}$ are not invariant under Milne boosts, so that the covariant derivative is not Milne-invariant. This deficiency can be somewhat ameliorated by a tensorial redefinition of $\Gamma$ using terms that explicitly depend on $A_{\mu}$ rather than $F_{\mu\nu}$. Unfortunately that redefinition varies under $U(1)$ gauge transformations.

One way to proceed is to build a $d+1$-dimensional Lorentzian manifold with a null isometry from the Newton-Cartan data, as outlined in Section 3 of~\cite{Jensen:2014aia}. Here the connection $A_{\mu}$ is the graviphoton of the reduction, and the Milne boosts correspond to an ambiguity in the identification of the Newton-Cartan data from the $d+1$-dimensional metric. That is, tensorial data on $\M$ can then be obtained by constructing tensors from the metric and null isometry in one higher dimension.

There is another way which will prove much more useful in this work. Suppose that we also equip our spacetime with a Milne-invariant vector field $u^{\mu}$ which is everywhere timelike, meaning $n_{\mu}u^{\mu} >0$. Normalize $u^{\mu}$ so that $n_{\mu}u^{\mu} = 1$. In hydrodynamics, we will see shortly that Nature graciously provides just such a vector field in the fluid velocity.

We can use $u^{\mu}$ to construct a Milne-invairant covariant derivative as we now show. Using $h_{\mu\nu}$ to define $u_{\mu} = h_{\mu\nu}u^{\nu}$ along with $u^2 = u_{\mu}u^{\mu}$, these objects inherit transformation properties under Milne boosts from the variation~\eqref{E:hvMilne} of $h_{\mu\nu}$,
\begin{align}
\begin{split}
(u')_{\mu} &= u_{\mu} - P_{\mu}^{\nu} \psi_{\nu} + n_{\mu}h^{\nu\rho} \left( \psi_{\nu}\psi_{\rho} -  u_{\nu}  \psi_{\rho}\right)\,.
\\
(u')^2 & = u^2 + \psi^2 - 2 h^{\mu\nu}u_{\mu} \psi_{\nu}\,,
\end{split}
\end{align}
so that the combination $-u_{\mu} + \frac{1}{2}n_{\mu} u^2$ varies as
\beq
\label{E:theMagicMilne}
\left( -u_{\mu} +\frac{1}{2} n_{\mu}u^2\right)' = \left( -u_{\mu} + \frac{1}{2}n_{\mu} u^2\right) + P_{\mu}^{\nu} \psi_{\nu} - \frac{1}{2}n_{\mu} \psi^2\,.
\eeq
Note that this covector transforms under Milne boosts in the same way as $A_{\mu}$. Indeed, we can use this covector to define a new gravitational connection $\tilde{\Gamma}$ and $U(1)$ connection $\tilde{A}$ which are invariant under Milne boosts
\begin{align}
\begin{split}
\label{E:milneD}
\tilde{\Gamma}^{\mu}{}_{\nu\rho} &= \Gamma^{\mu}{}_{\nu\rho} + h^{\mu\sigma} \left\{ u_{\sigma} \partial_{[\rho}n_{\nu]}  - \left( u_{\nu} - \frac{1}{2}n_{\nu}u^2\right) \partial_{[\rho}n_{\sigma]} - \left( u_{\rho} - \frac{1}{2}n_{\rho} u^2\right)\partial_{[\nu}n_{\sigma]}\right\} \,,
\\
\tilde{A}_{\mu} & = A_{\mu} + u_{\mu} - \frac{1}{2}n_{\mu} u^2\,.
\end{split}
\end{align} 
Both of these objects clearly transform as connections under diffeomorphisms and $U(1)$ transformations, as they are the sum of a connection and a gauge-invariant tensor. The covariant derivative $\tilde{D}$ defined with the connections $\tilde{\Gamma}$ and $\tilde{A}$ has the property that its derivative of Milne-invariant tensors is also a Milne-invariant tensor. So we refer to $\tilde{D}$ as the \emph{Milne covariant derivative}.\footnote{Essentially the same construction is at play in~\cite{Hoyos:2011ez}, where those authors implicitly define a non-local $u^{\mu}$ in $2+1$ dimensions by demanding $\tilde{F}_{\mu\nu}u^{\nu} = 0$ and $u^{\mu}n_{\mu}=1$.}

What are the derivatives of $n_{\mu}$ and $h^{\mu\nu}$ under $\tilde{D}$? To efficiently proceed, we first observe that from the data $(n_{\mu},h^{\mu\nu},u^{\mu})$ one can define a Milne-invariant version of $h_{\mu\nu}$,
\beq
\tilde{h}_{\mu\nu} \equiv h_{\mu\nu} - \left( u_{\mu}n_{\nu} + u_{\nu}n_{\mu}\right) + u^2 n_{\mu}n_{\nu}\,,
\eeq
which satisfies
\beq
\label{E:huProjection}
\tilde{h}_{\mu\nu} u^{\nu} = 0\,, \qquad \tilde{h}_{\mu\rho}h^{\nu\rho} \equiv \tilde{P}_{\mu}^{\nu} =  \delta_{\mu}^{\nu} - u^{\nu}n_{\mu} \,.
\eeq
From this we find that $\tilde{\Gamma}$ can be simply written in the same form as the Newton-Cartan connection $\Gamma$ in~\eqref{E:gamma}, upon substituting the Milne-non-invariant data $(v^{\mu},h_{\mu\nu},A_{\mu})$ for the Milne-invariant data $(u^{\mu},\tilde{h}_{\mu\nu},\tilde{A}_{\mu})$. That is,
\begin{align}
\begin{split}
\label{E:simpleGammau}
\tilde{\Gamma}^{\mu}{}_{\nu\rho} &= u^{\mu}\partial_{\rho}n_{\nu} + \frac{1}{2}h^{\mu\sigma}\left( \partial_{\nu}\tilde{h}_{\rho\sigma} + \partial_{\rho} \tilde{h}_{\nu\sigma} - \partial_{\sigma}\tilde{h}_{\nu\rho}\right) + h^{\mu\sigma} n_{(\nu}\tilde{F}_{\rho)\sigma}\,,
\\
\tilde{T}^{\mu}{}_{\nu\rho} & = \tilde{\Gamma}^{\mu}{}_{\nu\rho}-\tilde{\Gamma}^{\mu}{}_{\rho\nu} = u^{\mu} F^n_{\rho\nu}\,,
\end{split}
\end{align}
where we have denoted the field strength of $\tilde{A}$ as $\tilde{F}=d\tilde{A}$ and defined the torsion of $\tilde{\Gamma}$ as $\tilde{T}$. The calculation that shows this is mechanical and a little laborious, but let us show the highlights. We write out $\tilde{\Gamma}$ in \eqref{E:simpleGammau} as
\begin{align}
\begin{split}
\tilde{\Gamma}^{\mu}{}_{\nu\rho}&=   \Gamma^{\mu}{}_{\nu\rho} + h^{\mu\sigma} \left\{ u_{\sigma}\partial_{[\rho}n_{\nu]}  - \left( u_{\nu} -n_{\nu}u^2\right)\partial_{[\rho}n_{\sigma]} - \left( u_{\rho}-n_{\rho}u^2\right)\partial_{[\nu}n_{\sigma]} \right.
\\
 & \qquad \qquad  \left. - n_{\nu}\partial_{[\rho} u_{\sigma]} - n_{\rho} \partial_{[\nu}u_{\sigma]} - \frac{1}{2}n_{\nu}n_{\rho} \partial_{\sigma}u^2\right.
 \\
  & \qquad\qquad  \left. + n_{\nu}\partial_{[\rho}u_{\sigma]} + n_{\rho}\partial_{[\nu}u_{\sigma]} - \frac{1}{2}n_{\nu}u^2 \partial_{[\rho}n_{\sigma]} - \frac{1}{2}n_{\rho}u^2 \partial_{[\nu}n_{\sigma]} + \frac{1}{2}n_{\nu}n_{\rho}\partial_{\sigma}u^2\right\}
  \\
  & = \Gamma^{\mu}{}_{\nu\rho} + h^{\mu\sigma} \left\{ u_{\sigma}\partial_{[\rho}n_{\nu]} - \left( u_{\nu}-\frac{1}{2}n_{\nu}u^2\right)\partial_{[\rho}n_{\sigma]} - \left( u_{\rho} - \frac{1}{2}n_{\rho}u^2\right)\partial_{[\nu}n_{\sigma]}\right\}\,,
\end{split}
\end{align}
which coincides with the expression for $\tilde{\Gamma}$ given in~\eqref{E:milneD}. The terms with $u$ in the first and second lines of the equation above come from intelligently rewriting the first and second terms in~\eqref{E:simpleGammau}, while the third line comes from writing out the terms with $\tilde{F}$.

Since $\tilde{\Gamma}$ is of the same form as $\Gamma$ in terms of $(n_{\mu},h^{\mu\nu},u^{\mu},\tilde{h}_{\mu\nu},\tilde{A}_{\mu})$, we can immediately borrow the various properties of $\Gamma$ (e.g.~\eqref{E:nhConstant} and~\eqref{E:Dv}), giving
\beq
\tilde{D}_{\mu}n_{\nu}  = 0\,, \qquad \tilde{D}_{\mu}h^{\nu\rho} = 0\,, \qquad u^{\nu}\tilde{D}_{\nu} u^{\mu}  = - \tilde{F}^{\mu}{}_{\nu}u^{\nu}\,, \qquad \tilde{D}^{\mu}u^{\nu} - \tilde{D}^{\nu}u^{\mu} = \tilde{F}^{\mu\nu}\,,
\eeq
and
\beq
2\tilde{D}^{(\mu}u^{\nu)} = h^{\mu\rho}h^{\nu\sigma}\pounds_u \tilde{h}_{\rho\sigma}\,.
\eeq

The output of this is the following. A general Milne-invariant tensor built from the Newton-Cartan data and $u^{\mu}$ may be constructed from the data $(n_{\mu},h^{\mu\nu},u^{\mu},\tilde{h}_{\mu\nu},\tilde{A}_{\mu})$ and the Milne-covariant derivative~\eqref{E:milneD}. In the context of hydrodynamics, this result will allow us to efficiently express the constitutive relations in a manifestly Milne-invariant way.

Before moving on, there are two more useful results which we may as well establish here. Recall that spacetime stress tensor $\mathcal{T}^{\mu\nu}$ is Milne-invariant, but the energy current $\mathcal{E}^{\mu}$ is not. Its Milne variation was given in~\eqref{E:deltaEMilne}. Using~\eqref{E:theMagicMilne}, we can then construct a Milne-invariant energy current $\tilde{\mathcal{E}}^{\mu}$ as
\beq
\label{E:boostEnergy}
\tilde{\mathcal{E}}^{\mu} \equiv \mathcal{E}^{\mu} - \left( u_{\nu} - \frac{1}{2}n_{\nu}u^2\right)\mathcal{T}^{\mu\nu}\,.
\eeq
We now rewrite the Ward identities~\eqref{E:ward} in terms of it and the Milne covariant derivative. We first define $\tilde{\mathcal{G}}_{\mu}$ from the torsion for $\tilde{D}_{\mu}$ in analogy with the definition of $\mathcal{G}_{\mu}$,
\beq
\tilde{\mathcal{G}}_{\mu} = \tilde{T}^{\nu}{}_{\mu\nu} = - F^n_{\mu\nu}u^{\nu}\,.
\eeq
It is easiest to begin with the stress tensor Ward identity. We straightforwardly obtain
\begin{align}
\begin{split}
&\left( D_{\nu}-\mathcal{G}_{\nu}\right)  \mathcal{T}^{\mu\nu} + (F^n)^{\mu}{}_{\nu}  \mathcal{E}^{\nu}-\left\{  \left( \tilde{D}_{\nu} - \tilde{\mathcal{G}}_{\nu}\right) \mathcal{T}^{\mu\nu} +(F^n)^{\mu}{}_{\nu}\tilde{\mathcal{E}}^{\nu}\right\}
\\
& \quad =  \left[\left( \Gamma^{\mu}{}_{\rho\nu} -\tilde{\Gamma}^{\mu}{}_{\rho\nu}\right)- (F^n)^{\mu}{}_{\nu}\left( u_{\rho}-\frac{1}{2}n_{\rho}u^2\right)\right]  \mathcal{T}^{\nu\rho} = 0\,,
\end{split}
\end{align}
Next, using~\eqref{E:dontWorryAboutDivergence}, which here implies
\beq
\label{E:divergence}
\left( D_{\mu}-\mathcal{G}_{\mu}\right)\mathfrak{v}^{\mu} = \frac{1}{\sqrt{\gamma}}\partial_{\mu}\left( \sqrt{\gamma}\mathfrak{v}^{\mu}\right) = \left( \tilde{D}_{\mu} - \tilde{\mathcal{G}}_{\mu}\right)\mathfrak{v}^{\mu}\,,
\eeq
along with the stress tensor Ward identity, we find that the Ward identity for the energy current becomes
\begin{align}
\begin{split}
&\left( D_{\mu} - \mathcal{G}_{\mu}\right)\mathcal{E}^{\mu} -\mathcal{G}_{\mu}\mathcal{E}^{\mu} + h_{\rho(\mu}D_{\nu)}v^{\rho} \mathcal{T}^{\mu\nu}
\\
& \quad = \left( \tilde{D}_{\mu}-\tilde{\mathcal{G}}_{\mu}+F^n_{\mu\rho}v^{\rho}\right)\left[ \tilde{\mathcal{E}}^{\mu} + \left( u_{\nu} - \frac{1}{2}n_{\nu}u^2\right) \mathcal{T}^{\mu\nu}\right]+ h_{\rho(\mu}D_{\nu)}v^{\rho} \mathcal{T}^{\mu\nu} 
\\
& \quad = \left( \tilde{D}_{\mu}-\tilde{\mathcal{G}}_{\mu} \right)\tilde{ \mathcal{E}}^{\mu} + F^n_{\mu\nu}u^{\nu}\tilde{\mathcal{E}}^{\mu}
\\
& \qquad \qquad + \left( D_{(\mu}u_{\nu)} + h_{\rho(\mu}D_{\nu)}v^{\rho} -\frac{1}{2}n_{(\mu}D_{\nu)}u^2+\left( u_{(\mu} -\frac{1}{2} n_{(\mu}u^2 \right)F^n_{\nu)\rho}u^{\rho}\right) \mathcal{T}^{\mu\nu}
\\
& \qquad =  \left( \tilde{D}_{\mu}-\tilde{G}_{\mu}\right) \tilde{\mathcal{E}}^{\mu} + F^n_{\mu\nu}u^{\nu} \tilde{\mathcal{E}}^{\mu} + \tilde{h}_{\rho(\mu}\tilde{D}_{\nu)}u^{\rho} \mathcal{T}^{\mu\nu}\,.
\end{split}
\end{align}
Putting the pieces together, the Ward identities~\eqref{E:ward} may be expressed in a manifestly Milne-invariant way as
\begin{align}
\begin{split}
\label{E:ward2}
\left( \tilde{D}_{\mu}-\tilde{\mathcal{G}}_{\mu}\right)\tilde{\mathcal{E}}^{\mu} & = \tilde{\mathcal{G}}_{\mu}\tilde{\mathcal{E}}^{\mu} - \tilde{h}_{\rho(\mu}\tilde{D}_{\nu)} u^{\rho}\mathcal{T}^{\mu\nu}\,,
\\
\left( \tilde{D}_{\nu}-\tilde{\mathcal{G}}_{\nu}\right) \mathcal{T}^{\mu\nu} &= - (F^n)^{\mu}{}_{\nu}\tilde{\mathcal{E}}^{\nu}\,,
\end{split}
\end{align}

\subsection{Including a magnetic moment} 
\label{S:gs}

Recently, Son~\cite{Son:2013rqa} (and Son with collaborators in~\cite{Geracie:2014nka}) has added a magnetic moment coupling to the free field action~\eqref{E:Sfree} in a way that is invariant under his non-relativistic ``general covariance.'' In~\cite{Jensen:2014aia} we showed that this modified theory can be written in a manifestly reparameterization and $U(1)$-invariant way as
\beq
S_g = \int d^dx \sqrt{\gamma} \left\{ \frac{i v^{\mu}}{2}\Psi^{\dagger}\overleftrightarrow{D}_{\mu}\Psi - \frac{1}{2m}\left( h^{\mu\nu} + \frac{ig_s}{2}\varepsilon^{\mu\nu}\right) D_{\mu}\Psi^{\dagger}D_{\nu}\Psi\right\}\,,
\eeq
where $\varepsilon^{\mu\nu} = \varepsilon^{\rho\mu\nu}n_{\rho}$ and $\varepsilon^{\mu\nu\rho} = \frac{\epsilon^{\mu\nu\rho}}{\sqrt{\gamma}}$ with $\epsilon^{\mu\nu\rho}$ the epsilon symbol satisfying $\epsilon^{012}=+1$. This theory is Milne-invariant provided that the Milne variation of $A_{\mu}$ is modified as
\beq
\label{E:modifiedMilne}
(A')_{\mu} = A_{\mu} + P_{\mu}^{\nu} \psi_{\nu} - \frac{1}{2}n_{\mu} \psi^2 + n_{\mu} \frac{g_s}{4m}\varepsilon^{\nu\rho\sigma}\partial_{\nu}\left( n_{\rho}P_{\sigma}^{\alpha}\psi_{\alpha}\right)\,.
\eeq

We will not consider the theory with $g_s\neq 0$ for much of this article. However here we point out that using the same logic of the previous Subsection, we can use $u^{\mu}$ to define a Milne covariant derivative. The Milne-invariant gravitational and $U(1)$ connections are given by
\begin{align}
\label{E:gsMilneD}
(A_g)_{\mu} & = A_{\mu} + P_{\mu}^{\nu}u_{\nu} - \frac{1}{2}n_{\mu} u^2 + n_{\mu} \frac{g_s}{4m}\varepsilon^{\nu\rho\sigma}\partial_{\nu}\left( n_{\rho} u_{\sigma}\right)\,,
\\
(\Gamma_g)^{\mu}{}_{\nu\rho} & = u^{\mu}\partial_{\rho}n_{\nu} + \frac{1}{2}h^{\mu\sigma}\left( \partial_{\nu} \tilde{h}_{\rho\sigma} + \partial_{\rho}\tilde{h}_{\nu\sigma} - \partial_{\sigma}\tilde{h}_{\nu\rho}\right) + h^{\mu\sigma} n_{(\nu}(F_g)_{\rho)\sigma}\,,
\end{align}
where $F_g = dA_g$.

\section{Basics of life at $T > 0$} 
\label{S:Tneq0}

We now turn our attention to general features of hot Galilean field theory. We begin with Euclidean thermal field theory, paying careful attention to the geometry on which the thermal partition function depends. Using these results we pave the way to formulating Galilean hydrodynamics in a coordinate-independent way.

\subsection{Euclidean field theory and thermal circles} 
\label{S:euclidean}

Consider a Galilean theory in flat space without global symmetries (the extension of our work to theories with global symmetries is straightforward). The flat background is specified by
\beq
n = dx^0\,, \qquad h_{\mu\nu}dx^{\mu}\otimes dx^{\nu} = \delta_{ij}dx^i\otimes dx^j\,, \qquad A = 0\,.
\eeq
Now turn on a temperature $T = 1/\beta$ and chemical potential $\mu_0$ in the rest frame in which time is $x^0$. Denoting the generator of time translations at $\mu_0 = 0$ as $H$ and the generator of particle number as $M$, the thermal partition function of the theory is
\beq
\label{E:ZE}
\mathcal{Z}_E = \text{tr}\left( e^{-\beta (H-\mu_0 M)}\right)\,.
\eeq
For theories with a functional integration representation, $\mathcal{Z}_E$ is the functional integral on the Euclideanized spacetime $\mathbb{S}^1\times\mathbb{R}^{d-1}$ in which we Wick-rotate $x^0 =-i x_E$, compactify imaginary time as $x_E \sim x_E + \beta$, and put a $U(1)$ holonomy around the circle.\footnote{Actually, this is not entirely correct. The functional integral usually differs from the sum over states by a term which only depends on the background fields. This term arises when passing from the canonical to grand canonical ensembles. It is important for instance when relating any torus partition function of a two-dimensional conformal field theory with chemical potentials to a sum over states.} In order to get the partition function~\eqref{E:ZE}, the holonomy experienced by a particle of charge $q$ transported around the circle is $\exp(\beta \mu_0 q)$. This can be done by turning on a constant $A = \mu_0 dx^0 = - i \mu_0 dx_E$, so that
\beq
\exp\left( i q\int_{\mathcal{C}} A\right) = \exp\left( \beta q K^{\mu}A_{\mu}\right) = \exp \left( \beta \mu_0 q\right)\,.
\eeq
Note that the tangent vector $K_E^{\mu}$ to the thermal circle is $\partial_E = i \partial_0$. It is useful to define $K^{\mu} = - i K_E^{\mu} $, which is the real, timelike (in the sense that $n_{\mu}K^{\mu} >0$) vector which we get from $K_E^{\mu}$ after ``un-Wick-rotating'' back to ordinary time. In this case we have $K^{\mu}\partial_{\mu} = \partial_0$ so that trivially $n_{\mu}K^{\mu} = 1$. 

In what follows, $K^{\mu}$ will play a pivotal role. We use it to construct the relevant Euclidean geometry for a more general equilibrium in two steps. First, we Wick-rotate the affine parameter $\tau$ along the integral curves of $K^{\mu}$ (the curves to which $K^{\mu}$ is tangent). In this case the affine parameter is just ordinary time $x^0$. Then we compactify the Wick-rotated affine parameter, $\tau_E$, with periodicity $\beta$. Provided that $K^{\mu}$ generates a symmetry of the background, then there is an conserved charge $\mathcal{H}_{\tau}$ which generates translations along $\tau$, and the functional integral on this Euclidean geometry gives
\beq
\mathcal{Z}_E = \text{tr} \left( e^{-\beta\mathcal{H}_{\tau}}\right)\,.
\eeq
In this instance, the operator which generates translations in $\tau$ at nonzero $\mu_0$ is $H - \mu_0 Q$, so that this partition function indeed matches~\eqref{E:ZE}.

Na\"ively the temperature of the thermal state is $1/\beta$. This is almost true. Suppose that we take the vector $K^{\mu}$ to be $K^{\mu} \partial_{\mu} = c\partial_0$ rather than $ \partial_0$. Then running through the construction above, we find that $x_E/c$ has period $\beta$, so $x_E$ has period $c\beta$ from which we would identify the temperature to be $T = 1/(c\beta)$. In general, the physical temperature is
\beq
T = \frac{1}{\beta \, n_{\mu} K^{\mu}}\,,
\eeq
or the inverse of the integral of $n_{\mu}$ around the thermal circle. Similarly, the physical chemical potential is defined through
\beq
\frac{\mu_0}{T} = \ln \exp\left( i \int_{\mathcal{C}}A\right)\,.
\eeq
which implies that in this time-independent gauge,
\beq
\mu_0 = u^{\mu}A_{\mu}\,, \qquad u^{\mu} \equiv \frac{K^{\mu}}{n_{\nu}K^{\nu}}\,.
\eeq
Here we have defined a normalized vector $u^{\mu}$ which satisfies $u^{\mu}n_{\mu}=1$. Shortly, we will see that this definition of the chemical potential is not boost-invariant (which is not surprising as $A_{\mu}$ varies under Milne boosts), after which we will abandon this definition in favor of a boost-invariant version of $\mu$.

A corollary of this result is that a rescaling of $\beta$ can be absorbed into a rescaling of $ K^{\mu}$ in such a way as to keep the physical temperature fixed. That is, observables only depend upon $\beta$ as well as the overall normalization of $K^{\mu}$ through the invariant combination $T=\frac{1}{\beta n_{\mu}K^{\mu}}$. So we find it convenient to parameterize the dependence of observables on $\beta$ and $K^{\mu}$ through the physical temperature $T$ and ``velocity vector'' $u^{\mu} = \frac{K^{\mu}}{n_{\nu}K^{\nu}}$ satisfying $u^{\mu}n_{\mu} = 1$.

Since there are many symmetries of the flat Newton-Cartan structure (those symmetries are generated by the centrally extended Galilean algebra), we can study more general partition functions in flat space. Let us discuss these first from the point of view of algebra, and then from Euclidean field theory.

From the algebraic perspective, the most general thermal partition function in flat space has a Boltzmann weight which is constructed from a linear combination of the generators of the Galilean algebra. Those generators are $(H,P_i, R_{ij}, K_i,M)$. We have already discussed $H$ and $M$; the $P_i$ generate translations, $R_{ij}$ spatial rotations, and $K_i$ the Galilean boosts. For instance, a Boltzmann weight $\exp\left( - \beta (H-\mu M - \omega^{ij} R_{ij})\right)$ corresponds to thermal field theory at a temperature $T = 1/\beta$, chemical potential $\mu$, with a chemical potential $\omega^{ij}$ for rotation.\footnote{Note that unlike in the relativistic case, a rigidly rotating non-relativistic fluid is perfectly consistent with causality, so that one need not put the theory on a compact spatial manifold in order to consistently study thermal equilibria with $\omega^{ij} \neq 0$.} Because $[R_{ij},P_k]$ and $[K_i,P_j]$ are nonzero, chemical potentials for rotations and boosts break translational invariance. So the most general Bolzmann weight for a translationally-invariant flat-space equilibrium is
\beq
\label{E:flatBoltzmann}
\exp \left\{-\beta \left[H+ u^i P_i - \left( \mu_0+\frac{u^2}{2}\right) M\right]\right\}\,.
\eeq
The reason for the redefinition of the chemical potential $\mu_0 \to \mu_0 + \frac{u^2}{2}$ will be clear shortly.

Now recall that in the Galilean algebra we have
\beq
[H,K_i] = -i P_i\,, \qquad [P_i,K_j] = - i \delta_{ij} M\,,
\eeq
with $M$ central. Then
\beq
e^{ -\beta \left[ H + u^i P_i - \left( \mu_0+ \frac{u^2}{2}\right) M\right]} = e^{-i u^i K_i} e^{-\beta \left[ H - \mu_0 M\right]}e^{i u^j K_j}\,,
\eeq
so the equilibrium described by~\eqref{E:flatBoltzmann} is just a boosted equilibrium with Boltzmann weight $\exp\left\{ -\beta \left[ H -\mu_0 M\right]\right\}$. Note that this result tells us that a rest frame chemical potential $\mu_0$ is perceived to be a chemical potential $\mu_0 + \frac{u^2}{2}$ in a relatively boosted frame. That is, the boost-invariant chemical potential is $\mu = \mu_0 + \frac{u^2}{2}$.

Now for Euclidean field theory. As we mentioned above, the Euclidean background corresponding to some partition function is constructed from a real-time background using the operator $\mathcal{H}$ which generates translations in some symmetry direction. In this setting, a symmetry is generated by a combination of a vector field $K^{\mu}$, a Milne boost $\psi^K$, and a gauge transformation $\Lambda_K$, which we collectively denote as $K = (K^{\mu},\psi^K,\Lambda_K)$. We denote the action of $K$ as $\delta_K$. For instance, $K$ acts on the gauge field $A_{\mu}$ as
\beq
\delta_K A_{\mu} = \pounds_K A_{\mu} + P_{\mu}^{\nu}\psi^K_{\nu} + \partial_{\mu}\Lambda_K\,,
\eeq
for $\pounds_K$ the Lie derivative along $K^{\mu}$. These transformations generate an algebra. It is easy to find the $K$ which fix the flat Newton-Cartan structure, and using this algebra, these $K$ are exactly the generators of the centrally extended Galilean algebra. See e.g. Subsection 2.4 of~\cite{Jensen:2014aia} for details. The $K$ which correspond to $H$ and the $P_i$ are
\beq
H = (-\partial_0,0,0), \qquad P_i = (-\partial_i,0,0)\,, \qquad M = (0,0,1)\,.
\eeq

What is the Euclidean thermal field theory version of~\eqref{E:flatBoltzmann}? On the one hand, we know that it is the functional integral on a Euclidean spacetime in which we take $K^{\mu}$ to be a linear combination of $\partial_0$ and the $\partial_i$. On the other hand, our algebraic discussion shows that it is just a boosted version of the partition function on $\mathbb{S}^1\times \mathbb{R}^{d-1}$ as described above. Let us see how both of these versions work, starting with the background corresponding to
\begin{equation*}
K^{\mu}\partial_{\mu} =u^{\mu}\partial_{\mu}= \partial_0 + u^i\partial_i\,. 
\end{equation*}
Following our discussion above, the Euclidean background is constructed by Wick-rotating the affine parameter along the integral curves of $K^{\mu}$, which here is $x^0+u_i x^i$, and compactifying with periodicity $\beta$. The physical temperature is
\begin{equation*}
T = \frac{1}{\beta n_{\mu}K^{\mu}}=\frac{1}{\beta}\,,
\end{equation*}
but to get the right chemical potential $\mu_0+ \frac{u^2}{2}$ we need
\beq
\label{E:boostedA}
A = \left( \mu_0 - \frac{u^2}{2}\right)dx^0 + u_i dx^i\,,
\eeq
up to a constant covector which annihilates $u^{\mu}$. To see this, the holonomy experienced by a charge $q$ particle transported around the thermal circle $\mathcal{C}$ is
\beq
\exp\left( i q \int_{\mathcal{C}} A\right) = \exp\left( \beta q K^{\mu}A_{\mu}\right) = \exp \left[ \beta q \left( \mu_0 + \frac{u^2}{2}\right)\right]\,,
\eeq
as it ought to be.

We can get this same background by a boost. Starting with the flat background above before Wick-rotation, a Galilean boost is a combination of a coordinate reparamterization and a Milne boost,
\beq
x'^0 = x^0\,, \qquad x'^i = x^i + u^i x^0\,, \qquad \psi_i = u_i\,.
\eeq
Under this transformation the geometric data $(n_{\mu},h_{\nu\rho})$ is invariant,
\beq
n = dx^0 = dx'^0\,, \qquad h_{\mu\nu}dx^{\mu}\otimes dx^{\nu} = \delta_{ij} dx^i \otimes dx^j = \delta_{ij} dx'^i \otimes dx'^j\,,
\eeq
while the gauge field $A=\mu_0 dx^0$ shifts as
\beq
A' = \left(\mu_0- \frac{u^2}{2}\right) dx'^0 + u_i dx'^i\,,
\eeq
which recovers~\eqref{E:boostedA} in the primed coordinates. Previously, the tangent vector to the thermal circle was ($i$ times) the analytic continuation of $K^{\mu}\partial_{\mu} = \partial_0$. After the coordinate transformation we have $K^{\mu} \partial_{\mu} = \frac{\partial}{\partial x'^0} + u^i \frac{\partial}{\partial x'^i}$. So instead of Wick-rotating $x^0$, we Wick-rotate $x'^0 + u_i x'^i$ and compactify with periodicity $\beta$. 

We would like a boost-invariant definition of the chemical potential. Fortunately, there is a boost-invariant version of the holonomy of $A$, namely the holonomy of the Milne-invariant $\tilde{A}$ around the thermal circle. Recall that $\tilde{A}$ is constructed from $A$ and $u^{\mu}$ as in~\eqref{E:milneD},
\begin{equation*}
\tilde{A}_{\mu} = A_{\mu} + u_{\mu} - \frac{1}{2}n_{\mu}u^2\,.
\end{equation*}
Note that with $A = \mu_0 dx^0$ and $u_{\mu} = u_i dx^i$, we have
\beq
\tilde{A} = \left( \mu_0 - \frac{u^2}{2}\right) dx^0 + u_i dx^i\,,
\eeq
which is just the $U(1)$ connection~\eqref{E:boostedA} that appeared in a relatively boosted frame. The physical, boost-invariant, chemical potential $\mu$ is constructed from the holonomy of $\tilde{A}$ as
\beq
\label{E:defmuT}
\frac{\mu}{T}= \ln \exp\left( i \int_{\mathcal{C}} \tilde{A}\right)\,,
\eeq
which in this time-independent gauge is just
\beq
\mu = u^{\mu} \tilde{A}_{\mu}\,.
\eeq

To summarize, the translationally-invariant flat-space thermal equilibria of a Galilean-invariant field theory are specified by a temperature $T$, a boost-invariant chemical potential $\mu$ for particle number, and a fluid velocity $u^{\mu}$ satisfying $u^{\mu}n_{\mu} = 1$. In Euclidean thermal field theory, $u^{\mu}$ analytically continues to the (normalized) tangent vector to the thermal circle, the inverse temperature is the integral of $n_{\mu}$ around the circle, and $\mu/T$ is the logarithm of the Milne-invariant holonomy around the circle.

\subsection{Hydrodynamic variables} 
\label{S:basic}

We are now in a position to begin recasting non-relativistic fluid mechanics in a manifestly covariant way. Fluid mechanics describes long wavelength, low frequency fluctuations around thermal equilibrium. In the usual way we assume that the near-equilibrium state can be described by a local temperature $T(x)$, chemical potential $\mu(x)$, and fluid velocity $u^{\mu}(x)$. That is, we promote the parameters that classify the equilibrium state to classical fields. We refer to $(T,\mu,u^{\mu})$ as the ``hydrodynamic variables'' or ``fluid variables.'' 

In this work we also couple the underlying theory to a non-trivial but weakly curved spacetime $\M$, more precisely to the Newton-Cartan structure $(n_{\mu},h_{\mu\nu},A_{\mu})$. Non-relativistic hydrodynamics will be an effective description of the low-energy physics on this spacetime in terms of these collective variables.

\subsection{Constitutive relations and the gradient expansion} 
\label{S:constitutive}

Unlike in conventional effective field theory where the dynamics is specified by a Lagrangian in terms of the low-energy degrees of freedom, in hydrodynamics one continues by supplying constitutive relations for conserved quantities in terms of the hydrodynamic variables. That is, one expresses the spacetime stress tensor $\mathcal{T}^{\mu\nu}$ and energy current $\mathcal{E}^{\mu}$ in terms of the $(T,\mu,u^{\mu})$, the background fields $(n_{\mu},h_{\mu\nu},A_{\mu})$, and gradients of both. The resulting expressions are known as constitutive relations. In conventional effective field theory, the Lagrangian is written in a gradient expansion of the low-energy variables as well as the background. The analogous statement here is that the constitutive relations are organized in a gradient expansion. We count derivatives in the following way. We take the fluid variables and background fields to be $\sim \mathcal{O}(\partial^0)$, so that the field strength of $A_{\mu}$ and the connection $\Gamma$ are both $\sim \mathcal{O}(\partial^1)$. This is the power counting which is appropriate in order to compute hydrodynamic correlation functions in the source-free thermal state. The term $n^{th}$ order hydrodynamics refers to constitutive relations whose terms possess at most $n$ derivatives. In this work we consider zeroth (or ideal) and first order hydrodynamics.

The microscopic field theories which we couple to spacetime are invariant under reparameterizations of coordinates, $U(1)$ gauge transformations, and Milne boosts. The effective hydrodynamic description is invariant under the same symmetries. So, the constitutive relations for $\mathcal{T}^{\mu\nu}$ and $\mathcal{E}^{\mu}$ ought to be expressed in terms of $U(1)$-invariant contravariant symmetric tensors and vectors respectively. Since $\mathcal{T}^{\mu\nu}$ is Milne-invariant, it can be expressed in a basis of Milne-invariant tensors, whereas $\mathcal{E}^{\mu}$ should be specified in such a way that it varies as~\eqref{E:deltaEMilne} under Milne boosts,
\begin{equation*}
( \mathcal{E}')^{\mu} = \mathcal{E}^{\mu} - \left( P_{\nu}^{\rho}\psi_{\rho} - \frac{1}{2}n_{\nu}\psi^2\right) \mathcal{T}^{\mu\nu}\,.
\end{equation*}

The Milne boost symmetry is awkward to enforce by brute force. So we use the machinery we developed in Subsection~\ref{S:milneD}. Namely, rather than constructing tensors with a fixed number of gradients out of $(T,\mu,u^{\mu};n_{\mu},h^{\mu\nu},v^{\mu},A_{\mu})$, we build manifestly Milne-invariant tensors from the data $(T,\mu,u^{\mu},n_{\mu},h^{\mu\nu},\tilde{A}_{\mu})$ and the Milne covariant derivative $\tilde{D}_{\mu}$ defined from~\eqref{E:milneD}. We express the Milne-invariant $\mathcal{T}^{\mu\nu}$ in terms of these manifestly Milne-invariant tensors, but what of the energy current? We separate it into a manifestly Milne-invariant part $\tilde{\mathcal{E}}^{\mu}$, for which we supply constitutive relations in terms of manifestly Milne-invariant tensors, and a part constructed algebraically from $\mathcal{T}^{\mu\nu}$ and the fluid data. In an equation, we have
\beq
\label{E:fromMilneEtoE}
\mathcal{E}^{\mu} = \tilde{\mathcal{E}}^{\mu} + \left( u_{\nu} -\frac{1}{2}n_{\nu}u^2\right)\mathcal{T}^{\mu\nu} \,,
\eeq
where we remind the reader that $u_{\mu} = h_{\mu\nu}u^{\nu}$ and $u^2 = u^{\mu} u_{\mu}$.

Exploiting the residual rotational symmetry which fixes $u^{\mu}$, we write the most general constitutive relations as
\begin{align}
\begin{split}
\label{E:constitutive}
\tilde{\mathcal{E}}^{\mu} & = \mathcal{E} u^{\mu} +\eta^{\mu}\,,
\\
\mathcal{T}^{\mu\nu} & = \mathcal{P} h^{\mu\nu} + \mathcal{N} u^{\mu}u^{\nu} +u^{\mu}q^{\nu} + u^{\nu}q^{\mu} + \tau^{\mu\nu}\,,
\end{split}
\end{align}
where $\eta^{\mu}, q^{\mu}$, and $\tau^{\mu\nu}$ are transverse to $n_{\mu}$ and $\tau^{\mu\nu}$ is traceless. By~\eqref{E:fromMilneEtoE} the total energy current is
\beq
\label{E:totalEnergy}
\mathcal{E}^{\mu} = \left( \mathcal{E} + \frac{1}{2}\mathcal{N} u^2 + u_{\nu} q^{\nu}\right) u^{\mu} + \eta^{\mu} + \mathcal{P} \,P_{\nu}^{\mu} u^{\nu} +\frac{1}{2}u^2 q^{\mu}+ \tau^{\mu\nu}u_{\nu}\,.
\eeq
The constitutive relations of $n^{th}$ order hydrodynamics are expressions of the scalars, vectors, and tensor in~\eqref{E:constitutive} in a basis of $U(1)$ and Milne-invariant scalars, vectors, and tensors with up to $n$ derivatives.

Having described the low-energy variables and the constitutive equations, it only remains to specify equations of motion. These will fix the dynamical fields (and therefore the one-point functions of operators) as functionals of the background fields and the boundary conditions. We take the equations of motions to be the Ward identities~\eqref{E:ward}. However, in practice it will be convenient to rewrite the Ward identities  in terms of manifestly boost-invariant quantities as in~\eqref{E:ward2}.

In what follows it will be useful to classify the inequivalent one-derivative Milne-invariant data. Because the Milne-invariant field strength $\tilde{F}_{\mu\nu}$ can be constructed from the derivative of $u^{\mu}$, all one-derivative tensor data comes from the derivative of the velocity. We decompose the derivative into the various representations of the residual rotational symmetry which fixes $u^{\mu}$ in the same way as in~\eqref{E:Dv},
\beq
\label{E:Du}
\tilde{D}_{\mu} u^{\nu} = - n_{\mu} E^{\nu} + \frac{1}{2}B_{\mu}{}^{\nu} +\tilde{h}_{\mu\rho} \sigma^{\nu\rho} + \frac{1}{d-1}\tilde{P}^{\nu}_{\mu} \vartheta\,,
\eeq
where we have decomposed the field strength $\tilde{F}_{\mu\nu}$ into electric and magnetic parts with respect to $u^{\mu}$,
\beq
E_{\mu} = \tilde{F}_{\mu\nu}u^{\nu}\,, \qquad \tilde{F}_{\mu\nu} = E_{\mu}n_{\nu} - n_{\mu}E_{\nu} + B_{\mu\nu}\,,
\eeq
and indices are raised with $h^{\mu\nu}$. We have also implicitly defined the expansion $\vartheta$ and shear tensor $\sigma^{\mu\nu}$ via
\beq
\vartheta = \tilde{D}_{\mu}u^{\mu}\,, \qquad \sigma^{\mu\nu} = \frac{1}{2}\left( \tilde{D}^{\mu}u^{\nu} + \tilde{D}^{\nu}u^{\mu} - \frac{2}{d-1}h^{\mu\nu}\vartheta\right)\,,
\eeq
and we again remind the reader that indices are raised with $h^{\mu\nu}$. As a result the various tensors defined here are transverse to $n_{\mu}$ and $u^{\mu}$. We also decompose the derivative of $n$, $F^n_{\mu\nu}=\partial_{\mu}n_{\nu}-\partial_{\nu}n_{\mu}$, into electric and magnetic parts as
\beq
E^n_{\mu} = F^n_{\mu\nu}u^{\nu}\,, \qquad F^n_{\mu\nu} = E^n_{\mu}n_{\nu} - E^n_{\nu}n_{\mu} + B^n_{\mu\nu}\,,
\eeq
so that
\beq
\tilde{\mathcal{G}}_{\mu} = - F^n_{\mu\nu}u^{\nu} = - E^n_{\mu}\,.
\eeq

In writing the various one-derivative tensors, we have done so in a manifestly Milne-invariant way. This is useful, but to some extent it obscures physics. For example, the electric field $E^{\mu}$ is a rather non-trivial combination of the electromagnetic field which couples to particle number, the fluid velocity, and the remaining Newton-Cartan data. As a result we think it is instructive to decompose the various one-derivative tensors into their constituents in order to appreciate what the Milne symmetry does. We find
\begin{align}
\begin{split}
E_{\mu} &= \tilde{F}_{\mu\nu}u^{\nu} = F_{\mu\nu}u^{\nu}+ (D_{\mu}u_{\nu}-D_{\nu}u_{\mu})u^{\nu} - \frac{1}{2}E^n_{\mu}u^2 + \frac{1}{2}\left( n_{\mu}\dot{u}^2 - \partial_{\mu}u^2\right)\,,
\\
B^{\mu\nu} & = h^{\mu\rho}h^{\nu\sigma}\tilde{F}_{\rho\sigma} = F^{\mu\nu} + D^{\mu}\left( P^{\nu}_{\rho}u^{\rho}\right) - D^{\nu}\left( P^{\mu}_{\rho}u^{\rho}\right) - \frac{1}{2}u^2 (B^n)^{\mu\nu}\,,
\\
\vartheta & = \tilde{D}_{\mu}u^{\mu} = (D_{\mu}-\mathcal{G}_{\mu})u^{\mu} = \frac{1}{\sqrt{\gamma}}\partial_{\mu}\left( \sqrt{\gamma}u^{\mu}\right)\,,
\\
\sigma^{\mu\nu} & = \frac{1}{2}h^{\mu\rho}h^{\nu\sigma}\pounds_u \tilde{h}_{\rho\sigma}- \frac{h^{\mu\nu}}{d-1}\vartheta \,.
\end{split}
\end{align}
Note that $\vartheta$ and $\sigma^{\mu\nu}$ do not depend on $A_{\mu}$.

\subsection{Positivity of entropy production}
\label{S:entropy}

The last ingredient in formulating hydrodynamics is to demand a local version of the second Law of thermodynamics. That is, we mandate the existence of an entropy current $S^{\mu}$ whose divergence is semi-positive-definite for fluid flows which solve the hydrodynamic equations.

Now recall~\eqref{E:divergence}, which tells us that
\begin{equation*}
\left( \tilde{D}_{\mu}-\tilde{\mathcal{G}}_{\mu}\right) \mathfrak{v}^{\mu} = \left( D_{\mu} - \mathcal{G}_{\mu}\right) \mathfrak{v}^{\mu} = \frac{1}{\sqrt{\gamma}}\partial_{\mu} \left( \sqrt{\gamma}\mathfrak{v}^{\mu}\right)\,,
\end{equation*}
for any vector field $\mathfrak{v}^{\mu}$, so that the divergence defined through $\tilde{D}_{\mu}-\tilde{G}_{\mu}$ is equal to the usual notion of a divergence (recall that $d^dx\sqrt{\gamma}$ is the covariant volume element). We then write the entropy criterion as
\beq
\left( \tilde{D}_{\mu} - \tilde{G}_{\mu}\right)S^{\mu} \geq 0\,.
\eeq
There are various ways of implementing this condition, as we discuss below.

\section{Reformulating non-relativistic hydrodynamics} 
\label{S:NRhydro}

In Subsections~\ref{S:basic}-\ref{S:entropy} we laid down a modern presentation of non-relativistic hydrodynamics in a gradient expansion. To construct $n^{th}$ order hydrodynamics, we follow a two step program:
\begin{enumerate}
\item Specify the most general constitutive relations for $\mathcal{T}^{\mu\nu}$ and $\tilde{\mathcal{E}}^{\mu}$ with tensors containing up to $n$ derivatives of the fluid variables and background fields.
\item Demand the existence of an entropy current.
\end{enumerate}

In the remainder of this Section, we go through this algorithm for the zeroth and first order hydrodynamics of normal fluids. Before tackling first-order hydrodynamics, we discuss the role of field redefinitions of the fluid variables, as well as two different methods for solving the entropy constraint.

\subsection{Ideal hydrodynamics} 
\label{S:ideal}

We begin with ideal hydrodynamics, that is hydrodynamics at zeroth order in gradients. The most general constitutive relations at this order are
\begin{align}
\begin{split}
\label{E:idealConstitutive}
\mathcal{T}^{\mu\nu} & = P \, h^{\mu\nu} + \rho \, u^{\mu}u^{\nu}\,,
\\
\tilde{\mathcal{E}}^{\mu} & = \varepsilon\, u^{\mu}\,,
\end{split}
\end{align}
where $P$ is the pressure, $\rho$ the particle number density, and $\varepsilon$ the energy density. They are related by
\beq
dP = s \,dT + \rho\, d\mu\,, \qquad \varepsilon = - P + T s + \mu \rho\,,
\eeq
where $s$ is the entropy density.~\eqref{E:idealConstitutive} and the Ward identities~\eqref{E:ward2} completely specify ideal non-relativistic hydrodynamics. Later in Subsection~\ref{S:Z0derivative} we will obtain these one-point functions from general properties of the thermal partition function. 

As the presentation~\eqref{E:idealConstitutive} differs somewhat from the usual textbook presentation of ideal hydrodynamics, we pause to verify that~\eqref{E:idealConstitutive} is simply a repackaging of the canonical expressions for the constitutive relations, at least in flat space. Using the definition~\eqref{E:calT}  of the spacetime stress tensor as well as the Milne Ward identity~\eqref{E:milneWard}, we deduce the particle number and momentum currents along with the spatial stress tensor,
\beq
J^{\mu} = \rho \,u^{\mu}\,, \qquad \mathcal{P}_{\mu} = \rho\, u_{\mu}\,, \qquad T_{\mu\nu} = P\,h_{\mu\nu} + \rho \,u_{\mu}u_{\nu}\,.
\eeq 
The full energy current obtained from $\mathcal{T}^{\mu\nu}$ and $\tilde{\mathcal{E}}^{\mu}$ is
\beq
\label{E:idealE}
\mathcal{E}^{\mu} = \left( \varepsilon + \frac{1}{2}\rho u^2\right) u^{\mu} + P \, P_{\nu}^{\mu}u^{\nu}\,,
\eeq
Now we specialize to fluids in flat space. Recall that the flat Newton-Cartan structure is (up to a constant boost)
\beq
n_{\mu}dx^{\mu} = dx^0\,, \qquad h^{\mu\nu}\partial_{\mu}\otimes \partial_{\nu} = \delta^{ij}\partial_i\otimes \partial_j \,, \qquad v^{\mu}\partial_{\mu} = \partial_0\,, \qquad A = 0\,.
\eeq
Then the fluid velocity is
\beq
u^{\mu}\partial_{\mu} = \partial_0 + u^i\partial_i\,.
\eeq
In this case the constitutive relations~\eqref{E:idealConstitutive} then take their more familiar form
\begin{align}
\nonumber
J^0 & = \rho\,, & J^i &  =\rho\, u^i\,,
\\
\mathcal{E}^0 &= \varepsilon + \frac{1}{2}\rho u^2\,, & \mathcal{E}^i & =\left(  \varepsilon + P + \frac{1}{2}\rho u^2\right) u^i\,,
\\
\nonumber
\mathcal{P}_i & = \rho \, u_i\,, & T_{ij} & = P \,\delta_{ij} + \rho \, u_i u_j\,,
\end{align}
and the Ward identities~\eqref{E:ward} become
\beq
\partial_{\mu} J^{\mu} = 0\,, \qquad \partial_{\mu} \mathcal{E}^{\mu} = 0\,, \qquad \dot{\mathcal{P}}_i + \partial_j T_{ij} = 0\,.
\eeq
Note that all of the ``strange'' terms in the energy current (the $\frac{1}{2}\rho u^2$ as well as the pressure term) arise from the relation~\eqref{E:fromMilneEtoE} between the boost-invariant energy current and the physical energy current. 

Before going on to consider hydrodynamics at higher order in the gradient expansion, we verify that the ideal constitutive relations~\eqref{E:idealConstitutive} are consistent with the existence of an entropy current. To do so we consider a linear combination of the Ward identities for the number and energy currents From~\eqref{E:ward2}, these are
\beq
\left( \tilde{D}_{\mu}-\tilde{\mathcal{G}}_{\mu}\right)J^{\mu} = 0\,, \qquad \left( \tilde{D}_{\mu}-2\tilde{\mathcal{G}}_{\mu}\right)\tilde{\mathcal{E}}^{\mu} + \tilde{h}_{\rho(\mu}\tilde{D}_{\nu)}u^{\rho}\mathcal{T}^{\mu\nu} = 0\,.
\eeq
Massaging, we get
\begin{align}
\begin{split}
\label{E:idealEntropy}
0&=\frac{1}{T}\left\{ \left[ \left( \tilde{D}_{\mu}-2\tilde{\mathcal{G}}_{\mu}\right)\tilde{\mathcal{E}}^{\mu} +\tilde{h}_{\rho(\mu}\tilde{D}_{\nu)} u^{\rho} \mathcal{T}^{\mu\nu}\right] - \mu\left( \tilde{D}_{\mu}-\tilde{\mathcal{G}}_{\mu}\right)J^{\mu}\right\}
\\
&=\frac{1}{T}\left\{ \tilde{D}_{\mu}(\varepsilon u^{\mu} )+  \frac{1}{d-1}\tilde{h}_{\mu\nu}\vartheta \left(P h^{\mu\nu}+\rho u^{\mu}u^{\nu}\right) - \mu \tilde{D}_{\mu} (\rho u^{\mu})\right\}
\\
& = \frac{1}{T}\left\{ u^{\mu} \left( \mu \partial_{\mu} \rho + T \partial_{\mu} s \right) + (\varepsilon+P-\mu\rho)\vartheta - \mu u^{\mu}\partial_{\mu}\rho \right\}
\\
& = \left( \tilde{D}_{\mu}-\tilde{\mathcal{G}}_{\mu}\right)(su^{\mu})\,,
\end{split}
\end{align}
where we have used
\beq
d\varepsilon = T ds + \mu d\rho\,,
\eeq
along with the decomposition of the velocity~\eqref{E:Du} and that $\tilde{\mathcal{G}}_{\mu} = - E^n_{\mu}$ is transverse. So there is an entropy current $S^{\mu} = s u^{\mu}$ which is identically conserved for fluid flows which solve the equations of motion of ideal hydrodynamics. 

\subsection{Hydrodynamic frame transformations and frame-invariants} 
\label{S:frame}

Temperature, chemical potential, and fluid velocity are uniquely defined in flat space, translationally-invariant thermal states. Correspondingly, there is no unique notion of e.g. temperature near equilibrium, or even in a hydrostatic equilibrium. In hydrodynamics, this is merely the statement that $(T,\mu,u^{\mu})$ are dynamical fields, on which we are free to perform field redefinitions.

A hydrodynamic frame transformation is a field redefinition of $(T,\mu,u^{\mu})$ by terms involving at least one gradient, so that the $(T,\mu,u^{\mu})$ of the flat-space, translationally-invariant equilibrium are unchanged. Rather than leaving a Lagrangian invariant, hydrodynamic frame transformations leave the one-point functions $\mathcal{T}^{\mu\nu}$ and $\mathcal{E}^{\mu}$ invariant. A particular choice of $(T,\mu,u^{\mu})$ is known as a choice of hydrodynamic frame. When specifying the constitutive relations, one implicitly does so in a choice of frame. Of course, physics is independent of this choice, and so it is useful to either (i.) fix the frame, or (ii.) work in a frame-independent way. We discuss both in this work.

In $d$ spacetime dimensions, there are two scalar fluid variables and a spatial vector's worth of fluid variables, and so one can completely fix the frame by fixing two scalars and a spatial vector in the constitutive relations. For instance, we can choose to work in the ``Landau frame''
\beq
\label{E:landau}
\mathcal{E} = \varepsilon\,, \qquad \mathcal{N} =\rho\,, \qquad q^{\mu} = 0\,,
\eeq
or in the ``Eckhart frame''\footnote{More precisely, the terms ``Landau frame'' and ``Eckhart frame'' refer to frames in relativistic hydrodynamics in which either the energy flux or charge flux vanish. We borrow this terminology for non-relativistic fluids. It is amusing to note that, under this convention, Landau himself uses the ``Landau frame'' in his textbook discussion of relativistic hydrodynamics~\cite{LL6}, but the ``Eckhart frame'' for non-relativistic hydrodynamics.}
\beq
\label{E:eckhart}
\mathcal{E} = \varepsilon\,, \qquad \mathcal{N} = \rho\,, \qquad \eta^{\mu} = 0\,,
\eeq
where $\varepsilon$ and $\rho$ are the thermodynamic energy and number densities respectively. There is another frame, the ``adiabatic frame,'' which will prove especially useful later in this work.

Of course, it is nice to work in a way that is independent of the choice of frame. To do so, we must first compute the variation of the quantities $\mathcal{E},\mathcal{N}$, \&c appearing in the constitutive relations~\eqref{E:constitutive} under frame transformations. We can then obtain invariant combinations. 

So consider a redefinition of the hydrodynamic variables,
\beq
\label{E:frameChange}
T \to T + \delta T\,, \qquad \mu \to \mu + \delta \mu\,, \qquad u^{\mu} \to u^{\mu} + \delta u^{\mu}\,,
\eeq
where $(\delta T,\delta\mu,\delta u^{\mu})$ have at least one derivative built out of the hydrodynamic variables and background fields. Note that since $n_{\mu} u^{\mu} =1$, we require $n_{\mu} \delta u^{\mu} = 0$, i.e. $\delta u^{\mu}$ is transverse. We also need to know that
\begin{align}
\begin{split}
\mathcal{E} & = \varepsilon + \mathcal{O}(\partial)\,, \qquad \mathcal{N} = \rho + \mathcal{O}(\partial)\,, \qquad \mathcal{P} = P + \mathcal{O}(\partial)\,, \\
q^{\mu} & = \mathcal{O}(\partial)\,, \hspace{.545in} \eta^{\mu} = \mathcal{O}(\partial)\,, \hspace{.47in} \tau^{\mu\nu} = \mathcal{O}(\partial)\,.
\end{split}
\end{align}
Under the transformation~\eqref{E:frameChange} and using~\eqref{E:constitutive},~\eqref{E:totalEnergy}, we then have
\begin{align}
\nonumber
\mathcal{T}^{\mu\nu} & = \left( \mathcal{N} + \delta \rho\right) u^{\mu}u^{\nu} + \left( \mathcal{P} + \delta P\right) h^{\mu\nu} + u^{\mu}\left( q^{\nu} + \rho \delta u^{\nu}\right) + u^{\nu}\left( q^{\mu} + \rho \delta u^{\mu}\right) + \tau^{\mu\nu} + \mathcal{O}(\partial^2)\,,
\\
\nonumber
\mathcal{E}^{\mu} & = \left( \mathcal{E} + \frac{1}{2}\mathcal{N}u^2 + u_{\nu}q^{\nu} + \delta \varepsilon + \frac{1}{2}\delta \rho\, u^2 + \rho u_{\nu}\delta u^{\nu}\right)u^{\mu} + \eta^{\mu} + \left( \varepsilon+P + \frac{1}{2}\rho u^2\right)\delta u^{\mu}
\\
\label{E:frameTE}
& \qquad \qquad + (\mathcal{P} + \delta P )\, P^{\mu}_{\nu}u^{\nu} + \frac{1}{2}u^2q^{\mu} + \tau^{\mu\nu}u_{\nu} + \mathcal{O}(\partial^2)\,,
\end{align}
where we have defined
\beq
\delta P = \rho \delta \mu + s \delta T\,, \qquad \delta \rho = \frac{\partial \rho}{\partial T}\delta T + \frac{\partial\rho}{\partial\mu}\delta \mu\,, \qquad \delta \varepsilon = \frac{\partial\varepsilon}{\partial T}\delta T + \frac{\partial \varepsilon}{\partial\mu}\delta \mu\,.
\eeq

By~\eqref{E:fromMilneEtoE}, the boost-invariant energy current becomes\footnote{Note that this differs from what we would obtain if we fixed $\tilde{\mathcal{E}}^{\mu}$, which is constructed with the fluid variables, rather than the physical energy current $\mathcal{E}^{\mu}$. If we fixed $\tilde{\mathcal{E}}^{\mu}$, we would find
\begin{equation*}
\tilde{\mathcal{E}}^{\mu} = \left( \mathcal{E} + \delta \varepsilon\right)u^{\mu} + \eta^{\mu} + \varepsilon \delta u^{\mu} + \mathcal{O}(\partial^2)\,
\end{equation*}
instead of~\eqref{E:frameME}.}
\beq
\label{E:frameME}
\tilde{\mathcal{E}}^{\mu} = \left( \mathcal{E} + \delta \varepsilon\right)u^{\mu} + \eta^{\mu} + \left( \varepsilon + P\right)\delta u^{\mu} + \mathcal{O}(\partial^2)\,.
\eeq
Comparing~\eqref{E:frameTE} and~\eqref{E:frameME} with~\eqref{E:constitutive}, we see that the variations of $\mathcal{E},\mathcal{N},$ \&c under frame transformation properties are
\begin{align}
\nonumber
\mathcal{E} &\to \mathcal{E} + \delta\varepsilon + \mathcal{O}(\partial^2)\,, &\mathcal{N} & \to \mathcal{N} + \delta \rho + \mathcal{O}(\partial^2)\,,
\\
\label{E:frameTransformationOfConstitutive}
\mathcal{P} &\to \mathcal{P} + \delta P + \mathcal{O}(\partial^2)\,, & q^{\mu} &\to q^{\mu} + \rho \delta u^{\mu}+\mathcal{O}(\partial^2)\,, 
\\
\nonumber
\eta^{\mu} &\to \eta^{\mu} + \left( \varepsilon+ P\right)\delta u^{\mu}+\mathcal{O}(\partial^2)\,,&\tau^{\mu\nu}&\to \tau^{\mu\nu}+\mathcal{O}(\partial^2)\,.
\end{align}
Using that we can take $\delta \rho$ and $\delta \varepsilon$ to be the independent scalar redefinitions so that
\beq
\delta P = \left( \frac{\partial P}{\partial\varepsilon}\right)_{\rho} \delta \varepsilon + \left( \frac{\partial P}{\partial \rho}\right)_{\varepsilon}\delta\rho\,,
\eeq
we can deduce the frame-invariant combinations of $\mathcal{E},\mathcal{N},$ \&c from~\eqref{E:frameTransformationOfConstitutive}, at least to first order in gradients. They are
\begin{align}
\begin{split}
\label{E:frameInvariants}
\mathscr{S} & = \mathcal{P}-P  - \frac{\partial P}{\partial\varepsilon}\left( \mathcal{E}-\varepsilon\right)  - \frac{\partial P}{\partial\rho}\left( \mathcal{N} -\rho\right)+ \mathcal{O}(\partial^2)\,,
\\
\mathscr{V}^{\mu} & = q^{\mu} - \frac{\rho}{\varepsilon + P}\eta^{\mu} + \mathcal{O}(\partial^2)\,,
\\
\mathscr{T}^{\mu\nu} & = \tau^{\mu\nu} + \mathcal{O}(\partial^2)\,,
\end{split}
\end{align}
where we use $\mathscr{S}, \mathscr{V}^{\mu}$, and $\mathscr{T}^{\mu\nu}$ to denote the scalar, transverse vector, and transverse traceless tensor frame-invariants.

\subsection{The canonical entropy current and adiabaticity} 
\label{S:canonical}

We now undertake a general discussion of the entropy criterion. Our presentation will differ slightly from the usual textbook treatment, and instead will follow the spirit of~\cite{Loganayagam:2011mu}.

In any physical process, the total number $M$ and rest-frame energy $\tilde{E}$ are unchanged (modulo Lorentz forces or energy injection),
\begin{equation*}
\delta M = 0\,, \qquad \delta \tilde{E} = 0\,,
\end{equation*}
so that one can trivially rewrite the second Law as
\begin{equation*}
T \delta S + \mu \delta M - \delta \tilde{E} \geq 0\,.
\end{equation*}
In addition to encoding the second Law for physical processes, this relation implicitly encodes the change of entropy during an \emph{adiabatic} process. Recall that in an adiabatic process, $M$ and $\tilde{E}$ can change, but the entropy $S$ also changes in such a way that the thermodynamic relation $T\delta S + \mu \delta M = \delta \tilde{E}$ holds.

The local version of this inequality is
\beq
\label{E:entropy}
T \left( \tilde{D}_{\mu} - \tilde{\mathcal{G}}_{\mu}\right) S^{\mu} + \mu \left( \tilde{D}_{\mu}-\tilde{\mathcal{G}}_{\mu}\right)J^{\mu} - \left( \tilde{D}_{\mu}-2\tilde{\mathcal{G}}_{\mu}\right)\tilde{\mathcal{E}}^{\mu} - \tilde{h}_{\rho(\mu}\tilde{D}_{\nu)}u^{\rho}\mathcal{T}^{\mu\nu}
\geq 0\,,
\eeq
where $S^{\mu}$ is the entropy current and the other terms are $\mu$ times the number Ward identity and (minus) the energy Ward identity. This condition is equivalent to the usual notion of positivity of entropy production. A physical process is one which solves the hydrodynamic equations, in which case~\eqref{E:entropy} becomes
\beq
\left( \tilde{D}_{\mu}-\tilde{\mathcal{G}}_{\mu}\right)S^{\mu}\geq 0\,.
\eeq

Let us massage~\eqref{E:entropy} into a more instructive form. We begin with the linear combination of the Ward identities, which we rewrite in terms of the constitutive relations~\eqref{E:constitutive}
\begin{align}
\begin{split}
&\mu \left( \tilde{D}_{\mu}-\tilde{\mathcal{G}}_{\mu}\right)J^{\mu} - \left( \tilde{D}_{\mu}-2\tilde{\mathcal{G}}_{\mu}\right)\tilde{\mathcal{E}}^{\mu} - \tilde{h}_{\rho(\mu}\tilde{D}_{\nu)}u^{\rho}\mathcal{T}^{\mu\nu}
\\
 = & T \left( \tilde{D}_{\mu} - \tilde{\mathcal{G}}_{\mu}\right) \left\{ \frac{\mu}{T}J^{\mu} - \frac{1}{T}\tilde{\mathcal{E}}^{\mu}\right\} - \mathcal{P}\vartheta  - \frac{\mathcal{E}}{T}\dot{T} - \mathcal{N}T \dot{\left( \frac{\mu}{T}\right)}+ q^{\mu} \left( E_{\mu} - T \tilde{D}_{\mu}\left( \frac{\mu}{T}\right)\right)
 \\
 & \quad - \eta^{\mu}\left( E^n_{\mu} + \frac{\tilde{D}_{\mu}T}{T}\right) - \tau^{\mu\nu}\sigma_{\mu\nu}\,,
\end{split}
\end{align}
where we have lowered the indices of $\sigma^{\mu\nu}$ with $\tilde{h}_{\mu\nu}$ and have defined
\beq
\dot{\mathcal{X}} = u^{\mu}\partial_{\mu}\mathcal{X}\,,
\eeq
for any scalar $\mathcal{X}$. We then separate the entropy current into a ``canonical'' part and a ``non-canonical'' part,
\begin{align}
\begin{split}
\label{E:canonicalEntropy}
S^{\mu}& = S_{canon}^{\mu} + S^{\mu}_{non}\,,
\\
S_{canon}^{\mu}& \equiv \frac{P}{T} u^{\mu} - \frac{\mu}{T}J^{\mu} + \frac{1}{T}\tilde{\mathcal{E}}^{\mu}\,.
\end{split}
\end{align}
Note that
\begin{equation*}
S_{canon}^{\mu} = su^{\mu} + \mathcal{O}(\partial)\,,
\end{equation*}
so that we require
\begin{equation*}
S_{non}^{\mu} = \mathcal{O}(\partial)\,.
\end{equation*}
Using that
\beq
\left( \tilde{D}_{\mu} - \tilde{\mathcal{G}}_{\mu}\right) \left( \frac{P}{T}u^{\mu}\right) = P\vartheta + \frac{\varepsilon}{T}\dot{T} + \rho T \dot{\left( \frac{\mu}{T}\right)}\,,
\eeq
\eqref{E:entropy} becomes
\begin{align}
\begin{split}
\label{E:entropy2}
T \left(\tilde{D}_{\mu}-\tilde{\mathcal{G}}_{\mu}\right) S^{\mu}_{non} - \left( \mathcal{P} - P\right)\vartheta - \frac{\mathcal{E}-\varepsilon}{T}\dot{T} - \left( \mathcal{N}- \rho\right) T \dot{\left( \frac{\mu}{T}\right)} & 
\\
+ q^{\mu}\left( E_{\mu} - T \tilde{D}_{\mu}\left( \frac{\mu}{T}\right)\right) - \eta^{\mu}\left( E^n_{\mu} + \frac{\tilde{D}_{\mu}T}{T}\right) - \tau^{\mu\nu}\sigma_{\mu\nu} & \geq 0\,,
\end{split}
\end{align}
which will be the form that is more useful for us in Section~\ref{S:2dFluids}. Note that by our definition of the canonical entropy current~\eqref{E:canonicalEntropy}, each term in this inequality has at least two gradients.

Because~\eqref{E:entropy2} is linear in both $S^{\mu}_{non}$ and constitutive relations, we can decompose $S_{non}^{\mu}$ and the constitutive relations into sums of a dissipative part and a dissipationless (or perhaps more aptly, adiabatic) part. That is,
\begin{align}
\begin{split}
S_{non}^{\mu} &= S_{diss}^{\mu} + S_{non-diss}^{\mu}\,,
\\
\tilde{\mathcal{E}}^{\mu} & =  \tilde{\mathcal{E}}_{diss}^{\mu} + \tilde{\mathcal{E}}_{non-diss}^{\mu}\,,
\\
\mathcal{T}^{\mu\nu}& = \mathcal{T}^{\mu\nu}_{diss} + \mathcal{T}^{\mu\nu}_{non-diss}\,,
\end{split}
\end{align}
with
\begin{align}
\begin{split}
\tilde{\mathcal{E}}^{\mu}_{non-diss} &= \varepsilon u^{\mu} + \mathcal{O}(\partial)\,, 
\\
\mathcal{T}^{\mu\nu}_{non-diss} & = P h^{\mu\nu} + \rho u^{\mu}u^{\nu} + \mathcal{O}(\partial)\,,
\end{split}
\end{align}
and the dissipative parts are at least $\mathcal{O}(\partial)$. The crucial feature of the dissipationless part is that it solves the \emph{adiabaticity equation}
\begin{align}
\begin{split}
\label{E:adiabaticity}
T \left( \tilde{D}_{\mu} - \tilde{\mathcal{G}}_{\mu}\right)S_{non-diss}^{\mu} - \left( \mathcal{P}_{non-diss} - P\right)\vartheta - \frac{\mathcal{E}_{non-diss}-\varepsilon}{T}\dot{T} - \left( \mathcal{N}_{non-diss}-\rho\right)T \dot{\left( \frac{\mu}{T}\right)}&
\\
+q^{\mu}_{non-diss}\left( E_{\mu} - T \tilde{D}_{\mu}\left( \frac{\mu}{T}\right) \right) - \eta^{\mu}_{non-diss} \left( E^n_{\mu} + \frac{\tilde{D}_{\mu}T}{T}\right) - \tau^{\mu\nu}_{non-diss}\sigma_{\mu\nu} & = 0\,,
\end{split}
\end{align}
which one should think of as the local version of thermodynamic adiabaticity $T \delta S + \mu \delta M - \delta \tilde{E} = 0$. Then~\eqref{E:entropy2} tells us that the dissipative part satisfies a similar relation, except that the $=0$ on the RHS is replaced with $\geq 0$.

Since the adiabaticity equation~\eqref{E:adiabaticity} is linear in the entropy current and constitutive relations, we can classify its solutions at any order in the gradient expansion, independent of hydrodynamics at any other order. That is, we can decompose
\begin{align*}
S_{non-diss}^{\mu} & = (S_{non-diss}^{(1)})^{\mu} + (S_{non-diss}^{(2)})^{\mu} + \hdots\,, 
\end{align*}
and similarly for the constitutive relations, where $S_{non-diss}^{(n)}$ has $n$ derivatives and so on. Then $S_{non-diss}^{(n)}, \mathcal{E}_{non-diss}^{(n)}$, \&c satisfy the adiabaticity equation~\eqref{E:adiabaticity} independently of terms at any other order in derivatives.

The same cannot be said for the dissipative part. For example, one can imagine a solution to the dissipative part in which the RHS is a positive-definite scalar formed from a square of a sum of scalars containing different numbers of derivatives.

In any case, the solutions to adiabaticity are exact to all orders in the gradient expansion. In the next two Sections, we will see that these solutions are intimately related to the currents that follow from the existence of a hydrostatic partition function for the interesting case of a parity-violating fluid in two spatial dimensions.

Before moving on, we have developed the tools to also present the usual formulation of the entropy criterion. We do so presently. The crucial ingredient here is, when constructing hydrodynamics at $n^{th}$ order in gradients, to use the hydrodynamic equations at lower order in derivatives to deduce the ``on-shell'' inequivalent tensors at $\mathcal{O}(\partial^n)$. 

Let us show how this works for first-order hydrodynamics, wherein we require the zeroth order equations. These can be decomposed into two scalars and a single vector. Defining $\nu \equiv \frac{\mu}{T}$, the number and energy Ward identities~\eqref{E:ward2} give
\begin{align}
\begin{split}
0 & = \left( \tilde{D}_{\mu} - \tilde{G}_{\mu}\right) \left( \rho u^{\mu}\right)  + \mathcal{O}(\partial^2)= \dot{\rho} + \rho\vartheta +\mathcal{O}(\partial^2) = \frac{\partial \rho}{\partial T}\dot{T} + \frac{\partial\rho}{\partial\nu}\dot{\nu} + \rho \vartheta + \mathcal{O}(\partial^2)\,,
\\
0 & = \left( \tilde{D}_{\mu}- 2\tilde{G}_{\mu}\right)\left( \varepsilon u^{\mu}\right) + P \vartheta + \mathcal{O}(\partial^2) = \frac{\partial\varepsilon}{\partial T}\dot{T} + \frac{\partial\varepsilon}{\partial\nu}\dot{\nu} + \left( \varepsilon + P\right)\vartheta + \mathcal{O}(\partial^2)\,,
\end{split}
\end{align}
where the thermodynamic derivatives are taken at fixed $T$ and $\nu$ respectively. Using standard thermodynamic relations, it is not hard to show that (see e.g. Appendix B of~\cite{Bhattacharya:2011tra})
\begin{align}
\begin{split}
\label{E:0derivativeScalar}
\frac{\dot{T}}{T} &= \frac{\partial P}{\partial\varepsilon}\vartheta + \mathcal{O}(\partial^2)\,,
\\
 T\dot{\nu} & = \frac{\partial P}{\partial\rho}\vartheta + \mathcal{O}(\partial^2)\,,
\end{split}
\end{align}
where now the thermodynamic derivatives are taken at fixed $\varepsilon$ and $\rho$ respectively. The vector equation is
\begin{align}
\begin{split}
0 & = \tilde{P}^{\mu}_{\rho}\left\{ \left( \tilde{D}_{\nu} - \tilde{\mathcal{G}}_{\nu}\right) \left( P h^{\nu\rho} + \rho u^{\nu}u^{\rho}\right) + (F^n)^{\rho}{}_{\nu}(\varepsilon u^{\nu})\right\} + \mathcal{O}(\partial^2)
\\
& = \tilde{D}^{\mu}P + \left( \varepsilon+P\right)(E^n)^{\mu} - \rho E^{\mu} + \mathcal{O}(\partial^2)
\\
& = \rho\left\{ - \left( E^{\mu} - T \tilde{D}^{\mu}\left( \frac{\mu}{T}\right)\right) + \frac{\varepsilon+P}{\rho}\left( (E^n)^{\mu} + \frac{\tilde{D}^{\mu}T}{T}\right)\right\} + \mathcal{O}(\partial^2)\,,
\end{split}
\end{align}
that is
\beq
\label{E:0derivativeVector}
(E^n)^{\mu} + \frac{\tilde{D}^{\mu}T}{T} = \frac{\rho}{\varepsilon + P}\left( E^{\mu} - T \tilde{D}^{\mu}\left( \frac{\mu}{T}\right)\right) + \mathcal{O}(\partial^2)\,.
\eeq
We remind the reader that indices are raised with $h^{\mu\nu}$. 

So the ideal hydrodynamic equations relate the scalars $\dot{T}$ and $\dot{\nu}$ to the scalar $\vartheta$ and the vector $(E^n)^{\mu} + \tilde{D}^{\mu}T/T$ to the vector $E^{\mu} - T \tilde{D}^{\mu}(\mu/T)$. The resulting ``on-shell'' inequivalent tensor data at one-derivative order is given in Table~\ref{T:oneDerivativeWithP}. Plugging~\eqref{E:0derivativeScalar} and~\eqref{E:0derivativeVector} into the entropy condition~\eqref{E:entropy2} and using the definition of the frame-invariants~\eqref{E:frameInvariants},~\eqref{E:entropy2} becomes
\beq
\label{E:canonicalPositivity}
T\left( \tilde{D}_{\mu}-\tilde{\mathcal{G}}_{\mu}\right)S^{\mu}_{non} - \mathscr{S}\vartheta + \mathscr{V}^{\mu} \left( E_{\mu}- T \tilde{D}_{\mu} \left( \frac{\mu}{T}\right)\right) - \mathscr{T}^{\mu\nu}\sigma_{\mu\nu} + \mathcal{O}(\partial^3)\geq 0\,.
\eeq
Note that the constitutive relations now appear only through the frame-invariant scalar $\mathscr{S}$, vector $\mathscr{V}^{\mu}$, and tensor $\mathscr{T}^{\mu\nu}$. Equivalently, we can interpret this inequality as the rate at which entropy is produced in a physical fluid flow (i.e. one that solves the hydrodynamic equations)
\begin{align}
\begin{split}
T \left( \tilde{D}_{\mu}-\tilde{\mathcal{G}}_{\mu}\right)S^{\mu}|_{on-shell} = &T \left( \tilde{D}_{\mu}-\tilde{\mathcal{G}}_{\mu}\right)S^{\mu}_{non} - \mathscr{S}\vartheta
\\
&\qquad   +\mathscr{V}^{\mu}\left( E_{\mu} - T \tilde{D}_{\mu}\left( \frac{\mu}{T}\right)\right)  - \mathscr{T}^{\mu\nu}\sigma_{\mu\nu} + \mathcal{O}(\partial^3)\geq 0\,.
\end{split}
\end{align}
Consequently, the entropy current is a frame-dependent object but the entropy production is independent of the choice of frame.

\subsection{First-order hydrodynamics of normal, parity-preserving fluids} 
\label{S:1stOrder}

With all of this machinery behind us, we use it to work out non-relativistic hydrodynamics at first order in the gradient expansion. The result will be a covariant version of textbook first-order fluid mechanics. We will do so by the technique outlined at the end of the previous Subsection, wherein we use the hydrodynamic equations at zeroth order in derivatives so that the independent one-derivative data is given in Table~\ref{T:oneDerivativeWithP}, and we solve~\eqref{E:canonicalPositivity}.
\begin{table}
\begin{center}
\begin{tabular}{|c|c|c|c|}
\hline & 1 & 2 & 3 \\ \hline scalars & $\vartheta$ & & \\ \hline vectors & $V^{\mu}\equiv E^{\mu} - T \tilde{D}^{\mu}\left( \frac{\mu}{T}\right)$ & $E^{\mu}$ & $\tilde{D}^{\mu}T$ \\ \hline tensors & $\sigma^{\mu\nu}$ & & \\ \hline
\end{tabular}
\caption{\label{T:oneDerivativeWithP} The inequivalent parity-preserving tensor data at first order in derivatives, having used the ideal hydrodynamic equations.}
\end{center}
\end{table}

To proceed we parameterize the most general non-canonical entropy current to first order in derivatives,
\beq
S^{\mu}_{non} = S_1\vartheta + \Sigma_1 V^{\mu} + \Sigma_2 E^{\mu} + \Sigma_3 \tilde{D}^{\mu}T + \mathcal{O}(\partial^2)\,,
\eeq
where the vector $V^{\mu}$ is defined in Table~\ref{T:oneDerivativeWithP}. Schematically, we have
\begin{equation*}
\left( \tilde{D}_{\mu}-\tilde{G}_{\mu}\right)S^{\mu}_{non} = \left( \text{pure two derivative scalars}\right) + \left( \text{products of one-derivative tensors}\right) + \mathcal{O}(\partial^3)\,.
\end{equation*}
The pure two-derivative terms in this divergence must vanish in order to satisfy the positivity condition~\eqref{E:canonicalPositivity}.

We begin the entropy analysis with these two-derivative terms. Using $E^{\mu} = - u^{\nu}\tilde{D}_{\nu}u^{\mu}$, we readily find
\begin{align}
\begin{split}
\tilde{D}_{\mu} \left( \vartheta u^{\mu}\right) &= \dot{\vartheta} + \vartheta^2\,,
\\
\tilde{D}_{\mu} E^{\mu}& = -\left( \dot{\vartheta} + u^{\mu}u^{\nu}\tilde{R}_{\mu\nu} + \frac{1}{4}B_{\mu\nu}B^{\mu\nu} + \sigma_{\mu\nu}\sigma^{\mu\nu} + \frac{\vartheta^2}{d-1}\right)\,,
\\
\tilde{D}_{\mu}\left( \tilde{D}^{\mu}\mathcal{X}\right) & = h^{\mu\nu}\tilde{D}_{\mu}\tilde{D}_{\nu}\mathcal{X}\,,
\end{split}
\end{align}
where $\tilde{R}_{\mu\nu}=\tilde{R}^{\rho}{}_{\mu\rho\nu}$ is the Ricci curvature constructed from $\tilde{\Gamma}$ in~\eqref{E:milneD} and $\mathcal{X}$ is any invariant scalar.\footnote{Our convention is that the Riemann curvature is given through the commutator of covariant derivatives as $[\tilde{D}_{\mu},\tilde{D}_{\nu}]\mathfrak{v}^{\rho} \equiv \tilde{R}^{\rho}{}_{\sigma\mu\nu}\mathfrak{v}^{\sigma}$ for any vector field $\mathfrak{v}^{\mu}$. This gives the expression $\tilde{R}^{\mu}{}_{\nu\rho\sigma} = \partial_{\rho}\tilde{\Gamma}^{\mu}{}_{\nu\sigma} - \partial_{\sigma}\tilde{\Gamma}^{\mu}{}_{\nu\rho} + \tilde{\Gamma}^{\mu}{}_{\tau\rho}\tilde{\Gamma}^{\tau}{}_{\nu\sigma} - \tilde{\Gamma}^{\mu}{}_{\tau\sigma}\tilde{\Gamma}^{\tau}{}_{\nu\rho}$.} Then
\begin{align}
\begin{split}
\left( \tilde{D}_{\mu} - \tilde{G}_{\mu}\right) S^{\mu}_{non} = &\left( S_1 - \Sigma_1 - \Sigma_2\right) \dot{\vartheta} -\left( \Sigma_1 + \Sigma_2\right) u^{\mu}u^{\nu}\tilde{R}_{\mu\nu} - \frac{\Sigma_1}{T} h^{\mu\nu}\tilde{D}_{\mu}\tilde{D}_{\nu}\frac{\mu}{T}  
\\
& \,\,\, + \Sigma_3 h^{\mu\nu}\tilde{D}_{\mu}\tilde{D}_{\nu}T+ \left( \text{products of one-derivative data}\right) + \mathcal{O}(\partial^3)\,,
\end{split}
\end{align}
which immediately gives
\beq
S_1 = 0\,, \qquad \Sigma_i = 0\,.
\eeq
In other words, the non-canonical entropy current vanishes to first order in gradients,
\beq
S_{non}^{\mu} = \mathcal{O}(\partial^2)\,.
\eeq

The positivity of entropy production~\eqref{E:canonicalPositivity} then becomes
\beq
- \mathscr{S}\vartheta + \mathscr{V}^{\mu} V_{\mu} - \mathscr{T}^{\mu\nu}\sigma_{\mu\nu} + \mathcal{O}(\partial^3)\geq 0\,.
\eeq
The solution is
\beq
\label{E:oneDerivativePSummary}
\mathscr{S}  = - \zeta \vartheta\,, \qquad \mathscr{V}^{\mu} = \sigma V^{\mu}\,, \qquad \mathscr{T}^{\mu\nu} = - \eta \sigma^{\mu\nu}\,,
\eeq
where the bulk viscosity $\zeta$, conductivity $\sigma$ and shear viscosity $\eta$ are non-negative,
\beq
\zeta \geq 0\,, \qquad \sigma \geq 0\,, \qquad \eta \geq 0\,.
\eeq
For a physical fluid flow, the divergence of the entropy current $S^{\mu} = S_{canon}^{\mu} + \mathcal{O}(\partial^2)$ is then
\beq
\left( \tilde{D}_{\mu}-\tilde{\mathcal{G}}_{\mu}\right) S^{\mu} = \frac{\zeta}{T}\vartheta^2 + \frac{\sigma}{T}V^{\mu}V_{\mu} + \frac{\eta}{T}\sigma_{\mu\nu}\sigma^{\mu\nu} + \mathcal{O}(\partial^3)\,.
\eeq

It is instructive to parse these results by presenting the constitutive relations in a particular frame. We present the results above both in ``Landau frame''~\eqref{E:landau} and in ``Eckhart frame,''~\eqref{E:eckhart}, beginning with ``Landau frame.'' Using~\eqref{E:frameInvariants} we have
\begin{subequations}
\begin{align}
\begin{split}
\tilde{\mathcal{E}}^{\mu} &= \varepsilon u^{\mu} +\mathcal{O}(\partial^2)\,,
\\
\mathcal{T}^{\mu\nu} & = \left( P - \zeta \vartheta\right)h^{\mu\nu} + \rho u^{\mu}u^{\nu} + \sigma \left( u^{\mu}V^{\nu}+u^{\nu}V^{\mu}\right)- \eta \sigma^{\mu\nu} + \mathcal{O}(\partial^2)\,.
\end{split}
\end{align}
The total energy current~\eqref{E:totalEnergy} is
\beq
\mathcal{E}^{\mu} = \left( \varepsilon + \frac{1}{2}\rho u^2 + \sigma u_{\nu}V^{\nu}\right)u^{\mu} + \left( P- \zeta \vartheta\right)P^{\mu}_{\nu}u^{\nu} + \frac{1}{2}\sigma u^2 V^{\mu} - \eta \sigma^{\mu\nu}u_{\nu} + \mathcal{O}(\partial^2)\,.
\eeq
Decomposing $\mathcal{T}^{\mu\nu}$ into the number current, momentum current, and spatial stress tensor via~\eqref{E:calT}, we have
\begin{align}
\nonumber
J^{\mu} & = \rho u^{\mu} + \sigma \left( E^{\mu} - T\tilde{D}^{\mu}\left( \frac{\mu}{T}\right)\right) + \mathcal{O}(\partial^2)\,,
\\
\mathcal{P}_{\mu} & = \rho u_{\mu} + \sigma h_{\mu\nu}\left( E^{\nu} - T \tilde{D}^{\nu}\left( \frac{\mu}{T}\right)\right) + \mathcal{O}(\partial^2)\,,
\\
\nonumber
T_{\mu\nu} & = \left( P - \zeta \vartheta\right) h_{\mu\nu} + \rho u_{\mu}u_{\nu} + \sigma \left( u_{\mu}h_{\nu\rho} + u_{\nu}h_{\mu\rho}\right)\left[ E^{\rho} - T \tilde{D}^{\rho}\left( \frac{\mu}{T}\right)\right] - \eta h_{\mu\alpha}h_{\nu\beta} \sigma^{\alpha\beta} + \mathcal{O}(\partial^2)\,.
\end{align}
\end{subequations}
Inspecting the expression for the current, it is now clear why we labeled the transport coefficient multiplying $E^{\mu} - T \tilde{D}^{\mu}\left( \frac{\mu}{T}\right)$ as a ``conductivity'' $\sigma$. 

Now let us go to Eckhart frame~\eqref{E:eckhart}. To make contact with the usual presentation of first-order hydrodynamics in terms of a thermal conductivity $\kappa$ rather than number conductivity $\sigma$ (as found in e.g.~\cite{LL6}), let us use~\eqref{E:0derivativeVector} to exchange the vector $V^{\mu}$ for the vector $(E^n)^{\mu} + \frac{\tilde{D}^{\mu}T}{T}$. Then~\eqref{E:frameInvariants} now gives the boost-invariant energy current and spacetime stress tensor to be
\begin{subequations}
\label{E:eckhart1DerivativeWithP}
\begin{align}
\begin{split}
\tilde{\mathcal{E}}^{\mu} &= \varepsilon u^{\mu} - \frac{(\varepsilon + P)^2}{\rho^2 }\sigma \left( (E^n)^{\mu} + \frac{\tilde{D}^{\mu}T}{T}\right) +\mathcal{O}(\partial^2)\,,
\\
\mathcal{T}^{\mu\nu} & = \left( P - \zeta \vartheta\right)h^{\mu\nu} + \rho u^{\mu} u^{\nu} - \eta \sigma^{\mu\nu} + \mathcal{O}(\partial^2)\,.
\end{split}
\end{align}
The total energy current~\eqref{E:totalEnergy} is
\beq
\mathcal{E}^{\mu} = \left( \varepsilon + \frac{1}{2}\rho u^2\right)u^{\mu} +\left( P-\zeta \vartheta\right) P^{\mu}_{\nu}u^{\nu} - \eta \sigma^{\mu\nu}u_{\nu}  - \frac{(\varepsilon + P)^2}{\rho^2 }\sigma \left( (E^n)^{\mu} + \frac{\tilde{D}^{\mu}T}{T}\right)+ \mathcal{O}(\partial^2)\,.
\eeq
Decomposing $\mathcal{T}^{\mu\nu}$ into the number current, momentum current, and spatial stress tensor via~\eqref{E:calT}, we have
\begin{align}
\begin{split}
J^{\mu} & = \rho u^{\mu} \,,
\\
\mathcal{P}_{\mu} & = \rho u_{\mu} \,,
\\
T_{\mu\nu} & = \left( P-\zeta \vartheta\right) h_{\mu\nu} + \rho u_{\mu}u_{\nu} - \eta h_{\mu\alpha}h_{\nu\beta}\sigma^{\alpha\beta} + \mathcal{O}(\partial^2)\,.
\end{split}
\end{align}
\end{subequations}
(The Eckhart frame condition fixes $J^{\mu}=\rho u^{\mu}$ to all order in gradients, which in turn fixes the momentum via the Milne Ward identity.)

In flat space with zero gauge field, i.e.
\begin{equation*}
n = dx^0\,, \qquad h_{\mu\nu}dx^{\mu}\otimes dx^{\nu} = \delta_{ij}dx^i\otimes dx^j\,, \qquad A = 0\,, \qquad u^{\mu}\partial_{\mu} = \partial_0 + u^i\partial_i\,,
\end{equation*}
the Milne-invariant connection $\tilde{\Gamma}$~\eqref{E:milneD} vanishes, so that 
\begin{equation*}
\sigma_{ij} = \frac{1}{2}\left( \partial_i u_j + \partial_j u_i - \frac{2}{d-1}\delta_{ij} \partial_k u^k\right)\,, \qquad \vartheta = \partial_i u^i\,.
\end{equation*}
Then the Eckhart frame constitutive relations~\eqref{E:eckhart1DerivativeWithP} become
\begin{align}
\nonumber
\mathcal{E}^0 & = \varepsilon + \frac{1}{2}\rho u^2 + \mathcal{O}(\partial^2)\,, & \mathcal{E}^i & = \left( \varepsilon+ P + \frac{1}{2}\rho u^2\right)u^i + \tau^{ij}u_j - \frac{(\varepsilon + P)^2}{\rho^2T}\sigma \partial_i T+\mathcal{O}(\partial^2)\,,
\\
J^0 & = \rho \,, & J^i & = \rho u^i\,,
\\
\nonumber
T_{ij} &= P \delta_{ij} +\rho u_i u_j + \tau_{ij}+\mathcal{O}(\partial^2)\,, & \tau_{ij} & = - \frac{\eta}{2} \left( \partial_i u_j +\partial_j u_i - \frac{2}{d-1}\delta_{ij} \partial_k u^k\right) - \zeta \delta_{ij} \partial_k u^k\,.
\end{align}
These are the conventional expressions for the constitutive relations of first-order hydrodynamics, up to the fact that the transport coefficient multiplying $-\partial_iT$ is generally called $\kappa$, the thermal conductivity. Matching, we have
\beq
\kappa = \frac{(\varepsilon + P)^2}{\rho^2 T}\sigma\,.
\eeq

\section{Parity-violating two-dimensional fluids } 
\label{S:2dFluids}

We now have the machinery to tackle the much richer world where parity is broken. In this Section we construct the hydrodynamics of parity-violating fluids in two spatial dimensions to first order in the gradient expansion. As we mentioned in the Introduction, there are several recent works~\cite{Kaminski:2013gca,Banerjee:2014mka,Geracie:2014zha} which have also studied this problem. We find somewhat different results from those in~\cite{Kaminski:2013gca}, our work trivially matches~\cite{Banerjee:2014mka}, and~\cite{Geracie:2014zha} considers the hydrodynamics of a system coupled to an $\mathcal{O}(1)$ magnetic field. The latter is not analytically related to our results here, for the usual reason that the $B\to 0$ and low energy limits do not commute. So we postpone a detailed comparison to the Appendix.

Our approach here is somewhat different than that used in Subsection~\ref{S:1stOrder} in which we constructed the first-order hydrodynamics of parity-preserving fluids. We will not use the ideal hydrodynamic equations to relate one-derivative data, and correspondingly we will solve the entropy constraint in the form of~\eqref{E:entropy2}. As the reader will see in Subsection~\ref{S:Z1derivative}, the resulting solution for the hydrodynamics is closely related to the currents obtained from the hydrostatic partition function at first order in gradients.

\subsection{Constitutive relations} 
\label{S:2dconstitutive}

In this Subsection, we collect the inequivalent tensors with one derivative, from which we parameterize the constitutive relations of first-order parity-violating hydrodynamics.

There are a number of tensors that can be built out of a single derivative, the background fields, and the fluid variables. All such tensors can be constructed from the zero-derivative data
\begin{equation*}
n_{\mu}\,, \quad h^{\mu\nu}\,, \quad u^{\mu}\,, \quad \tilde{h}_{\mu\nu}\,,\quad T\,,  \quad \nu \equiv \frac{\mu}{T}\,, \quad \varepsilon^{\mu\nu\rho}\,,
\end{equation*}
and the first derivatives
\begin{align*}
\tilde{D}_{\mu}u^{\nu} &= - n_{\mu}E^{\nu} + \frac{1}{2}B_{\mu}{}^{\nu} + \sigma_{\mu}{}^{\nu} + \frac{\tilde{P}_{\mu}^{\nu}}{d-1}\vartheta\,, &\tilde{D}_{\mu} T &= n_{\mu}\dot{T} + \tilde{P}_{\mu}^{\nu} \tilde{D}_{\nu}T \,, \\
\tilde{D}_{\mu}\nu&= n_{\mu}\dot{\nu} + \tilde{P}_{\mu}^{\nu}\tilde{D}_{\nu}\nu\,, & F^n_{\mu\nu}& = E^n_{\mu}n_{\nu} - E^n_{\nu}n_{\mu} + B^n_{\mu\nu}\,,
\end{align*}
where we have defined $\nu$ for convenience. In two spatial dimensions we can dualize the two-forms $B_{\mu\nu}$ and $B^n_{\mu\nu}$ into pseudoscalars
\beq
\mathcal{B} \equiv \frac{1}{2}\varepsilon^{\mu\nu\rho}n_{\mu}B_{\nu\rho}= \frac{1}{2}\varepsilon^{\mu\nu\rho}n_{\mu}F_{\nu\rho}\,, \qquad \mathcal{B}^n\equiv \frac{1}{2}\varepsilon^{\mu\nu\rho}n_{\mu}B^n_{\nu\rho} = \varepsilon^{\mu\nu\rho}n_{\mu}\partial_{\nu}n_{\rho}\,.
\eeq
At this stage the inequivalent one-derivative vectors are $(E^{\mu},(E^n)^{\mu},\tilde{D}^{\mu}T,\tilde{D}^{\mu}\nu)$ and there is a single one-derivative traceless symmetric tensor $\sigma^{\mu\nu}$.

Now, any transverse vector $\mathfrak{v}^{\mu}$ can be dualized into a transverse pseudovector $\tilde{\mathfrak{v}}^{\mu}$ as
\beq
\label{E:dualizeVector}
\tilde{\mathfrak{v}}^{\mu} \equiv \varepsilon^{\mu\nu\rho}n_{\nu}\mathfrak{v}_{\rho}\,,
\eeq
where the index has been lowered with $\tilde{h}_{\mu\nu}$. Similarly, any transverse traceless tensor $\mathfrak{t}^{\mu\nu}$ can be dualized into a transverse traceless pseudotensor $\tilde{\mathfrak{t}}^{\mu\nu}$ via
\beq
\label{E:dualizeTensor}
\tilde{\mathfrak{t}}^{\mu\nu} \equiv \varepsilon^{\rho\sigma(\mu}n_{\rho}\mathfrak{t}_{\sigma}^{\nu)}\,.
\eeq
By these definitions we have
\beq
\tilde{\mathfrak{v}}^{\mu}\mathfrak{u}_{\mu} = -\tilde{\mathfrak{u}}^{\mu}\mathfrak{v}_{\mu}\,, \qquad \tilde{\mathfrak{v}}^{\mu}\mathfrak{v}_{\mu} = 0\,, \qquad \tilde{\mathfrak{t}}^{\mu\nu}\mathfrak{t}_{\mu\nu} =0\,,
\eeq
which will prove useful later. We then have four inequivalent pseudovectors $\tilde{V}^{\mu}_i$ and one pseudotensor $\tilde{\sigma}^{\mu\nu}$. All of the one-derivative pseudotensor data is collected in Table~\ref{T:oneDerivative2}.

\begin{table}
\begin{center}
\begin{tabular}{|c|c|c|c|c|}
\hline & 1 & 2 & 3 & 4 \\
\hline scalars ($s_i$) & $\vartheta$ & $\dot{T}$ & $\dot{\nu}$ & \\
\hline vectors ($V_i^{\mu}$) & $E^{\mu} - T \tilde{D}^{\mu}\nu$ & $(E^n)^{\mu} + \frac{\tilde{D}^{\mu}T}{T}$ & $E^{\mu}$ & $\tilde{D}^{\mu}T$ \\
\hline tensors & $\sigma^{\mu\nu}$ & & & \\
\hline
\end{tabular}
\caption{\label{T:oneDerivative1} The inequivalent parity-preserving tensor data at first order in derivatives, without using the ideal hydrodynamic equations.}
\end{center}
\end{table}

\begin{table}
\begin{center}
\begin{tabular}{|c|c|c|c|c|}
\hline & 1 & 2 & 3 & 4\\
\hline pseudoscalars & $\mathcal{B} = \frac{1}{2}\varepsilon^{\mu\nu\rho}n_{\mu}\tilde{F}_{\nu\rho}$ & $ \mathcal{B}^n = \varepsilon^{\mu\nu\rho}n_{\mu}\partial_{\nu}n_{\rho}$ & & \\
\hline pseudovectors ($\tilde{V}^{\mu}_i$) & $\varepsilon^{\mu\nu\rho}n_{\nu} \left( E_{\rho} - T \tilde{D}_{\rho}\nu\right)$ & $ \varepsilon^{\mu\nu\rho}n_{\nu}\left( E^n_{\rho} + \frac{\tilde{D}_{\rho}T}{T}\right)$ & $ \varepsilon^{\mu\nu\rho}n_{\nu}E_{\rho} $ & $ \varepsilon^{\mu\nu\rho}n_{\nu}\tilde{D}_{\rho}T$ \\ 
\hline pseudotensors & $\tilde{\sigma}^{\mu\nu}$  & & & \\
\hline
\end{tabular}
\caption{\label{T:oneDerivative2} The inequivalent pseudotensor data at first order in derivatives. The pseudovectors $\tilde{V}^{\mu}_i$ are the duals of the vectors $V^{\mu}_i$ (defined in Table~\ref{T:oneDerivative1}) via~\eqref{E:dualizeVector}. Similarly, the pseudotensor $\tilde{\sigma}^{\mu\nu}$ is the dual of the shear tensor $\sigma^{\mu\nu}$ via~\eqref{E:dualizeTensor}.}
\end{center}
\end{table}

We now have a complete basis of one-derivative scalars, vectors, and tensors, out of which we can form the most general constitutive relations and non-canonical entropy current to first order in the gradient expansion. Here we use the result of the previous Subsection that there is no frame-invariant parity-preserving dissipationless transport at first order in derivatives. This tells us that while we could solve adiabaticity with the most general parity-preserving one-derivative terms, the final result (after using the ideal hydrodynamic equations) has no frame-invariant adiabatic solution. 

So we proceed with the most general parity-violating non-canonical entropy current and dissipationless constitutive relations. We decompose $S_{non-diss}^{\mu}$ as
\beq
S_{non-diss}^{\mu} = S u^{\mu} + \Sigma^{\mu}\,,\qquad \Sigma^{\mu}n_{\mu} = 0\,,
\eeq
and parameterize the one-derivative dissipationless constitutive relations and entropy current as
\begin{subequations}
\begin{align}
\label{E:PviolatingConstitutive}
\mathcal{E}_{non-diss} -\varepsilon&=  \varepsilon_B \mathcal{B} + \varepsilon_n \mathcal{B}^n + \mathcal{O}(\partial^2)\,,
& \mathcal{N}_{non-diss} -\rho& =  \rho_B \mathcal{B} + \rho_n \mathcal{B}^n + \mathcal{O}(\partial^2)\,,
\\
\mathcal{P}_{non-diss}-P & = P_B \mathcal{B} + P_n \mathcal{B}^n + \mathcal{O}(\partial^2)\,,
& S_{non-diss} &= S_B \mathcal{B} + S_n \mathcal{B}^n + \mathcal{O}(\partial^2)\,,
\\
q^{\mu}_{non-diss} &= \sum_{i=1}^4 q_i \tilde{V}^{\mu}_i + \mathcal{O}(\partial^2)\,,
 & \eta^{\mu}_{non-diss} & = \sum_{i=1}^4 \eta_i \tilde{V}^{\mu}_i + \mathcal{O}(\partial^2)\,,
\\
\Sigma_{non-diss}^{\mu} & = \sum_{i=1}^4 \Sigma_i \tilde{V}^{\mu}_i + \mathcal{O}(\partial^2)\,,& \tau^{\mu\nu}_{non-diss} & = - \tilde{\eta} \tilde{\sigma}^{\mu\nu} + \mathcal{O}(\partial^2)\,. 
\end{align}
\end{subequations}

\subsection{Entropy and adiabaticity} 
\label{S:2dentropy}

We presently obtain the most general parity-violating solution to the adiabaticity equation~\eqref{E:adiabaticity} at one-derivative order. Before doing so, we collect some useful data. Using~\eqref{E:divergence}, we have
\beq
\left( \tilde{D}_{\mu} - \tilde{\mathcal{G}}_{\mu}\right) \varepsilon^{\mu\nu\rho}\partial_{\nu}\mathfrak{v}_{\rho} =\varepsilon^{\mu\nu\rho}\partial_{\mu}\partial_{\nu}\mathfrak{v}_{\rho}= 0\,, 
\eeq
for any one-form $\mathfrak{v}_{\mu}$. There are three results we need that follow from this identity. Letting $\mathfrak{v}_{\mu}$ be the Milne-invariant connection $\tilde{A}_{\mu}$, and decomposing
\beq
\varepsilon^{\mu\nu\rho}\partial_{\nu}\tilde{A}_{\rho} = \mathcal{B} u^{\mu} - \varepsilon^{\mu\nu\rho}n_{\nu}E_{\rho}\,,
\eeq
we find the useful identity
\beq
\left( \tilde{D}_{\mu}-\tilde{\mathcal{G}}_{\mu}\right)\varepsilon^{\mu\nu\rho}n_{\nu}E_{\rho} = \dot{\mathcal{B}} + \mathcal{B}\vartheta\,.
\eeq
Similarly, letting $\mathfrak{v}_{\mu} = n_{\mu}$, we obtain
\beq
\left( \tilde{D}_{\mu} - \tilde{\mathcal{G}}_{\mu}\right) \varepsilon^{\mu\nu\rho}n_{\nu}E^n_{\rho} = \dot{\mathcal{B}}^n + \mathcal{B}^n \vartheta\,.
\eeq
The third and final result is that the current
\beq
\label{E:trivialConserved}
\varepsilon^{\mu\nu\rho}\partial_{\nu}\left( f(T,\nu)n_{\rho}\right) = f \mathcal{B}^n u^{\mu} +\frac{1}{T}\frac{\partial f}{\partial \nu}\tilde{V}^{\mu}_1- f \tilde{V}_2^{\mu} -\frac{1}{T}\frac{\partial f}{\partial \nu} \tilde{V}^{\mu}_3 +\left( \frac{f}{T}- \frac{\partial f}{\partial T}\right)\tilde{V}^{\mu}_4
\eeq
is identically conserved for any $f$, and so the entropy current is only defined modulo $f$.

Then the divergence of $S_{non-diss}^{\mu}$ is
\begin{align}
\nonumber
\left( \tilde{D}_{\mu} - \tilde{\mathcal{G}}_{\mu}\right) S_{non-diss}^{\mu} =&  \left( S_B  +\Sigma_1+ \Sigma_3\right) \left( \dot{\mathcal{B}} + \mathcal{B}\vartheta\right) + \left( S_n + \Sigma_2\right)\left( \dot{\mathcal{B}}^n + \mathcal{B}^n\vartheta\right) 
\\
\nonumber
& + \frac{\partial S_B}{\partial T}\mathcal{B}\dot{T} 
+\left(  \frac{\partial S_n}{\partial T}+\frac{\Sigma_2}{T}+\Sigma_4\right)\mathcal{B}^n\dot{T} + \frac{\partial S_B}{\partial\nu}\mathcal{B}\dot{\nu} + \left( \frac{\partial S_n}{\partial\nu}-T\Sigma_1\right)\mathcal{B}^n\dot{\nu}
\\
& - \left( \Sigma_1 + \frac{1}{T}\frac{\partial\Sigma_2}{\partial\nu}\right)V_1\cdot \tilde{V}_2 - \frac{1}{T}\left(\frac{\partial\Sigma_1}{\partial\nu}+\frac{\partial\Sigma_3}{\partial\nu}\right)V_1 \cdot \tilde{V}_3
\\
\nonumber
& - \left( \frac{\partial\Sigma_1}{\partial T}+\frac{1}{T}\frac{\partial\Sigma_4}{\partial\nu}\right)V_1\cdot \tilde{V}_4 - \left( \Sigma_1 + \frac{1}{T}\frac{\partial\Sigma_2}{\partial\nu}\right) V_2\cdot \tilde{V}_3
\\
\nonumber
& + \left( \frac{\Sigma_2}{T}- \frac{\partial \Sigma_2}{\partial T}+ \Sigma_4\right) V_2 \cdot \tilde{V}_4 + \left(\frac{1}{T} \frac{\partial \Sigma_4}{\partial\nu}-\frac{\partial\Sigma_3}{\partial T}\right)V_3\cdot\tilde{V}_4\,,
\end{align}
where the partial derivatives are computed at constant $T$ or constant $\nu$. Comparing with~\eqref{E:adiabaticity}, this immediately gives
\beq
\label{E:constraint}
S_B = -(\Sigma_1+ \Sigma_3)\,, \qquad S_n = - \Sigma_2\,, \qquad \frac{\partial\Sigma_3}{\partial T}=\frac{1}{T}\frac{\partial\Sigma_4}{\partial\nu}\,.
\eeq
Now we use our freedom to add the trivially conserved current~\eqref{E:trivialConserved} to $S^{\mu}_{non-diss}$ in such a way as to set $\Sigma_4=0$. Then the last equality in~\eqref{E:constraint} gives $\Sigma_3 = \mathfrak{s}'(\nu)$. However, we can then set $\Sigma_3=0$ by adding a trivially conserved current~\eqref{E:trivialConserved} with $f = - T \mathfrak{s}$, which preserves $\Sigma_4 = 0$. Equivalently, one can redefine
\begin{align}
\begin{split}
\Sigma_1 & \to \Sigma_1 - \mathfrak{s}'\,,
\\
\Sigma_2 & \to \Sigma_2 + T \mathfrak{s}\,,
\\
S_n & \to S_n - T \mathfrak{s}\,,
\end{split}
\end{align}
so that the rest of~\eqref{E:constraint} becomes
\begin{equation*}
S_B = - \Sigma_1\,, \qquad S_n = - \Sigma_2\,,
\end{equation*}
and $\mathfrak{s}$ completely drops out of the divergence of $S^{\mu}_{non-diss}$, which is now
\begin{align}
\begin{split}
\label{E:divSnondiss}
\left( \tilde{D}_{\mu} - \tilde{\mathcal{G}}_{\mu}\right) S^{\mu}_{non-diss} =& \frac{\partial S_B}{\partial T}\mathcal{B} \dot{T} + \left( \frac{\partial S_n}{\partial T}- \frac{S_n}{T}\right) \mathcal{B}^n \dot{T} +  \frac{\partial S_B}{\partial\nu}\mathcal{B}\dot{\nu} + \left( \frac{\partial S_n}{\partial\nu} + T S_B\right)\mathcal{B}^n\dot{\nu}
\\
& + \left( \frac{1}{T}\frac{\partial S_n}{\partial\nu}+S_B\right)V_1\cdot\tilde{V}_2 + \frac{1}{T}\frac{\partial S_B}{\partial\nu}V_1\cdot\tilde{V}_3 + \frac{\partial S_B}{\partial T}V_1\cdot\tilde{V}_4
\\
& +\left( \frac{1}{T}\frac{\partial S_n}{\partial\nu}+S_B\right) V_2 \cdot\tilde{V}_3 + \left( \frac{\partial S_n}{\partial T}- \frac{S_n}{T}\right) V_2\cdot\tilde{V}_4\,. 
\end{split}
\end{align}
The rest of the terms in the adiabaticity equation coming from the constitutive relations give
\begin{align}
\begin{split}
\label{E:restOfAdiabaticity}
- (\mathcal{P}_{non-diss}&-P)\vartheta - \frac{\mathcal{E}_{non-diss}-\varepsilon}{T}\dot{T} - (\mathcal{N}_{non-diss}-\rho)T \dot{\nu} 
\\ &+ q^{\mu}_{non-diss}(V_1)_{\mu} - \eta^{\mu}_{non-diss}(V_2)_{\mu} - \tau_{non-diss}^{\mu\nu}\sigma_{\mu\nu} 
\\
& = - P_B \mathcal{B}\vartheta - P_n \mathcal{B}^n\vartheta - \frac{\varepsilon_B}{T}\mathcal{B}\dot{T} - \frac{\varepsilon_n}{T}\mathcal{B}^n\dot{T} - \rho_B T \mathcal{B}\dot{\nu} - \rho_nT\mathcal{B}^n\dot{\nu}
\\
& \qquad +(q_2+\eta_1)V_1\cdot \tilde{V}_2 + q_3 V_1\cdot \tilde{V}_3 + q_4 V_1\cdot \tilde{V}_4 - \eta_3 V_2\cdot \tilde{V}_3 - \eta_4 V_2\cdot \tilde{V}_4\,.
\end{split}
\end{align}
Here, the adiabaticity equation~\eqref{E:adiabaticity} is just that the sum of~\eqref{E:divSnondiss} and~\eqref{E:restOfAdiabaticity} vanishes. In order to make contact with the hydrostatic partition function, we exchange the two functions $S_B$ and $S_n$ for $\mathcal{M}$ and $\mathcal{M}_n$ via
\beq
\mathcal{M} \equiv T S_B \,, \qquad \mathcal{M}_n \equiv T S_n\,.
\eeq
We also convert $T$ and $\nu$ derivatives into $T$ and $\mu$ derivatives via
\beq
\left( \frac{\partial\mathcal{X}}{\partial T}\right)_{\nu} = \left( \frac{\partial\mathcal{X}}{\partial T}\right)_{\mu} + \frac{\mu}{T}\left(\frac{\partial\mathcal{X}}{\partial\mu}\right)_T \,, \qquad \left( \frac{\partial\mathcal{X}}{\partial\nu}\right)_T = T \left( \frac{\partial\mathcal{X}}{\partial\mu}\right)_T\,.
\eeq
Putting the pieces together, adiabaticity~\eqref{E:adiabaticity} fixes
\begin{align}
\nonumber
P_B & =P_n  = 0\,, 
\\
\nonumber
   \rho_B &=  -q_3 = \frac{\partial \mathcal{M}}{\partial\mu}\,, 
\\
\label{E:adiabaticityResult2d}
\rho_n & = -(q_2+\eta_1) = \eta_3 =  \frac{\partial \mathcal{M}_n}{\partial\mu} +\mathcal{M}\,,  
\\
\nonumber
\varepsilon_B & = - T q_4 = T\frac{\partial\mathcal{M}}{\partial T} + \mu\frac{\partial\mathcal{M}}{\partial\mu} - \mathcal{M}\,, 
\\
\nonumber
\varepsilon_n & = T\eta_4 = T \frac{\partial\mathcal{M}_n}{\partial T}+ \mu \frac{\partial\mathcal{M}_n}{\partial\mu}-2\mathcal{M}_n \,.
\end{align}
The transport coefficients which are unfixed by adiabaticity are $(q_1,\eta_2,\tilde{\eta})$. Meanwhile, the adiabatic part of the non-canonical entropy current is
\beq
S^{\mu}_{non-diss} = \frac{1}{T}\left\{ \mathcal{M}\left( \mathcal{B} u^{\mu} - \tilde{V}_1^{\mu}\right) + \mathcal{M}_n \left( \mathcal{B}^n u^{\mu} - \tilde{V}_2^{\mu}\right)\right\}+\varepsilon^{\mu\nu\rho}\partial_{\nu}\left( f n_{\rho}\right) + \mathcal{O}(\partial^2)\,,
\eeq
where $f$ is an arbitrary function of state.

The one-derivative parity-violating transport~\eqref{E:adiabaticityResult2d} is implicitly given in a choice of hydrodynamic frame. We call this an \emph{adiabatic} frame. In Section~\ref{S:thermal}, we will see that the currents obtained from the hydrostatic partition function are most naturally presented in this frame.

For completeness, we present the resulting constitutive relations~\eqref{E:constitutive} in an adiabatic frame to first order in gradients,
\begin{align}
\nonumber
\mathcal{E} & = \varepsilon + \left( T \frac{\partial\mathcal{M}}{\partial T}+ \frac{\partial\mathcal{M}}{\partial\mu} - \mathcal{M}\right)\mathcal{B} + \left( T \frac{\partial\mathcal{M}_n}{\partial T} + \mu \frac{\partial\mathcal{M}_n}{\partial\mu}-2\mathcal{M}_n\right)\mathcal{B}^n + \mathcal{O}(\partial^2)\,,
\\
\nonumber
\mathcal{N} & = \rho + \frac{\partial \mathcal{M}}{\partial\mu}\mathcal{B} + \left( \frac{\partial\mathcal{M}_n}{\partial\mu} + \mathcal{M}\right)\mathcal{B}^n + \mathcal{O}(\partial^2)\,,
\\
\nonumber
\mathcal{P} & = P - \zeta \vartheta + \mathcal{O}(\partial^2)\,,
\\
\nonumber
q^{\mu} & =\sigma \left(E^{\mu} - T \tilde{D}^{\mu}\left( \frac{\mu}{T}\right)\right) + q_1 \varepsilon^{\mu\nu\rho}n_{\nu}\left( E_{\rho} - T \tilde{D}_{\rho}\left( \frac{\mu}{T}\right)\right) +q_2 \varepsilon^{\mu\nu\rho}n_{\nu}\left( E^n_{\rho} + \frac{\tilde{D}_{\rho}T}{T}\right)
\\
& \qquad - \frac{\partial\mathcal{M}}{\partial\mu}\varepsilon^{\mu\nu\rho}n_{\nu}E_{\rho} - \frac{1}{T}\left( T \frac{\partial\mathcal{M}}{\partial T}+ \mu \frac{\partial\mathcal{M}}{\partial\mu}-\mathcal{M}\right)\varepsilon^{\mu\nu\rho}n_{\nu}\partial_{\rho}T + \mathcal{O}(\partial^2)\,,
\\
\nonumber
\eta^{\mu} & = - \left( q_2 + \frac{\partial\mathcal{M}_n}{\partial\mu}+\mathcal{M}\right)\varepsilon^{\mu\nu\rho}n_{\nu}\left( E_{\rho} -T \tilde{D}_{\rho}\left( \frac{\mu}{T}\right)\right) + \eta_2 \varepsilon^{\mu\nu\rho}n_{\nu}\left( E^n_{\rho} + \frac{\tilde{D}_{\rho}T}{T}\right)
\\
\nonumber
& \qquad + \left( \frac{\partial\mathcal{M}_n}{\partial\mu}+\mathcal{M}\right)\varepsilon^{\mu\nu\rho}n_{\nu}E_{\rho} + \frac{1}{T}\left( T \frac{\partial\mathcal{M}_n}{\partial T}+ \mu \frac{\partial\mathcal{M}_n}{\partial\mu} - 2\mathcal{M}_n\right)\varepsilon^{\mu\nu\rho}n_{\nu}\partial_{\rho}T + \mathcal{O}(\partial^2)\,,
\\
\nonumber
\tau^{\mu\nu} & = - \eta \sigma^{\mu\nu} - \tilde{\eta} \tilde{\sigma}^{\mu\nu} + \mathcal{O}(\partial^2)\,.
\end{align}

\subsection{Results in Landau and Eckhart frame}
\label{S:eckhart}

As in our analysis of parity-preserving hydrodynamics in Subsection~\ref{S:1stOrder}, it is instructive to translate our results for transport into results in both Landau and Eckhart frame.

First, let us exchange the pseudovector $\tilde{V}_2^{\mu}$ for $\tilde{V}^{\mu}_1$ using the ideal equations of motion~\eqref{E:0derivativeVector},
\beq
\tilde{V}^{\mu}_2 = \frac{\rho}{\varepsilon+P}\tilde{V}_1^{\mu} + \mathcal{O}(\partial^2)\,.
\eeq 
Then using~\eqref{E:adiabaticityResult2d}, the one-derivative frame-invariants~\eqref{E:frameInvariants} are
\begin{align}
\begin{split}
\label{E:frameInvariant2d}
\mathscr{S} & = - \zeta\vartheta +\tilde{\chi}_B \mathcal{B} + \tilde{\chi}_n \mathcal{B}_n + \mathcal{O}(\partial^2)\,,
\\
\mathscr{V}^{\mu} & = \sigma V_1^{\mu} + \tilde{\sigma} \tilde{V}_1^{\mu} + \tilde{\chi}_E \varepsilon^{\mu\nu\rho}n_{\nu}E_{\rho} + \tilde{\chi}_T \varepsilon^{\mu\nu\rho}n_{\nu}\tilde{D}_{\rho}T + \mathcal{O}(\partial^2)\,,
\\
\mathscr{T}^{\mu\nu} & = - \eta \sigma^{\mu\nu} - \tilde{\eta}\tilde{\sigma}^{\mu\nu} + \mathcal{O}(\partial^2)\,,
\end{split}
\end{align}
where\footnote{We cannot help but notice the similarities with the result for parity-violating relativistic hydrodynamics presented in~\cite{Jensen:2011xb}.}
\begin{align}
\begin{split}
\label{E:frameInvariantTransport2d}
\tilde{\chi}_B & = \frac{\partial P}{\partial\varepsilon}\left( T \frac{\partial\mathcal{M}}{\partial T}+ \mu \frac{\partial \mathcal{M}}{\partial \mu}-\mathcal{M}\right) + \frac{\partial P}{\partial\rho} \frac{\partial\mathcal{M}}{\partial\mu}\,,
\\
\tilde{\chi}_n & = \frac{\partial P}{\partial\varepsilon}\left( T \frac{\partial\mathcal{M}_n}{\partial T}+ \mu \frac{\partial\mathcal{M}_n}{\partial\mu}-2\mathcal{M}_n\right) + \frac{\partial P }{\partial\rho} \left( \frac{\partial\mathcal{M}_n}{\partial\mu}+ \mathcal{M}\right)\,,
\\
\tilde{\chi}_E & = - \left\{ \frac{\partial\mathcal{M}}{\partial\mu} + R \left( \frac{\partial \mathcal{M}_n}{\partial\mu}+ \mathcal{M}\right)\right\}\,,
\\
T \tilde{\chi}_T & = - \left\{ T \frac{\partial\mathcal{M}}{\partial T}+ \mu\frac{\partial\mathcal{M}}{\partial\mu}-\mathcal{M} + R \left( T\frac{\partial\mathcal{M}_n}{\partial T} + \mu \frac{\partial\mathcal{M}_n}{\partial\mu}-2\mathcal{M}_n\right)\right\}\,,
\end{split}
\end{align}
and we have defined
\beq
R = \frac{\rho}{\varepsilon + P}\,, \qquad \tilde{\sigma} = q_1 - R ( \eta_1-q_2) -R^2 \eta_2\,.
\eeq
Note that $\tilde{\sigma}$ is unconstrained. 

The free parameters governing the one-derivative transport are the Hall viscosity $\tilde{\eta}$, anomalous Hall conductivity $\tilde{\sigma}$, the adiabatic functions of state $\mathcal{M}$ and $\mathcal{M}_n$, and the ordinary viscosities and conductivity $(\zeta,\sigma,\eta)$. The latter are non-negative,
\beq
\zeta \geq 0\,, \qquad \sigma \geq 0\,, \qquad \eta\geq 0\,,
\eeq
while our analysis does not constrain the parity-violating transport,
\beq
\tilde{\eta} \in \mathbb{R}\,, \qquad \tilde{\sigma}\in \mathbb{R}\,, \qquad \mathcal{M}\in \mathbb{R}\,, \qquad \mathcal{M}_n\in \mathbb{R}\,.
\eeq

In Landau frame~\eqref{E:landau} we then find
\begin{align}
\begin{split}
\tilde{\mathcal{E}}^{\mu} & = \varepsilon u^{\mu}\,,
\\
\mathcal{T}^{\mu\nu} & = \left( P - \zeta \vartheta + \tilde{\chi}_B \mathcal{B} + \tilde{\chi}_n \mathcal{B}^n\right)h^{\mu\nu} + \rho u^{\mu}u^{\mu} + u^{\mu}\mathscr{V}^{\mu} + u^{\nu}\mathscr{V}^{\mu} 
\\
& \qquad \qquad - \eta \sigma^{\mu\nu}-\tilde{\eta}\tilde{\sigma}^{\mu\nu} + \mathcal{O}(\partial^2)\,,
\end{split}
\end{align}
and the total energy current~\eqref{E:totalEnergy} is
\begin{align}
\begin{split}
\mathcal{E}^{\mu} = & \left( \varepsilon + \frac{1}{2}\rho u^2 +u_{\nu}\mathscr{V}^{\nu}\right)u^{\mu} + \left( P -\zeta\vartheta + \tilde{\chi}_B \mathcal{B} + \tilde{\chi}_n \mathcal{B}^n\right)P_{\nu}^{\mu}u^{\nu}
\\ & \qquad + \frac{1}{2}u^2\mathscr{V}^{\mu} - \left( \eta \sigma^{\mu\nu} + \tilde{\eta}\tilde{\sigma}^{\mu\nu}\right)u_{\nu} + \mathcal{O}(\partial^2)\,.
\end{split}
\end{align}
The number current $J^{\mu} = \mathcal{T}^{\mu\nu}n_{\nu}$ that follows from this is
\beq
J^{\mu} = \rho u^{\mu} + \sigma V_1^{\mu} + \tilde{\sigma} \tilde{V}_1^{\mu} + \tilde{\chi}_E \varepsilon^{\mu\nu\rho}n_{\nu}E_{\rho} + \tilde{\chi}_T \varepsilon^{\mu\nu\rho}n_{\nu}\tilde{D}_{\rho}T + \mathcal{O}(\partial^2)\,,
\eeq
and recalling that
\begin{equation*}
\tilde{V}_1^{\mu} = \varepsilon^{\mu\nu\rho} n_{\nu}\left( E_{\rho} - T \tilde{D}_{\rho}\left( \frac{\mu}{T}\right)\right)\,,
\end{equation*}
it is clear why we termed $\tilde{\sigma}$ to be the anomalous Hall conductivity.

In the Eckhart frame~\eqref{E:eckhart} we instead have
\begin{align}
\begin{split}
\tilde{\mathcal{E}}^{\mu} & = \varepsilon u^{\mu} - \frac{\varepsilon+P}{\rho}\mathscr{V}^{\mu} + \mathcal{O}(\partial^2)\,,
\\
\mathcal{T}^{\mu\nu}& = \left( P - \zeta \vartheta + \tilde{\chi}_B \mathcal{B} + \tilde{\chi}_n\mathcal{B}^n\right)h^{\mu\nu} + \rho u^{\mu}u^{\nu} - \eta \sigma^{\mu\nu}-\tilde{\eta}\tilde{\sigma}^{\mu\nu}  + \mathcal{O}(\partial^2)\,.
\end{split}
\end{align}
Before going on, let us exchange $V_1^{\mu}$ and $\tilde{V}_1^{\mu}$ for $V_2^{\mu}$ and $\tilde{V}_2^{\mu}$ via~\eqref{E:0derivativeVector}, so that
\begin{equation*}
\mathscr{V}^{\mu} = \frac{\varepsilon + P}{\rho} \left( \sigma V_2^{\mu} +\tilde{\sigma}\tilde{V}_2^{\mu}\right) + \tilde{\chi}_E \varepsilon^{\mu\nu\rho}n_{\nu}E_{\rho} + \tilde{\chi}_T \varepsilon^{\mu\nu\rho}n_{\nu}\tilde{D}_{\rho}T\,.
\end{equation*}
Then the total energy current~\eqref{E:totalEnergy} is
\begin{align}
\begin{split}
\mathcal{E}^{\mu} = & \left( \varepsilon + \frac{1}{2}\rho u^2\right)u^{\mu} + \left( P - \zeta \vartheta + \tilde{\chi}_B \mathcal{B} + \tilde{\chi}_n \mathcal{B}^n\right)P^{\mu}_{\nu}u^{\nu} - \frac{(\varepsilon+P)^2}{\rho^2}\left( \sigma V_2^{\mu} + \tilde{\sigma}\tilde{V}_2^{\mu}\right)
\\
 &\qquad - \frac{\varepsilon+P}{\rho}\left( \tilde{\chi}_E \varepsilon^{\mu\nu\rho}n_{\nu}E_{\rho} + \tilde{\chi}_T \varepsilon^{\mu\nu\rho}n_{\nu}\tilde{D}_{\rho}T\right) - \left( \eta \sigma^{\mu\nu}+\tilde{\eta}\tilde{\sigma}^{\mu\nu}\right)u_{\nu} + \mathcal{O}(\partial^2)\,.
\end{split}
\end{align}
Using
\begin{equation*}
\tilde{V}_2^{\mu} = \varepsilon^{\mu\nu\rho}n_{\nu}\left( E^n_{\rho} + \frac{\tilde{D}_{\rho}T}{T}\right)\,,
\end{equation*}
we define the anomalous thermal conductivity $\tilde{\kappa}$ to be the coefficient multiplying $-\varepsilon^{\mu\nu\rho}n_{\nu}\tilde{D}_{\rho}T$, which gives
\beq
\tilde{\kappa}= \frac{(\varepsilon+P)^2}{\rho^2T}\tilde{\sigma}+ \frac{\varepsilon+P}{\rho}\tilde{\chi}_T\,.
\eeq

We conclude this Section with a summary of the constitutive relations and Ward identities in flat space, when the underlying theory is coupled to a slowly varying $A_{\mu}$. The Newton-Cartan background is
\beq
n_{\mu}dx^{\mu} = dx^0\,, \qquad h_{\mu\nu}dx^{\mu}\otimes dx^{\nu}= \delta_{ij}dx^i\otimes dx^j\,,
\eeq
and the fluid velocity $u^{\mu}$ is parameterized by
\beq
u^{\mu}\partial_{\mu} = \partial_0 + u^i\partial_i\,.
\eeq
In this background, the Milne-invariant $U(1)$ and gravitational connections become
\beq
\tilde{A}_0 = A_0 - \frac{1}{2}u^2\,, \qquad \tilde{A}_i = A_i + u_i\,, \qquad \tilde{\Gamma}^i{}_{0\mu}=\tilde{\Gamma}^i{}_{\mu 0} = \frac{1}{2}\tilde{F}_{\mu}{}^i\,,
\eeq
with all other components vanishing. The expansion and shear tensors are just
\beq
\vartheta = \partial_i u^i\,, \qquad \sigma^{ij} = \frac{1}{2}\left( \partial^i u^j + \partial^j u^i -  \delta^{ij} \vartheta\right)\,,
\eeq 
and
\begin{align}
\begin{split}
E_{\mu} &= F_{\mu \nu} u^{\nu} +\frac{1}{2}\partial_{\mu}u^2-u^{\nu}\partial_{\nu}\left( u_{\mu} + \frac{1}{2}\delta_{\mu}^0 u^2\right) \,,
\\
 \mathcal{B} &= \frac{1}{2}\varepsilon^{ij}F_{ij}+\varepsilon^{ij}\partial_iu_j  = B + \Omega\,.
\end{split}
\end{align}
The Ward identities~\eqref{E:ward} become
\beq
\partial_{\nu}\mathcal{T}^{\mu\nu} = F^{\mu}{}_{\nu}J^{\nu}\,, \qquad \partial_{\mu}\mathcal{E}^{\mu} = -F_{0\mu}J^{\mu}\,.
\eeq
In components, the Eckhart frame constitutive relations are
\begin{subequations}
\label{E:flatEckhart2d}
\begin{align}
\nonumber
\mathcal{E}^0 & =  \varepsilon + \frac{1}{2}\rho u^2 \,, & \mathcal{E}^i & = \left( \varepsilon + P -\zeta\vartheta+\tilde{\chi}_B\mathcal{B}+ \frac{1}{2}\rho u^2\right)u^i +\eta^i+ \tau^{ij}u_j\,,
\\
J^0 & = \rho\,, & J^i &= \rho u^i\,, 
\\
\nonumber
T_{ij} & = \left( P - \zeta\vartheta+\tilde{\chi}_B \mathcal{B}\right) \delta_{ij} + \rho u_iu_j +\tau_{ij}\,, & \tau^{ij}& = - \eta \sigma^{ij} - \tilde{\eta}\tilde{\sigma}^{ij}\,.
\end{align}
with
\beq
\eta^i = - \kappa \partial^i T +\tilde{\kappa} \epsilon^{ij}\partial_j T +\frac{\varepsilon+P}{\rho}\tilde{\chi}_E\varepsilon^{ij}E_j\,,
\eeq
\end{subequations}
The parity-violating part of the response is encoded in $\tilde{\eta}$, $\tilde{\kappa}$, $\mathcal{M}$, and $\mathcal{M}_n$, all of which are unconstrained. Note that $\mathcal{M}_n$ only appears through $\tilde{\chi}_E$.

\section{The geometry of non-relativistic thermal partition functions} 
\label{S:thermal}

In Subsection~\ref{S:euclidean} we considered the Euclidean thermal field theory of Galilean field theories in flat space. We now undertake a very similar discussion of Euclidean thermal field theory in a curved, time-independent spacetime. Denote the generator of time translations in this curved background as $\mathcal{H}_{\tau}$. The corresponding thermal partition function,
\beq
\label{E:generalZE}
\mathcal{Z}_E = \text{tr}\left( \exp\left( - \beta \mathcal{H}_{\tau}\right)\right)\,,
\eeq
computes static response. When the background spacetime is weakly curved over length scales much longer than the mean free path of the microscopic field theory, we call $ \mathcal{Z}_E$ the \emph{hydrostatic partition function} and its variations \emph{hydrostatic response}. 

In this Section, we will study general properties of the hydrostatic partition function, which simplifies enormously compared to the full partition function of interacting field theory in curved spacetime. Our analysis will closely parallel that of~\cite{Jensen:2012jh,Banerjee:2012iz} for relativistic field theories. The crux here is the same as in those works: a generic hot field theory has finite static screening length, meaning that equal-time thermal correlators of all operators fall off exponentially at long distance. The various distance scales that characterize the exponential falloffs are called screening lengths, and we refer to the longest such length as the static screening length, $\ell$, or equivalently the correlation length.\footnote{The notable examples of hot field theories with infinite static screening length are threefold: (i.) a superfluid phase, (ii.) theories tuned to a critical point, and in the relativistic case, (iii.) theories with a $U(1)$ gauge sector. However, we observe that there does not seem to be a Galilean-invariant version of $U(1)$ gauge theory.} Perhaps the most useful way to think about the static screening length is in terms of dimensional reduction. If one reduces a theory on the thermal circle, the field theory on the spatial slice would be zero-temperature field theory with a mass gap $m_{gap} = 1/\ell$. This immediately implies that, in a screened phase, the hydrostatic partition function can be written locally on the spatial slice in a gradient expansion of the background fields. This gradient expansion can at best be asymptotic, wherein its resummation reconstructs physics at the screening scale $\ell$.

This gradient expansion is rather useful. By it, all of the details of an arbitrarily complicated microscopic theory are subsumed into a finite set of real functions that characterize hydrostatic response up to a fixed order in gradients. This hydrostatic response is also computed by hydrodynamics. (More precisely, both the partition function and hydrodynamics can be used to compute Euclidean zero-frequency, small-momentum correlators.) In the relativistic case, there is strong evidence (see especially~\cite{Bhattacharyya:2013lha}) that the response computed from hydrodynamics can be matched to a hydrostatic partition function only if the equality-type constraints are satisfied. We find similar evidence in the non-relativistic case below.

Before going on, it is worth contrasting $\mathcal{Z}_E$ in a screened phase with the partition function for a gapped field theory in the vacuum. The partition function in each case can be written in a gradient expansion of background fields, the former on the spatial slice and the latter on all of spacetime. However, the gradient expansion in the latter case is unphysical, in the sense that the terms appearing in $\mathcal{Z}$ with $n$ derivatives are indistinguishable from local counterterms. This is not the case for the hydrostatic partition function, which only can be written locally on the spatial slice. Equivalently, $\mathcal{Z}_E$ can be written locally and covariantly on spacetime provided that one explicitly uses the symmetry data $K=(K^{\mu}\partial_{\mu},\psi^K,\Lambda_K)$. This is not to say that the terms in $\mathcal{Z}_E$ are completely physical. They are only mostly physical. For instance, the pressure $P$ can be shifted via a volume counterterm by a constant, so $P$ itself is unphysical. However the dependence of pressure on $T$ and $\mu$ cannot be modified by a counterterm and so is physical.

We proceed to discuss the geometry on which $\mathcal{Z}_E$ depends. Euclidean thermal field theory is the most natural framework for this. For theories with a functional integral description, the partition function~\eqref{E:generalZE} can be computed via a Euclidean functional integral on an appropriate analytic continuation of the spacetime. We discussed this procedure in detail in Subsection~\ref{S:euclidean} for Galilean theories in flat space, and here quote the corresponding result for a more general spacetime. The covariant version of a time-independent spacetime is a Newton-Cartan structure $(n_{\mu},h_{\nu\rho},A_{\sigma})$ on a manifold $\mathcal{M}$, where the NC data is invariant under the action of an infinitesimal coordinate transformation $K^{\mu}$, Milne boost $\psi^K$, and $U(1)$ gauge transformation $\Lambda_K$, which we collectively denote as $K = (K^{\mu}\partial_{\mu},\psi^K,\Lambda_K)$. We denote the action of $K$ as $\delta_K$, which acts on the Newton-Cartan data as
\begin{align}
\begin{split}
\label{E:deltaK}
\delta_K n_{\mu} & = \pounds_K n_{\mu} \,,
\\
\delta_K h_{\mu\nu} & = \pounds_K h_{\mu\nu} - \left( n_{\mu} \psi^K_{\nu} + n_{\nu}\psi^K_{\mu}\right) \,,
\\
\delta_K A_{\mu} & = \pounds_K A_{\mu} + \psi^K_{\mu} + \partial_{\mu}\Lambda_K\,,
\end{split}
\end{align}
where $\pounds_K$ is the Lie derivative along $K^{\mu}$. We also restrict $K^{\mu}$ to be ``time-like'' in the sense that $n_{\mu}K^{\mu}>0$. We can see that a spacetime invariant under the action of $K$, $\delta_K(\cdot)=0$, is time-independent in the following way. Locally, one can pick a set of coordinates and a Milne/$U(1)$ gauge such that $K^{\mu}\partial_{\mu} = \partial_0$ and $\psi^K=\Lambda_K=0$, in which case using~\eqref{E:deltaK} and the definition of the Lie derivative, $\delta_K(\cdot)=0$ becomes $\partial_0(\cdot)=0$, i.e. the NC data does not depend on $x^0$.

Anyway, we start constructing Euclidean thermal field theory with a spacetime $\mathcal{M}$ and a NC structure invariant under the action of some timelike $K$. Denoting the affine parameter along the integral curves of $K^{\mu}$ as $\tau$, one analytically continues $\tau = - i \tau_E$ and compactifies $\tau_E \sim \tau_E + \beta$. For theories with a functional integral description, the functional integral on this Euclidean manifold computes the $\mathcal{Z}_E$ in~\eqref{E:generalZE}. From here on out we will regard this Euclidean partition function as the fundamental object of consideration.

This Euclidean spacetime has the topology of a fiber bundle, wherein the thermal circle is fibered over the spatial base. See Fig.~\ref{F:fiber}.

As we mentioned above, $\mathcal{Z}_E$ can be written locally in a gradient expansion when the underlying theory is in a screened phase. That is, the hydrostatic generating functional 
\beq
W_{hydrostat} = - i \ln \mathcal{Z}_E\,,
\eeq
assumes the form
\beq
W_{hydrostat} = \sum_n W_n\,,
\eeq
where $W_n$ is a local integral build out of the spacetime background and symmetry data and its integrand contains $n$ derivatives. By assumption, the microscopic theory is reparameterization/Milne/$U(1)$-invariant, and so $W_{hydrostat}$ must be too. Consequently the $W_n$ are integrals of Milne and $U(1)$-invariant scalars.

From $W_{hydrostat}$ we define the currents through variation in the same way we described in Subsection~\ref{S:symmetries}. We remind the reader that we employ a constrained variational calculus detailed there. These currents automatically satisfy the Ward identities~\eqref{E:ward} on account of the fact that $K$ generates a symmetry and the $W_n$ are reparameterization/Milne/$U(1)$-invariant.

Below, we will construct the $W_n$ and hydrostatic response to first order in derivatives. Along the lines of our discussion of flat-space thermal field theory in Subsection~\ref{S:euclidean}, we define
\beq
\label{E:K0}
K_0 = K^{\mu}n_{\mu}\,,
\eeq
and then identify a local temperature $T$, chemical potential $\mu$, and fluid velocity $u^{\mu}$ from the symmetry data and background fields,
\beq
\label{E:equil0Derivative}
T = \frac{1}{\beta K_0}\,, \qquad \mu = \frac{K^{\mu} \tilde{A}_{\mu} + \Lambda_K}{K_0}=\frac{K^{\mu}A_{\mu}+\Lambda_K}{K_0}+\frac{u^2}{2} \,, \qquad u^{\mu} = \frac{K^{\mu}}{K_0}\,.
\eeq
Here we have used $u^{\mu}$ to build a Milne-invariant version of $A_{\mu}$ as in~\eqref{E:milneD} and defined $\mu/T$ through its holonomy in the same way as in~\eqref{E:defmuT}. The $W_n$ do not depend on $K_0$ or $\beta$ directly, but only on the physical combination $T$ as we explained there. The fluid velocity satisfies $u^{\mu}n_{\mu}=1$ and is moreover Milne-invariant. As a result we can use it to define a Milne-covariant derivative $\tilde{D}_{\mu}$ as in~\eqref{E:milneD}.

Note that there are exactly two invariant scalars which can be formed out of the symmetry data and background fields at zeroth order in derivatives: $T$ and $\mu$.

At $n^{th}$ order in derivatives, we can form manifestly invariant scalars in the same way as we did for hydrodynamics. One just takes $n$ Milne-covariant derivatives of the Milne/$U(1)$-invariant tensor data $(n_{\mu},u^{\mu},T,\mu,\tilde{h}_{\mu\nu},h^{\mu\nu})$ and forms a scalar. Notating the set of invariant scalars at $n^{th}$ order in derivatives as $s_i^{(n)}$, then the most general for $W_n$ is
\beq
\label{E:Wn}
W_n = \sum_i \int d^dx \sqrt{\gamma} f_i^{(n)}(T,\mu)s_i^{(n)}\,,
\eeq
for $f_i^{(n)}$ some functions.

Three technical comments are in order.
\begin{enumerate}
\item We need to be a little more careful about what we mean when we write~\eqref{E:Wn}. The precise meaning of~\eqref{E:Wn} is that one forms scalars out of the Euclideanized data on the Euclideanized spacetime, over which one then integrates. However, when computing the hydrostatic response, it is convenient to ``un-Wick-rotate'' the integrand back to ordinary signature. The resulting variations are automatically expressed in terms of the real NC and symmetry data. 

\item Later in this Section we will match the Euclidean correlators that follow from variation of $W_{hydrostat}$ to the real-time zero-frequency correlators obtained from hydrodynamics. These are a priori distinct observables, however one can show~\cite{Evans:1991ky} that all thermal correlators (whether Euclidean, fully retarded, partially symmetrized, \&c) coincide at zero frequency up to the correct factors of $i$. We are working in a convention so that the variations of $W_{hydrostat}$ exactly match the zero-frequency correlations computed in hydrodynamics with no extra factors of $i$.

\item Suppose that we were considering the hydrostatic partition function of a non-relativistic theory without boost invariance, like a theory with a Lifshitz scale symmetry. For definiteness, one can think of the theory of a free $z=2$ real scalar,
\begin{equation*}
S_{lif} = \int d^dx \left\{ (\partial_0\varphi)^2 + c (\partial_i\partial^i\varphi)^2\right\}\,.
\end{equation*}
Such theories do not necessarily have any global symmetries nor particle number. They naturally couple to a Newton-Cartan structure $(n_{\mu},h_{\nu\rho})$ without Milne invariance. The Euclidean thermal field theory for this type of system is very similar to what we described above with one crucial modification -- thermal states are characterized by a temperature computed from (the inverse of) the integral of $n$ around the thermal circle as above, as well as the non-Milne-invariant scalar $u^2=h_{\mu\nu}u^{\mu}u^{\nu}$. In turn, the thermal partition function and hydrodynamics will depend on both $T$ and $u^2$. There is some tension between this result and the approach to Lifshitz hydrodynamics (and thermodynamics) espoused in~\cite{Hoyos:2013eza} and subsequent work.

\end{enumerate}

In the remainder of this Section, we will study $W_{hydrostat}$ to one-derivative order in some detail. In the next Subsection, we show how $W_{hydrostat}$ can be written in an explicitly time-independent gauge, and then in Subsections~\ref{S:Z0derivative} and~\ref{S:Z1derivative} we construct the possible $W_0$ and $W_1$. The former matches to ideal non-relativistic hydrodynamics, and the latter is only non-vanishing is parity is violated. We also show how to construct a zero-energy, low-momentum superfluid effective action in Subsection~\ref{S:superfluid}.

\subsection{Static gauge}
\label{S:static}

One way to get a physical picture for these Euclidean backgrounds is to fix a gauge and set of coordinates in which all of the background fields are explicitly time-independent. We refer to this choice as a \emph{static gauge}. The authors of~\cite{Banerjee:2012iz} took this approach when constructing the hydrostatic partition functions of relativistic field theories. The authors of~\cite{Gromov:2014vla} performed a similar analysis for non-relativistic field theories coupled to Newton-Cartan geometry, although they did not impose Milne invariance. Here we show how this works for Galilean theories, and then switch to a completely covariant analysis for the rest of the Section.

In terms of the symmetry data $K=(K^{\mu},\psi^K_{\mu},\Lambda_K)$, a static gauge is specified by
\beq
K^{\mu} \partial_{\mu}= \partial_0, \qquad  \psi^K=0,\qquad  \Lambda_K = 0\,,
\eeq
in which case the most general Newton-Cartan background invariant under $\delta_K$ is
\begin{align}
\begin{split}
\label{E:staticBackground}
n_{\mu}dx^{\mu} & = n_0(dx^0+\mathfrak{a})\,,
\\
h_{\mu\nu}dx^{\mu}\otimes dx^{\nu} & = V^2(dx^0+\mathfrak{a})^2 - (dx^0+\mathfrak{a})\otimes V -V\otimes (dx^0+\mathfrak{a})+ \hat{h}_{ij}dx^i\otimes dx^j\,,
\\
A_{\mu}dx^{\mu} & = A_0 (dx^0+\mathfrak{a}) + \hat{A}\,,
\end{split}
\end{align}
where
\beq
V = V_i dx^i\,, \qquad V^2 = \hat{h}^{ij} V_iV_j\,, \qquad \mathfrak{a}= \mathfrak{a}_i dx^i\,, \qquad \hat{A} = \hat{A}_idx^i\,,
\eeq
and crucially the component functions depend on the $x^i$ but not on $x^0$. Here, $\hat{h}_{ij}$ is an invertible spatial tensor with inverse $\hat{h}^{ij}$, and one can check that $h$ has rank $d-1$. So $W_{hydrostat}$ can be thought of as a functional of the fields appearing in~\eqref{E:staticBackground}, $W_{hydrostat}[n_0,A_0,\mathfrak{a}_i,\hat{A}_i,V_i,\hat{h}_{ij}]$. The covariant volume element becomes
\beq
\sqrt{\gamma} = n_0 \sqrt{\hat{h}}\,,
\eeq
and since nothing depends on $x^0$, $W_{hydrostat}$ can be expressed as an integral over the spatial slice,
\beq
W_{hydrostat} =  \int d^{d-1}x \,\sqrt{\hat{h}}\left\{ \mathcal{L}_0 + \mathcal{L}_1 + \hdots\right\}\,,
\eeq 
where the terms in $\mathcal{L}_n$ have $n$ derivatives. One can think of $W_{hydrostat}$ as the result of a dimensional reduction on the thermal circle in the limit that the circle is small compared to gradients on the spatial slice.

One can reconstruct $v^{\mu}$ and $h^{\mu\nu}$ from~\eqref{E:staticBackground}. They are
\begin{align}
\begin{split}
v^{\mu}\partial_{\mu} &= \frac{1}{n_0}\left\{ \left( 1-\mathfrak{a}^iV_i\right)\partial_0 +V^i\partial_j\right\}\,, 
\\
h^{\mu\nu}\partial_{\mu}\otimes\partial_{\nu} &=\mathfrak{a}^2\partial_0\otimes \partial_0 -\mathfrak{a}^i \left( \partial_0\otimes \partial_i + \partial_i\otimes\partial_0\right) + \hat{h}^{ij}\partial_i\otimes\partial_j\,,
\end{split}
\end{align}
where we have raised spatial indices with $\hat{h}^{ij}$.

The static gauge is not unique. The time-independent background~\eqref{E:staticBackground} is transformed into another time-independent background under (i.) time-independent gauge transformations, (ii.) spatial reparameterizations of time, (iii.) spatial reparameterizations of space, and (iv.) time-independent Milne boosts. Let us see how these transformations act on the various fields in~\eqref{E:staticBackground}, beginning with gauge transformations. Under these, the only field which transforms is $\hat{A}$,
\beq
\hat{A}_i\to \hat{A}_i + \partial_i\Lambda\,,
\eeq
so we identify $\hat{A}$ as a spatial $U(1)$ connection. Under spatial reparameterizations of time $x^0 \to x^0 + f(x^i)$, the only field which transforms is $\mathfrak{a}$,
\beq
\mathfrak{a}_i \to \mathfrak{a}_i + \partial_i f\,,
\eeq
so $\mathfrak{a}$ is a Kaluza-Klein connection. Under spatial reparameterizations of space, $(n_0,A_0)$ are scalars, $(\mathfrak{a}_i,\hat{A}_i,V_i)$ transform as one-forms, and $\hat{h}_{ij}$ as a symmetric covariant tensor. Finally, under time-independent Milne boosts, $n_0,\mathfrak{a}$, and $\hat{h}_{ij}$ are invariant but we have
\begin{align}
\nonumber
\psi_{\mu}dx^{\mu} &= \frac{1}{n_0}\left\{ - \frac{V^i\psi_i}{1-\mathfrak{a}_jV^j}dx^0 + \psi_i dx^i\right\}\,,
\\
\nonumber
(V')_i & = V_i + \psi_i + \frac{V^j\psi_j}{1-\mathfrak{a}_kV^k}\mathfrak{a}_i \,,
\\
A_0' & = A_0 -\frac{1}{n_0}\left\{ \frac{V^i \psi_i}{1-\mathfrak{a}_jV^j}+\frac{\mathfrak{a}^2}{2}\left( \frac{V^i\psi_i}{1-\mathfrak{a}_jV^j}\right)^2 +\frac{\mathfrak{a}^iV^j\psi_i\psi_j}{1-\mathfrak{a}_kV^k} +\frac{ \psi^i\psi_i}{2}\right\}
\\
\nonumber
& = A_0 - \frac{(V')^2-V^2}{2n_0}\,,
\\
\nonumber
(\hat{A}')_i & = \hat{A}_i + \frac{1}{n_0}\left\{ \psi_i + \frac{V^j\psi_j}{1-\mathfrak{a}_kV^k}\mathfrak{a}_i \right\} = \hat{A}_i + \frac{(V')_i-V_i}{n_0}\,,
\end{align}
where we have included a factor of $1/n_0$ in the boost for convenience and $(V')^2 = \hat{h}^{ij}(V')_i(V')_j$.

Writing $W_{hydrostat}=W_{hydrostat}[n_0,A_0,\mathfrak{a}_i,\hat{A}_i,V_i,\hat{h}_{ij}]$, we see that $W_{hydrostat}$ must be invariant under the four time-independent symmetries mentioned above. So the $\mathcal{L}_n$ are $U(1)$/Kaluza-Klein/Milne-invariant scalars.

Now, we proceed to match the data above to the local temperature, \&c, that characterize the equilibrium. Matching to~\eqref{E:K0} and~\eqref{E:equil0Derivative} we immediately find
\beq
K_0 = n_0\,, \qquad T = \frac{1}{\beta n_0}\,, \qquad u^{\mu}\partial_{\mu} = \frac{1}{n_0}\partial_0\,.
\eeq
Then using~\eqref{E:staticBackground} we have
\beq
u_{\mu}dx^{\mu} = \frac{1}{n_0}\left\{ V^2(dx^0+\mathfrak{a}) - V\right\}\,, \qquad u^2 = \frac{V^2}{n_0^2}\,, \qquad \tilde{h}_{\mu\nu}dx^{\mu}\otimes dx^{\nu} = \hat{h}_{ij} dx^i\otimes dx^j\,.
\eeq
The Milne-invariant $U(1)$ connection $\tilde{A}_{\mu}=A_{\mu} + u_{\mu} - \frac{1}{2}n_{\mu}u^2$ is
\beq
\tilde{A}_{\mu}dx^{\mu} = \left(A_0+\frac{V^2}{2n_0}\right)(dx^0+\mathfrak{a})+\hat{A}-\frac{V}{n_0}\,.
\eeq
Its scalar component gives the chemical potential
\beq
\mu = \frac{A_0}{n_0} + \frac{V^2}{2n_0^2}\,,
\eeq
and the rest gives a Milne-invariant spatial $U(1)$ connection
\beq
\bar{A} \equiv \hat{A} - \frac{V}{n_0}\,.
\eeq

Let us summarize. The Milne-non-invariant background fields are $(n_0,A_0,\mathfrak{a}_i,\hat{A}_i,V_i,\hat{h}_{ij})$, but the Milne-invariant combinations are
\begin{equation*}
n_0\,, \qquad A_0 + \frac{V^2}{2n_0}\,, \qquad \mathfrak{a}_i\,, \qquad \hat{A}_i-\frac{V_i}{n_0}\,, \qquad \hat{h}_{ij}\,.
\end{equation*}
Equivalently, there are two scalars which we can normalize to be $(T,\mu)$\footnote{Note that there would be three independent scalars $(n_0,A_0,V^2)$ if we did not impose Milne invariance. The authors of~\cite{Gromov:2014vla} study the hydrostatic partition function of non-relativistic theories without boost invariance, but they miss $V^2$ in their classification of zero-derivative scalars.}, the Kaluza-Klein connection $\mathfrak{a}$, the Milne-invariant spatial $U(1)$ connection $\bar{A}$, and a spatial metric $\hat{h}_{ij}=\tilde{h}_{ij}$. It is interesting to note that this is the same data (two scalars, a Kaluza-Klein and $U(1)$ connection, spatial metric) that one finds in the reduction of a relativistic field theory with a $U(1)$ symmetry on a circle.

We now have enough information to see that at zeroth order in gradients
\beq
W_{hydrostat} = \int d^{d-1}x \sqrt{\hat{h}}\left\{ p\left(\beta n_0,\frac{A_0}{n_0} + \frac{V^2}{2n_0^2}\right) + \mathcal{O}(\partial)\right\}\,.
\eeq

It is probably clear that while this construction is useful, it is a little cumbersome when it comes to computing variations of $W_{hydrostat}$. So for the remainder of this Section we will work with a completely equivalent, covariant formalism along the lines of~\cite{Jensen:2012jh,Jensen:2012jy}. The static gauge will briefly rear its head in a $(3+1)$-dimensional, parity-violating analysis in Subsection~\ref{S:Z1derivative} in the form of Chern-Simons terms on the spatial slice.

\subsection{One-derivative interrelations} 
\label{S:1derivativeRelations}

In thermal equilibrium, the derivatives of $(T,\mu,u^{\mu})$ defined in~\eqref{E:equil0Derivative} are related to the other background fields. In a static gauge, this is just because $(T,\mu,u^{\mu})$ are absorbed into the parameterization of the spacetime background. How does one obtain the interrelations in a coordinate-independent way? The easiest way to do so is to use that $\delta_K$ annihilates the background and $(T,\mu,u^{\mu})$, as in~\cite{Jensen:2013kka}. 

We begin with $n_{\mu}$, finding
\begin{align}
\begin{split}
\delta_K n_{\mu} & = \pounds_K n_{\mu} = K^{\nu}\partial_{\nu} n_{\mu} + n_{\nu} \partial_{\mu} K^{\nu}
\\
 & = - (\partial_{\mu} n_{\nu} - \partial_{\nu} n_{\mu} )K^{\nu} + \partial_{\mu} K_0 = 0\,,
\end{split}
\end{align}
which upon using~\eqref{E:equil0Derivative} and recalling $E^n_{\mu} = F^n_{\mu\nu}u^{\nu}$ immediately gives
\beq
\label{E:equilInterrelation1}
\left( \partial_{\mu} + E^n_{\mu}\right) T = 0\,.
\eeq
We can think of this as an interrelation between the local temperature and the ``energy electric field'' $E^n_{\mu}$ that holds in equilibrium. Note that the longitudinal component of this equation gives
\begin{equation*}
u^{\mu}\left( \partial_{\mu} + E^n_{\mu}\right)T = \dot{T} = 0\,.
\end{equation*}

We will now compute the remaining one-derivative relations. It is somewhat cumbersome to use this approach on the Milne-non-invariant data $(v^{\mu},A_{\mu})$. We will instead use that $\delta_K$ annihilates the symmetry data and instead consider the variations of the Milne-invariant combination $\tilde{A}_{\mu}$. (We do not consider the variation of $u^{\mu}=K^{\mu}/K_0$, which vanishes if $\delta_K n_{\mu}$ does.) The variation of $\tilde{A}_{\mu}$ gives
\begin{align}
\begin{split}
\delta_K \tilde{A}_{\mu} & = \pounds_K \tilde{A}_{\mu} + \partial_{\mu} \Lambda_K = K^{\nu} \partial_{\nu} \tilde{A}_{\mu} + \tilde{A}_{\nu}\partial_{\mu} K^{\nu} + \partial_{\mu} \Lambda_K 
\\
& =   K^{\nu} \left( \partial_{\nu}\tilde{A}_{\mu} - \partial_{\mu} \tilde{A}_{\nu}\right) + \partial_{\mu} \left( K^{\nu} \tilde{A}_{\nu} + \Lambda_K\right)
\\
& = K_0 \left\{ - \tilde{F}_{\mu\nu}u^{\nu} + T\,\partial_{\mu}\left( \frac{\mu}{T}\right)\right\} \,,
\end{split}
\end{align}
so that
\beq
E_{\mu} - T \tilde{D}_{\mu}\left( \frac{\mu}{T}\right)=0\,,
\eeq
whose scalar component is $\propto\dot{\left( \frac{\mu}{T}\right)}$ and whose vector component is $V_1^{\mu}$ which we defined in Table~\ref{T:oneDerivative1}.

For $h^{\mu\nu}$, we have
\begin{align}
\begin{split}
\delta_K h^{\mu\nu} & = \pounds_K h^{\mu\nu} = K^{\rho}\partial_{\rho} h^{\mu\nu} - h^{\rho\nu}\partial_{\rho} K^{\mu} - h^{\mu\rho}\partial_{\rho}K^{\nu}
\\
& =  - h^{\rho\nu}\tilde{D}_{\rho} K^{\mu} - h^{\mu\rho}\tilde{D}_{\rho} K^{\nu} + \left( \tilde{T}^{\mu}{}_{\alpha\rho}h^{\rho\nu} + \tilde{T}^{\nu}{}_{\alpha\rho}h^{\rho\mu}\right)K^{\alpha}
\\
& = -\beta \left\{  h^{\rho\nu}\tilde{D}_{\rho}\left( \frac{u^{\mu}}{T}\right) + h^{\mu\rho}\tilde{D}_{\rho}\left( \frac{u^{\nu}}{T}\right) - \frac{u^{\mu}}{T}(E^n)^{\nu} - \frac{u^{\nu}}{T}(E^n)^{\mu}\right\}\,,
\end{split}
\end{align}
where $\tilde{T}^{\mu}{}_{\nu\rho}$ is the torsion of $\tilde{D}_{\mu}$ defined in~\eqref{E:simpleGammau}. Using~\eqref{E:equilInterrelation1} this becomes
\begin{align}
\begin{split}
\delta_K h^{\mu\nu} & =-  \left( u^{\mu} h^{\nu\rho} + u^{\nu}h^{\mu\rho}\right) \delta_K n_{\rho} - K_0 \left\{ h^{\mu\rho} \tilde{D}_{\rho} u^{\nu} + h^{\nu\rho} \tilde{D}_{\rho} u^{\mu}\right\} 
\\
& = -  \left( u^{\mu} h^{\nu\rho} + u^{\nu}h^{\mu\rho}\right) \delta_K n_{\rho} - K_0 \left\{ \sigma^{\mu\nu} + \frac{2}{d-1}h^{\mu\nu} \vartheta\right\} = 0\,,
\end{split}
\end{align}
where we have decomposed the derivative of $u^{\mu}$ as in~\eqref{E:Du}.

Putting all of the pieces together, we find that the one-derivative interrelations satisfied in equilibrium are
\begin{subequations}
\label{E:1derivativeEquil}
\begin{align}
\dot{T} & = 0 \,, & \dot{\left(\frac{\mu}{T}\right)} & = 0\,, & \vartheta & = 0\,,
\\
V_1^{\mu} = E^{\mu} -T \tilde{D}^{\mu}\left( \frac{\mu}{T}\right) & = 0\,, & V_2^{\mu} = (E^n)^{\mu}+ \frac{\tilde{D}^{\mu}T}{T}& = 0\,, & \sigma^{\mu\nu} & = 0\,.
\end{align}
\end{subequations}
Note that all of these quantities appear multiplying the constitutive relations in the divergence of the entropy current~\eqref{E:entropy2}, so we conclude that there is no entropy production in hydrostatic equilibrium, which is a nice cross-check on our approach.

\subsection{The partition function and currents at zeroth order in derivatives} 
\label{S:Z0derivative}

At zeroth order in derivatives, the most general scalar is a function of $T$ and $\mu$, which gives
\beq
W_0 = \int d^dx \sqrt{\gamma} \, P(T,\mu)\,,
\eeq
where $P$ is the pressure.

We proceed to compute the equilibrium currents that follow from $W_0$ in some detail, so the reader gets the idea of how these calculations work. If all goes as it ought, these currents will be precisely the constitutive relations of ideal non-relativistic hydrodynamics,~\eqref{E:idealConstitutive}. To do so we vary it
\beq
\delta W_0 =  \int d^dx \sqrt{\gamma} \left\{\frac{\delta \sqrt{\gamma}}{\sqrt{\gamma}}P + \delta T\frac{\partial P}{\partial T} + \delta \mu\frac{\partial P}{\partial \mu}\right\}\,.
\eeq
We use~\eqref{E:deltavh} to compute
\begin{align}
\nonumber
\frac{\delta \sqrt{\gamma}}{\sqrt{\gamma}} &= v^{\mu} \delta n_{\mu} - h_{\mu\nu}\frac{\delta \bar{h}^{\mu\nu}}{2}\,,
\\
\nonumber
\frac{\delta K_0}{K_0} & = \frac{\delta (n_{\mu}K^{\mu})}{K_0} = u^{\mu} \delta n_{\mu}\,,
\\
\label{E:0derivativeVariations}
\delta T & = \delta \left( \frac{1}{\beta K_0}\right) = - \frac{1}{\beta K_0}\frac{\delta K_0}{K_0} = - T u^{\mu} \delta n_{\mu}\,,
\\
\nonumber
\delta \mu & = \delta \left( \frac{K^{\mu} A_{\mu} + \Lambda_K}{K_0} + \frac{u^2}{2}\right) = u^{\mu} \delta A_{\mu} + \frac{\delta h_{\mu\nu}}{2}u^{\mu}u^{\nu} - \left( \mu+ \frac{u^2}{2}\right)\frac{\delta K_0}{K_0} 
 \\
 \nonumber
 & = u^{\mu} \delta A_{\mu} - u_{\mu} \delta \bar{v}^{\mu} - u_{\mu}u_{\nu}\frac{\delta \bar{h}^{\mu\nu}}{2} - \left( \mu + \frac{u^2}{2}\right)u^{\mu} \delta n_{\mu}\,.
\end{align}
Putting it all together we have
\begin{align}
\begin{split}
\delta W_0 =& \int d^dx \sqrt{\gamma} \left\{ \frac{\partial P}{\partial \mu}\left( u^{\mu}\delta A_{\mu}- u_{\mu} \delta \bar{v}^{\mu}\right)  -  \frac{\delta \bar{h}^{\mu\nu}}{2}\left( \frac{\partial P}{\partial \mu}u_{\mu}u_{\nu} + P h_{\mu\nu}\right) \right.
\\
 & \qquad \qquad \qquad \qquad \left. - \delta n_{\mu}\left[ \left( T \frac{\partial P}{\partial T}+ \mu \frac{\partial P}{\partial \mu} + \frac{1}{2}\frac{\partial P}{\partial \mu}u^2\right) u^{\mu} - P v^{\mu}\right]  \right\}\,.
\end{split}
\end{align}
From the definition of the currents~\eqref{E:defineCurrents}, we read off
\begin{align}
\begin{split}
\label{E:idealCurrents}
J^{\mu}  &= \rho u^{\mu}\,,
\\
 \mathcal{P}_{\mu} & = \rho u_{\mu}\,,
\\
\mathcal{E}^{\mu}  & =\left( \varepsilon + \frac{1}{2}\rho u^2 \right)u^{\mu} + P\, P^{\mu}_{\nu}u^{\nu}\,,
\\
T_{\mu\nu} & = \rho u_{\mu}u_{\nu} + P\, h_{\mu\nu}\,,
\end{split}
\end{align}
where we have defined the charge, entropy, and energy densities through
\begin{equation*}
\rho = \frac{\partial P}{\partial T}\,, \qquad s = \frac{\partial P}{\partial T}\,, \qquad \varepsilon = T s + \mu \rho - P\,.
\end{equation*}
The spacetime stress tensor $\mathcal{T}^{\mu\nu}$~\eqref{E:calT} and boost-invariant energy current~\eqref{E:boostEnergy} that follow from these one-point functions are
\beq
\tilde{\mathcal{E}}^{\mu} = \varepsilon u^{\mu}\,, \qquad \mathcal{T}^{\mu\nu} = P h^{\mu\nu}+\rho u^{\mu}u^{\nu}\,,
\eeq
which trivially matches ideal hydrodynamics~\eqref{E:idealConstitutive}.

\subsection{The partition function and currents at first order in derivatives} 
\label{S:Z1derivative}

Now let us go on to classify the terms that may contribute to $W_1$. The one-derivative scalars which may be constructed from $(T,\mu,u^{\mu})$ and the background are $\dot{T}, \dot{\left( \frac{\mu}{T}\right)},$ and $\vartheta$, which however all vanish in equilibrium~\eqref{E:1derivativeEquil}. So $W_1$ simply vanishes in a parity-preserving theory, and there is no one-derivative hydrostatic response.

This immediately matches the results obtained for parity-preserving, first-order hydrodynamics in Subsection~\ref{S:1stOrder}. There, we found that the one-derivative transport can be summarized in the frame-invariant scalar, vector, and tensor~\eqref{E:oneDerivativePSummary} (in terms of the constitutive relations~\eqref{E:constitutive}, the frame-invariants were obtained in~\eqref{E:frameInvariants}),
\begin{equation*}
\mathscr{S} = - \zeta \vartheta\,, \qquad \mathscr{V}^{\mu} = \sigma \left( E^{\mu} - T \tilde{D}^{\mu}\left( \frac{\mu}{T}\right)\right)\,, \qquad \mathscr{T}^{\mu\nu} = - \eta \sigma^{\mu\nu}\,.
\end{equation*}
However, recall~\eqref{E:1derivativeEquil} that all three of these tensor structures vanish in equilibrium.

So only parity-violating theories can have nonzero hydrostatic response at first order in derivatives. We proceed to classify $W_1$ in two and three spatial dimensions. In higher dimension, $W_1=0$ identically.

\subsubsection{Parity-violating fluids in two spatial dimensions}

In this instance, the non-vanishing pseudotensor data with one derivative which can be built from the symmetry data and spacetime background is given in Table~\ref{T:2dEquilibrium}. There are two pseudoscalars, the local magnetic field $\mathcal{B}$ and ``energy magnetic field'' $\mathcal{B}^n$. Note that $\mathcal{B}$ is actually a linear combination of the magnetic field constructed from $A_{\mu}$, the fluid vorticity, and $\mathcal{B}^n$,
\beq
\mathcal{B} = \frac{1}{2}\varepsilon^{\mu\nu\rho}n_{\mu}\tilde{F}_{\nu\rho} = \frac{1}{2}\varepsilon^{\mu\nu\rho}n_{\mu}F_{\nu\rho} + \varepsilon^{\mu\nu\rho} n_{\mu}\partial_{\nu}u_{\rho} - \frac{1}{2}u^2 \varepsilon^{\mu\nu\rho}n_{\mu}\partial_{\nu}n_{\rho}\,.
\eeq
In any case, the most general $W_1$ is
\beq
W_1 = \int d^3x \sqrt{\gamma}\left\{ \tilde{f}_1 \mathcal{B} + \tilde{f}_2\mathcal{B}^n\right\}\,
\eeq
where the $\tilde{f}_i$ are arbitrary functions of $T$ and $\mu$.

\begin{table}
\begin{center}
\begin{tabular}{|c|c|c|}
\hline & 1 & 2 \\
\hline pseudoscalars & $\mathcal{B} =\frac{1}{2} \varepsilon^{\mu\nu\rho}n_{\mu}\tilde{F}_{\nu\rho}$ & $\mathcal{B}^n = \varepsilon^{\mu\nu\rho}n_{\mu}\partial_{\nu}n_{\rho}$ \\
\hline pseudovectors & $\tilde{V}_3^{\mu} = \varepsilon^{\mu\nu\rho}n_{\nu}E_{\rho}$ & $\tilde{V}_4^{\mu} = \varepsilon^{\mu\nu\rho}n_{\nu}\tilde{D}_{\rho}T$ \\
\hline
\end{tabular}
\caption{\label{T:2dEquilibrium} The non-vanishing pseudotensor data built the symmetry data $(T,\mu,u^{\mu})$ in~\eqref{E:equil0Derivative} and the time-independent Newton-Cartan background. Here we follow the conventions in our hydrodynamic analysis of two-dimensional fluids, and in particular the classification of tensor structures in Table~\ref{T:oneDerivative2}. Correspondingly, the pseudovectors $\tilde{V}_1^{\mu}$ and $\tilde{V}_2^{\mu}$ along with the pseudotensor $\tilde{\sigma}^{\mu\nu}$ vanish in equilibrium.}
\end{center}
\end{table}

Now we vary $W_1$ to obtain the hydrostatic response. Here, recall that $\varepsilon^{\mu\nu\rho}=\frac{\epsilon^{\mu\nu\rho}}{\sqrt{\gamma}}$ with $\epsilon^{\mu\nu\rho}$ the epsilon symbol so that the factors of $\sqrt{\gamma}$ cancel in $W_1$. Then using the variations of $T$ and $\mu$ in~\eqref{E:0derivativeVariations} along with
\beq
\delta \tilde{A}_{\mu} = \delta A_{\mu} -h_{\mu\nu}\delta \bar{v}^{\nu} - \left\{ u_{\mu}u^{\nu} + \frac{u^2}{2}\left( \tilde{P}_{\mu}^{\nu}- n_{\mu}u^{\nu}\right)\right\}\delta n_{\nu}- \left( h_{\mu\nu}u_{\rho}+h_{\mu\rho}u_{\nu} - n_{\mu}u_{\nu}u_{\rho}\right)\frac{\delta \bar{h}^{\nu\rho}}{2}\,,
\eeq
and the definition of the currents~\eqref{E:defineCurrents} we find that the spacetime stress tensor $\mathcal{T}^{\mu\nu}$~\eqref{E:calT} is to $\mathcal{O}(\partial)$
\begin{align}
\begin{split}
\label{E:TfromW12d}
\mathcal{T}^{\mu\nu} = & P h^{\mu\nu} + \rho u^{\mu}u^{\nu} + \left\{ \frac{\partial\tilde{f}_1}{\partial\mu}\mathcal{B} + \left( \frac{\partial \tilde{f}_2}{\partial\mu}+\tilde{f}_1\right)\mathcal{B}^n\right\}u^{\mu}u^{\nu}
\\
& \qquad \qquad - 2 u^{(\mu}\left\{ \frac{\partial\tilde{f}_1}{\partial\mu} \tilde{V}_3^{\nu)} + \frac{1}{T}\left( T \frac{\partial\tilde{f}_1}{\partial T}+\mu\frac{\partial\tilde{f}_1}{\partial\mu}-\tilde{f}_1\right)\tilde{V}_4^{\nu)}\right\}\,,
\end{split}
\end{align}
and the boost-invariant energy current~\eqref{E:boostEnergy} is
\begin{align}
\begin{split}
\label{E:EfromW12d}
\tilde{\mathcal{E}}^{\mu} = & \varepsilon u^{\mu} + \left\{ \left( T \frac{\partial\tilde{f}_1}{\partial T}+\mu \frac{\partial\tilde{f}_1}{\partial\mu}-\tilde{f}_1\right)\mathcal{B} + \left( T \frac{\partial\tilde{f}_2}{\partial T}+\mu\frac{\partial\tilde{f}_2}{\partial\mu}-2\tilde{f}_2\right)\mathcal{B}^n\right\}u^{\mu}
\\
& \qquad \qquad + \left( \frac{\partial\tilde{f}_2}{\partial\mu}+\tilde{f}_1\right)\tilde{V}_3^{\mu} + \frac{1}{T}\left( T\frac{\partial\tilde{f}_2}{\partial T}+\mu\frac{\partial\tilde{f}_2}{\partial\mu}-2\tilde{f}_2\right)\tilde{V}_4^{\mu}
\end{split}
\end{align}

Now compare~\eqref{E:TfromW12d} and~\eqref{E:EfromW12d} with the stress tensor and energy current~\eqref{E:PviolatingConstitutive} and~\eqref{E:adiabaticityResult2d} obtained from solving the adiabaticity equation (where $\mathcal{T}^{\mu\nu}$ and $\tilde{\mathcal{E}}^{\mu}$ have been decomposed as in~\eqref{E:constitutive}), and recall that $\tilde{\sigma}^{\mu\nu}$, $\tilde{V}_1^{\mu}$, and $\tilde{V}_2$ vanish in equilibrium. The two results precisely agree with the identification
\beq
\tilde{f}_1 = \mathcal{M}\,, \qquad \tilde{f}_2 = \mathcal{M}_n\,.
\eeq
That is, the equality-type relations predicted by adiabaticity are mandated by the existence of the hydrostatic partition function.

This result also gives a physical interpretation for the parameters $\mathcal{M}$ and $\mathcal{M}_n$. They are the magnetic and ``energy magnetic'' susceptibilities in the source-free thermal state.

\subsubsection{Parity-violating fluids in three spatial dimensions}

We move on to classify $W_1$ in three spatial dimensions. It is easy to show that the only one-derivative pseudotensors which can be built from $(T,\mu,u^{\mu})$ and the spacetime background are the transverse pseudovectors
\beq
\mathcal{B}^{\mu} = \frac{1}{2}\varepsilon^{\mu\nu\rho\sigma}n_{\nu}\tilde{F}_{\rho\sigma}\,, \qquad \omega^{\mu} = \varepsilon^{\mu\nu\rho\sigma}n_{\nu}\partial_{\rho}n_{\sigma}\,,
\eeq
which are just the duals of $B_{\mu\nu}$ and $B^n_{\mu\nu}$ and so may be non-vanishing in equilibrium.

Na\"ively it follows that $W_1=0$, but this is not quite the case. The easiest way to see this is to go back to our discussion of the static gauge in Subsection~\ref{S:static}, and to think of $W_1$ as a three-dimensional integral over the spatial slice. While there are no one-derivative invariant scalars that can be formed from the symmetry data and background, there are three independent Chern-Simons terms with one derivative
\begin{equation*}
W_1 = \int \left\{ c_1 \bar{A} \wedge d\bar{A}+ \frac{c_2}{\beta}\mathfrak{a}\wedge d\bar{A} + \frac{c_3}{\beta^2} \mathfrak{a} \wedge d\mathfrak{a} \right\}\,,
\end{equation*}
where we have included the factors of $\beta$ by dimensional analysis. The Kaluza-Klein and $U(1)$ gauge invariances restrict the $c_i$'s to be constants.\footnote{The situation is almost identical for relativistic field theory in $(3+1)$ dimensions~\cite{Banerjee:2012iz}. When the microscopic field theory has anomalies, the potential $W_1$ includes the three Chern-Simons terms above as well as a term which accounts for $U(1)^3$ anomalies. CPT forbids $c_1$ and $c_3$, and $c_2$ is secretly fixed by the gravitational anomalies~\cite{Jensen:2012kj}. Varying $W_1$ and matching to hydrodynamics, this analysis recovers the anomaly-induced transport obtained via the entropy argument by Son and Surowka~\cite{Son:2009tf} ($c_1$ and $c_2$ do not appear there, but were found later in~\cite{Neiman:2010zi}, whilst $c_3$ was only discovered in~\cite{Banerjee:2012iz,Jensen:2012jy}). The relation between $c_2$ and the gravitational anomaly was conjectured in~\cite{Landsteiner:2011cp} based on free-field computations and was proven for theories in a screened phase in~\cite{Jensen:2012kj} using some mild analyticity assumptions on the thermal partition function. See also~\cite{Golkar:2012kb} which proved the relation for theories with a functional integral description.} A quick computation shows that the $c_i$ are all parity-violating but time-reversal-preserving.

We would like to characterize these Chern-Simons terms in our covariant formalism, and to thence compute the hydrostatic currents. It turns out that the $c_1$ and $c_2$ terms can be described in a four-dimensional covariant way as outlined in~\cite{Jensen:2012kj}, but to capture $c_3$ we need to lift to five dimensions. In fact, all three terms can be lifted to terms in five dimensions using the technology of~\cite{Jensen:2013kka,Jensen:2013rga}

We begin with the four-dimensional version of $c_1$ and $c_2$. From here until the end of this Subsection, we will require the extensive use of differential forms. Using
\beq
F^n=dn = E^n\wedge n + B^n\,,
\eeq
and defining $\hat{A}_T \equiv -T n$, we note that
\beq
\hat{F}_T = d\hat{A}_T=-d(Tn) =  -(d+E^n)T \wedge n - T B^n = -T B^n\,,
\eeq
in equilibrium. In other words, $\hat{F}_T$ is a transverse closed two-form. Similarly, define $\hat{A} = \tilde{A} - \mu n$, whose field strength in equilibrium is
\beq
\hat{F} = d\hat{A} =( E-d\mu-\mu E^n) \wedge n + B -\mu B^n = B - \mu B^n\,,
\eeq
which is also a transverse closed two-form. Now consider
\beq
-T n \wedge \tilde{A} \wedge \left( k_1 \hat{F} + k_2 \hat{F}_T\right)\,,
\eeq
where $k_1$ and $k_2$ are constants. This is a Milne-invariant top form, whose gauge variation is a boundary term on account of the fact that $\hat{F}$ and $\hat{F}_T$ are transverse so that $\hat{F}\wedge \hat{F}_T$ and $\hat{F}_T \wedge \hat{F}_T$ both vanish in equilibrium. So we can include this four-form in $W_1$. Writing it out in static gauge, one finds that the $k_1$ and $k_2$ are linear combinations of $c_1$ and $c_2$.

Now for the five-dimensional description. Extend spacetime to a five-dimensional manifold with $\mathcal{M}$ as its boundary. Correspondingly extend the four-dimensional data to fields in higher dimension. The five-form
\beq
I_T= - k_1 \hat{A}_T \wedge \hat{F} \wedge \hat{F} - k_2\hat{A}_T  \wedge \hat{F}\wedge\hat{F}_T - k_3 \hat{A}_T\wedge \hat{F}_T \wedge \hat{F}_T\,,
\eeq
is closed in five dimensions for the same reason as above. So $I_T$ can be locally represented as $dW_T$, and our $W_1$ is just
\beq
W_1 = \int W_T\,.
\eeq
In static gauge, the $k_i$ correspond to linear combinations of the $c_i$, so this construction gives all three Chern-Simons terms.

To compute the currents, we use the machinery of~\cite{Jensen:2013kka,Jensen:2013rga} which we do not review here. Instead we refer the reader to Appendix E of~\cite{Jensen:2013kka} which gives the basic idea. The resulting one-derivative contribution to the charge and energy currents is
\begin{align}
\begin{split}
q^{\mu} & = -2k_1T \mathcal{B}^{\mu} + \left( 2k_1T\mu - k_2T^2\right)\omega^{\mu}\,,
\\
\eta^{\mu} & = -\left( 2k_1T\mu - k_2T^2\right)\mathcal{B}^{\mu} + 2\left( k_1T\mu^2 + k_2T^2\mu+k_3T^3\right)\omega^{\mu}\,.
\end{split}
\end{align}
Presumably this transport can also be obtained by an adiabaticity analysis as in Section~\ref{S:2dFluids}, although we have not done so.

\subsection{Ideal superfluids} 
\label{S:superfluid}

Let us switch gears and briefly show how this formalism can also be used to describe the hydrostatic response in a superfluid phase. Our approach here is the Galilean version of the analysis in~\cite{Jensen:2012jh,Bhattacharyya:2012xi}.

Consider a Galilean-invariant theory at nonzero temperature in a phase where particle number is spontaneously broken. Denoting the Goldstone mode of the symmetry breaking as $\varphi$ (ignoring various subtleties that arise in the physics of Goldstone modes in non-relativistic systems), the static correlation length of the theory is infinite rather than finite. Rather than $W_{hydrostat}$, we consider $S_{hydrostat}$, which is the zero-energy, low-momentum effective action for the Matsubara zero mode of $\varphi$ coupled to the time-independent background~\eqref{E:staticBackground}. As before $S_{hydrostat}$ may be written in a gradient expansion of $\varphi$, the symmetry data, and the spacetime background.

Under gauge transformations, $\varphi$ transforms as
\beq
\varphi \to \varphi - \Lambda\,,
\eeq
so that the Milne covariant derivative acting on $\varphi$ is
\beq
\xi_{\mu} = \tilde{D}_{\mu}\varphi = \partial_{\mu}\varphi + \tilde{A}_{\mu}\,.
\eeq
We count $\xi_{\mu}$ as $\mathcal{O}(\partial^0)$ in the gradient expansion. In the usual way, we identify the chemical potential as the longitudinal component of the derivative
\beq
\mu = u^{\mu}\tilde{D}_{\mu}\varphi\,,
\eeq
and the spatial part gives a new scalar at zeroth order in derivatives,
\beq
\xi^2 = h^{\mu\nu}\xi_{\mu}\xi_{\nu}\,.
\eeq

The leading contribution to $S_{hydrostat} = S_0 + S_1 + \hdots$ is then
\beq
S_0 = \int d^dx \sqrt{\gamma}\,P\left( T,\mu,\xi^2\right)\,,
\eeq
and $P$ is the pressure. Varying, we find that the consequent one-point functions of $\mathcal{T}^{\mu\nu}$ and $\tilde{\mathcal{E}}^{\mu}$ are 
\begin{align}
\begin{split}
\tilde{\mathcal{E}}^{\mu} & = \varepsilon u^{\mu} -f\mu \xi^{\mu}\,,
\\
\mathcal{T}^{\mu\nu} & = P h^{\mu\nu} + \rho u^{\mu}u^{\nu} +f  \xi^{\mu}\xi^{\nu}-f\left( u^{\mu}\xi^{\nu}+u^{\nu}\xi^{\mu}\right) \,,
\end{split}
\end{align}
where
\beq
dP = sdT+\rho d\mu - \frac{1}{2}f d\xi^2\,, \qquad \varepsilon = T s + \mu \rho - P\,, \qquad \xi^{\mu}=h^{\mu\nu}\xi_{\nu}\,.
\eeq
These may be regarded as the constitutive relations in a covariant formulation of ideal superfluid hydrodynamics.

\subsection{The story at $g_s\neq 0$} 
\label{S:spin}

In Subsection~\ref{S:gs} we reviewed the introduction of a magnetic moment $g_s$ in two spatial dimensions. Crucially, the action of the Milne boosts on $A_{\mu}$ was modified as in~\eqref{E:modifiedMilne}.

Here we wish to briefly show that $g_s$ can be easily incorporated in the hydrostatic partition function. We do not see an obstruction to introducing $g_s$ in hydrodynamics, although it will end up being rather cumbersome. So we elect to study the partition function in lieu of a hydrodynamic analysis.

To proceed we recall that we can use $u^{\mu}$ to define a Milne-invariant $U(1)$ and gravitational connection via~\eqref{E:gsMilneD}. We use this covariant derivative in order to manifest boost invariance. From the Milne-invariant $U(1)$ connection $A_g$ we define a modified chemical potential\footnote{Note that $A_g$ and $\mu_g$ include terms with zero and one derivative, so that the Milne symmetry acts on terms at different order in the gradient expansion. A reasonable gradient expansion can still be formed by classifying tensors with a minimum number of derivatives.}
\beq
\mu_g = \frac{K^{\mu}(A_g)_{\mu}+\Lambda_K}{K_0} = \mu + \frac{g_s}{4m}\varepsilon^{\mu\nu\rho}\partial_{\mu}\left( n_{\nu}u_{\rho}\right)\,.
\eeq
The Milne-invariant data is then $(T,\mu_g,u^{\mu},n_{\mu},h^{\mu\nu},\tilde{h}_{\mu\nu})$, and we take derivatives in a manifestly invariant way via $D_g$.

We decompose
\beq
(D_g)_{\mu}u^{\nu} = - n_{\mu}E_g^{\nu} + \frac{1}{2}(B_g)_{\mu}{}^{\nu} + \tilde{h}_{\mu\rho}\sigma_g^{\mu\rho} + \frac{\tilde{P}^{\mu}_{\nu}}{d-1}\vartheta_g\,,
\eeq
and define the local electric field $E_g^{\mu}$, \&c in the same way as above. (It is easy to see that $\sigma_g^{\mu\nu}=\sigma^{\mu\nu}$ and $\vartheta_g=\vartheta$.) One can use that $\delta_K$ fixes the background to deduce various interrelations like those in~\eqref{E:1derivativeEquil}, 
\begin{align*}
\dot{T} & = 0\,, &\dot{\left( \frac{\mu_g}{T}\right)} & = 0\,, & \vartheta_g & = 0\,,
\\
(E^n)^{\mu} + \frac{D_g^{\mu}T}{T}& = 0\,, & E_g^{\mu}-TD_g^{\mu}\left( \frac{\mu_g}{T}\right) & = 0\,, & \sigma_g^{\mu\nu} & = 0\,.
\end{align*}
To first order in gradients, we then have
\beq
W = \int d^3x \sqrt{\gamma}\left\{ P + \tilde{f}_1 \mathcal{B}_g + \tilde{f}_2\mathcal{B}_n + \mathcal{O}(\partial^2)\right\}\,,
\eeq
where $P$ and the $\tilde{f}_i$'s are functions of $T$ and $\mu_g$ and $\mathcal{B}_g=\frac{1}{2}\varepsilon^{\mu\nu\rho}n_{\mu}(F_g)_{\nu\rho}$.

\section{Discussion} 
\label{S:discuss}

In this manuscript we have discussed various aspects of non-relativistic hydrodynamics and hydrostatics. To conclude, we outline some natural lines of work which are suggested by our analysis.

\begin{enumerate}

\item As we pointed out in the Introduction, there is a very practical reason for coupling hydrodynamics to a background spacetime. Namely, by solving the fluid equations in a weakly curved background, one gets access to all fully retarded correlation functions, at least in the hydrodynamic limit. It would be interesting to study the linear response one gets for parity-violating fluids in two spatial dimensions, and in particular to obtain Kubo formulae for the parity-violating response coefficients. As this paper was already getting too long, we have elected to do this in future work.

\item Another natural direction is to study the hydrodynamics of systems coupled to an $\mathcal{O}(1)$ background magnetic field, along the lines of the analysis in~\cite{Geracie:2014nka} for lowest Landau level fluids. For now, we will only say that some care must be taken when constructing such a magnetohydrodynamics. It is well known that the $B\to 0$ and $\omega\to 0$ limits do not commute. One manifestation of this result is that the in-plane fluctuations of the velocity are gapped at $B\neq0$ (they become ``cyclotron modes''), in which case the low-energy effective description does not include them. Any proper magnetohydrodynamics, like that presented in~\cite{Geracie:2014nka}, should account for this.

\item In Subsection~\ref{S:Z1derivative}, we studied the one-derivative part of the hydrostatic partition function for theories in three spatial dimensions. There, we found that the partition function can include three distinct Chern-Simons terms on the spatial slice, which lead to hydrodynamic response parameterized by three parity-violating constants. This situation is eerily similar to that of relativistic hydrodynamics in three spatial dimensions. There, one can also have three constants which govern one-derivative, parity-violating response, but CPT kills two of the constants and the remaining one is related to the gravitational anomalies~\cite{Jensen:2012kj} (at least in a normal phase). What happens here for non-relativistic theories? Can we find simple theories for which any of these constants is nonzero? Are there corresponding anomalies?

\item Using our machinery, it should be straightforward to classify the hydrodynamics of normal fluids at second order in derivatives as in~\cite{Bhattacharyya:2012nq}, as well as superfluids at first order in derivatives as in~\cite{Bhattacharya:2011tra}. Here, we simply note that in the relativistic setting, the complete theory of second-order hydrodynamics (as well as first order superfluid hydrodynamics) is much richer than had been anticipated previously.

\item In this work, we have not considered further constraints that arise for the hydrodynamics of a conformal fluid, like unitary Fermi gases. Conformal non-relativistic theories are invariant under a ``Weyl'' symmetry (see~\cite{Jensen:2014aia} for a discussion in terms of Newton-Cartan geometry), much like relativistic conformal theories. It should be straightforward to manifest the Weyl symmetry in hydrodynamics. The simplest approach is probably that of~\cite{Loganayagam:2008is}, by redefining the covariant derivative to be Weyl-invariant using the fluid variables.

\end{enumerate}

\section*{Acknowledgments}

We thank A.~Abanov and D.~Son for useful discussions. We are especially grateful to K.~Balasubramanian, A.~Gromov, and A.~Karch for the same. The author would like to thank the organizers of the ``2014 Simons Summer Workshop in Mathematics and Physics'' at the Simons Center for Geometry and Physics for their hospitality during which most of this work was completed. The author was supported in part by National Science Foundation under grant PHY-0969739.

\begin{appendix}

\section{A comparison with previous results}
\label{A:comparison}

In this Appendix we compare our results for parity-violating hydrodynamics in two spatial dimensions with those previously reported in~\cite{Kaminski:2013gca} and~\cite{Banerjee:2014mka}. We find that the work of~\cite{Kaminski:2013gca} is incommensurate with our results, and that the work of~\cite{Banerjee:2014mka} trivially matches ours. 

\subsection{Comparing with Kaminski and Moroz}
\label{A:kaminski}

We begin with~\cite{Kaminski:2013gca}. Those authors do several things. Here, we compare with their study of the large $c$ limit of first-order, parity-violating relativistic hydrodynamics~\cite{Jensen:2011xb}. They work in a flat background and a nonzero $\mathcal{O}(\partial^0)$ background $A_{\mu}$. In the language of our work, the resulting non-relativistic hydrodynamics is coupled to a flat Newton-Cartan geometry
\beq
\label{E:flat}
n_{\mu}dx^{\mu} = dx^0\,, \qquad h_{\mu\nu}dx^{\mu}\otimes dx^{\nu} = \delta_{ij}dx^i\otimes dx^j\,, 
\eeq
with non-vanishing $A$. To convert from their notation to ours, we convert the spatial velocity vector $u^i$ into a spacetime vector via
\beq
u^{\mu}\partial_{\mu} = \partial_0 + u^i\partial_i\,.
\eeq
Their Ward identities can be written as
\beq
\label{E:wardKamMor}
\partial_{\mu}\mathcal{E}^{\mu} = v^{\mu}F_{\mu\nu}J^{\nu}\,, \qquad \partial_{\nu}\mathcal{T}^{\mu\nu}  = F^{\mu}{}_{\nu}J^{\nu}\,, 
\eeq
where $F^{\mu}{}_{\nu}=h^{\mu\rho}F_{\rho\nu}$. Their constitutive relations can be written in the language of our work (our $\mathcal{B}$ is their sum of external magnetic field and vorticity, $B+\Omega_{nr}$) as
\begin{align}
\nonumber
J^{\mu} & =\left(  \rho + \frac{\partial m}{\partial\mu}\mathcal{B}\right) u^{\mu} -\frac{1}{T}\left( m+ \frac{\varepsilon+P}{\rho}\frac{\partial m}{\partial\mu} \right)\varepsilon^{\mu\nu\rho}n_{\nu}\partial_{\rho}T= \mathcal{N}u^{\mu} + q^{\mu}\,,
\\
\nonumber
\mathcal{T}^{\mu\nu} & = \left( P - \zeta \vartheta\right)h^{\mu\nu} + \left( \rho + \frac{\partial m}{\partial\mu}\mathcal{B}\right)u^{\mu}u^{\nu} - \eta \sigma^{\mu\nu}-\tilde{\eta}\tilde{\sigma}^{\mu\nu}\,,
\\
\label{E:KamMorConst}
\mathcal{E}^{\mu} & = \left( \varepsilon + \frac{1}{2}\rho^2 \right)u^{\mu} + \left( T \frac{\partial m}{\partial T}+\mu \frac{\partial m}{\partial\mu}-m + \frac{1}{2}u^2 \frac{\partial m}{\partial\mu}\right)\mathcal{B}u^{\mu} - \left( \eta \sigma^{\mu\nu}+\tilde{\eta}\tilde{\sigma}^{\mu\nu}\right)u_{\nu}
\\
\nonumber
& \qquad + \left( P - \zeta \vartheta\right) P^{\mu}_{\nu}u^{\nu} -\kappa \partial^{\mu}T+ \tilde{\kappa} \varepsilon^{\mu\nu\rho}n_{\nu}\partial_{\rho}T -u_{\nu}q^{\nu}u^{\mu} - \frac{1}{2}u^2q^{\mu} 
\\
\nonumber
& \qquad \qquad -\left( m+ \frac{\varepsilon+P}{\rho}\frac{\partial m}{\partial\mu} \right)\left\{ \varepsilon^{\mu\nu\rho}n_{\nu}F_{\rho\sigma}u^{\sigma}+\frac{1}{T}\varepsilon^{\mu\nu\rho}u_{\nu}\partial_{\rho}T\right\} \,,
\end{align}
where $m$ should be understood as the magnetic susceptibility of the source-free thermal state. They restrict $m$ so that
\beq
\label{E:theWeirdKamMorConstraint}
m+ \frac{\varepsilon+P}{\rho}\frac{\partial m}{\partial\mu} =f(T)\,,
\eeq
i.e. that combination is independent of $\mu$. So the dependence of $m$ on $\mu$ is completely fixed by thermodynamics.

These results differ in two fundamental ways from those we obtained in Section~\ref{S:2dFluids}, and we do not see how to reconcile their results with Newton-Cartan geometry and Milne boosts. The difficulties are:
\begin{enumerate}
\item The Ward identities~\eqref{E:wardKamMor} enforced by Kaminski and Moroz have the same schematic form as the Ward identities~\eqref{E:ward} that follow from coupling to Newton-Cartan geometry, at least in the flat background~\eqref{E:flat}. In deriving the latter, the longitudinal component of $\mathcal{T}^{\mu\nu}$ is the number current $J^{\mu}$ that appears on the right-hand-side of $\partial_{\nu}\mathcal{T}^{\mu\nu}=F^{\mu}{}_{\nu}J^{\nu}$~\cite{Jensen:2014aia}. However,~\eqref{E:KamMorConst} leads to
\beq
\mathcal{T}^{\mu\nu}n_{\nu} = \rho u^{\mu} \neq J^{\mu}\,.
\eeq
Consequently, their stress tensor Ward identity differs from the one obtained when demanding coordinate reparameterization invariance. More precisely, the longitudinal component of the stress tensor Ward identity coincides with conservation of $J^{\mu}$ on account of~\eqref{E:theWeirdKamMorConstraint}, but the spatial components differ.

\item The energy current in~\eqref{E:KamMorConst} transforms under Milne boosts in a way which is rather different from what one gets in Newton-Cartan geometry,~\eqref{E:deltaEMilne}. In particular, it cannot be written as the sum of a boost-invariant part and $\left( u_{\nu}-\frac{1}{2}n_{\nu}u^2\right)\mathcal{T}^{\mu\nu}$ as in~\eqref{E:totalEnergy}. The problematic terms are the $- u_{\nu}q^{\nu}u^{\mu} - \frac{1}{2}u^2q^{\mu}$ components in the second line of $\mathcal{E}^{\mu}$ (we saw earlier that both terms are present with the opposite sign in the total energy current~\eqref{E:totalEnergy}) and the entire last line of~\eqref{E:KamMorConst}.

\end{enumerate}

Because of these difficulties, we do not try to further match our results against those obtained in~\cite{Kaminski:2013gca}, although it is easy to see some similarities and discrepancies between the constitutive relations above~\eqref{E:KamMorConst} and ours~\eqref{E:flatEckhart2d}. However, we do note that the parity-preserving part of their result is completely consistent with our analysis.

To summarize, the results obtained by Kaminski and Moroz violate the symmetries that we claim should be imposed when coupling Galilean-invariant theories to a background (according to our earlier proposal~\cite{Jensen:2014aia}).

\subsection{Comparing with Banerjee, et al}
\label{A:banerjee}

We continue by reviewing the work of~\cite{Banerjee:2014mka}. They perform an analysis along the lines of that in~\cite{Rangamani:2008gi}, wherein they perform a null reduction of first-order relativistic hydrodynamics in $(3+1)$-dimensions. By null reduction, we mean that they couple their hydrodynamics to a background with a null symmetry, demand that all of the hydrodynamic variables do not depend on the symmetry direction, and then reduce to obtain a $(2+1)$-dimensional hydrodynamics. They consider general constitutive relations in the relativistic parent, only imposing the positivity of entropy production in the resulting non-relativistic hydrodynamics.

In our companion paper~\cite{Jensen:2014aia}, we showed in detail how such a null reduction leads to Newton-Cartan geometry and the Milne boosts. So we expect that this approach leads to a non-relativistic hydrodynamics consistent with the symmetries of the problem.

The relativistic parent hydrodynamics considered in~\cite{Banerjee:2014mka} is that of a theory with $N_f$ $U(1)$ symmetry currents $J_A^{\mu}$ in flat space, coupled to background gauge fields $(A_A)_{\mu}$. Crucially, the authors choose to work in the Landau frame. At first order in derivatives, the relativistic constitutive relations they consider are
\begin{align*}
T^{\mu\nu} &= \varepsilon u^{\mu}u^{\nu} + (P-\zeta \vartheta) \Delta^{\mu\nu} - \eta \sigma^{\mu\nu}\,,
\\
J_A^{\mu} & = \rho_A u^{\mu} + \sigma_{AB} \left( F_B^{\mu\nu}u_{\nu} - T \Delta^{\mu\nu}\partial_{\nu}\left( \frac{\mu_B}{T}\right)\right) + \frac{\xi_{AB}}{2} \varepsilon^{\mu\nu\rho\sigma}u_{\nu}(F_B)_{\rho\sigma} + \xi_A \varepsilon^{\mu\nu\rho\sigma}u_{\nu}\partial_{\rho}u_{\sigma}\,,
\end{align*}
where $\Delta^{\mu\nu}=g^{\mu\nu}+u^{\mu}u^{\nu}$ is the transverse projector and the $\mu_A$ are the various chemical potentials. The authors do not make any assumptions about the dependence of the transport coefficients on $T$ and the $\mu_A$.

Performing the light-cone reduction gives a non-relativistic fluid mechanics with the $U(1)$ particle number symmetry (whose charge is the momentum along the null symmetry) along with $N_f$ $U(1)$ global symmetries. In order to match to our work, where there are no additional global symmetries, we must take $N_f=0$. But in this case the only thing left is the stress tensor. This is just the usual stress tensor of neutral, first-order relativistic hydrodynamics that was already reduced in~\cite{Rangamani:2008gi}, and so the light-cone reduction at $N_f=0$ simply gives the first-order, parity-preserving hydrodynamics we discussed in Subsection~\ref{S:1stOrder}. In this sense, the work of~\cite{Banerjee:2014mka} trivially matches our results.

There is a way to modify the approach taken in~\cite{Banerjee:2014mka} which may lead to parity-violating non-relativistic hydrodynamics at $N_f=0$. Do not use the freedom to redefine the relativistic fluid variables (it is not clear to us if that freedom commutes with the null reduction), and so take the stress tensor to be
\beq
T^{\mu\nu}=\varepsilon u^{\mu}u^{\nu} +(P-\zeta\vartheta)\Delta^{\mu\nu} + 2\xi u^{(\mu}\varepsilon^{\nu)\rho\sigma\tau}u_{\rho}\partial_{\sigma}u_{\tau} - \eta \sigma^{\mu\nu}\,.
\eeq
Couple the hydrodynamics to the most general background with a null isometry,
\beq
G = 2n_{\mu}dx^{\mu}(dx^-+A) + h_{\mu\nu}dx^{\mu}dx^{\nu}\,,
\eeq
where the component functions depend on $x^{\mu}$ with $\mu=0,1,2$ and not $x^-$, and then reduce on $x^-$. Then impose the local second Law. Our na\"ive expectation is that $\xi$ will become the function $\mathcal{M}$ that parameterizes part of the transport we found in Section~\ref{S:2dFluids}.

\end{appendix}

\bibliography{ncRefs}
\bibliographystyle{JHEP}

\end{document}